\documentclass[12pt]{article}
\usepackage{amsmath}
\usepackage{graphicx,psfrag,epsf}
\usepackage{enumerate}
\usepackage{natbib}
\usepackage{url} 

\usepackage{amssymb}
\usepackage{bm}
\usepackage{threeparttable}
\usepackage{multirow}
\usepackage{enumerate}

\usepackage{ulem}
\usepackage{bbm}

\usepackage{adjustbox}
\usepackage{multirow}
\usepackage{booktabs}
\usepackage{graphicx}
\usepackage{lscape}

\usepackage{algorithm,setspace}
\usepackage{algorithmic}

\makeatletter
\newenvironment{breakablealgorithm}
{
	\begin{center}
		\refstepcounter{algorithm}
		\hrule height.8pt depth0pt \kern2pt
		\renewcommand{\caption}[2][\relax]{
			{\raggedright\textbf{\ALG@name~\thealgorithm} ##2\par}%
			\ifx\relax##1\relax 
			\addcontentsline{loa}{algorithm}{\protect\numberline{\thealgorithm}##2}%
			\else 
			\addcontentsline{loa}{algorithm}{\protect\numberline{\thealgorithm}##1}%
			\fi
			\kern2pt\hrule\kern2pt
		}
	}{
		\kern2pt\hrule\relax
	\end{center}
}
\makeatother

\def\beq{\begin{equation}}
\def\eeq{\end{equation}}
\def\beqr{\begin{eqnarray}}
\def\eeqr{\end{eqnarray}}
\def\beqrs{\begin{eqnarray*}}
\def\eeqrs{\end{eqnarray*}}
\def\bet{\begin{theorem}}
\def\eet{\end{theorem}}
\def\bel{\begin{lemma}}
\def\eel{\end{lemma}}
\def\bep{\begin{proposition}}
\def\eep{\end{proposition}}
\def\bg{\begin{figure}[tbph]\begin{center}}
\def\eg{\end{center}\end{figure}}

\def\bc{\begin{center}}
\def\ec{\end{center}}

\def\mR{\mathbb{R}}

\def\argmin{\mbox{argmin}}

\def\beq{\begin{equation}}
\def\eeq{\end{equation}}
\def\beqr{\begin{eqnarray}}
\def\eeqr{\end{eqnarray}}
\def\beqrs{\begin{eqnarray*}}
\def\eeqrs{\end{eqnarray*}}

\newtheorem{theorem}{Theorem}
\newtheorem{lemma}{Lemma}

\newtheorem{assumption}{Assumption}

\newcommand{\blind}{0}

\begin{document}

\def\spacingset#1{\renewcommand{\baselinestretch}%
{#1}\small\normalsize} \spacingset{1}


\if0\blind
{
  \title{\bf Network Model Averaging Prediction for Latent Space Models by $K$-Fold Edge Cross-Validation}
  \author{Yan Zhang\thanks{Yan Zhang, Jun Liao and Xinyan Fan are co-first authors and contributed equally to this work.} \hspace{.2cm}\\
    School of Economics, Xiamen University\\
    Jun Liao\thanks{
    Jun Liao’s work was partially supported by the National Natural Science Foundation of China (Grant No. 12001534).}\hspace{.2cm}\\
    School of Statistics, Renmin University of China\\
    Xinyan Fan\thanks{
    Xinyan Fan’s research was supported by the National Natural Science Foundation of China (NSFC, 12201626), the MOE Project of Key Research Institute of Humanities and Social Sciences (22JJD110001), and the Public Computing Cloud, Renmin University of China.}\hspace{.2cm}\\
    School of Statistics, Renmin University of China\\
    Kuangnan Fang\thanks{Corresponding author. E-mail address: xmufkn@xmu.edu.cn.
    Kuangnan Fang’s research was supported by the National Natural Science Foundation of China (72071169,72233002), and the Fundamental Research Funds for the Central Universities (20720231060).}\hspace{.2cm}\\
    School of Economics, Xiamen University\\
    and \\
    Yuhong Yang \\
    Yau Mathematical Sciences Center, Tsinghua University}
  \maketitle
} \fi

\if1\blind
{
  \bigskip
  \bigskip
  \bigskip
  \begin{center}
    {\LARGE\bf Network Model Averaging Prediction for Latent Space Models by $K$-Fold Edge Cross-Validation}
\end{center}
  \medskip
} \fi

\bigskip
\begin{abstract}
In complex systems, networks represent connectivity relationships between nodes through edges. Latent space models are crucial in analyzing network data for tasks like community detection and link prediction due to their interpretability and visualization capabilities. However, when the network size is relatively small, and the true latent space dimension is considerable, the parameters in latent space models may not be estimated very well. To address this issue, we propose a Network Model Averaging (NetMA) method tailored for latent space models with varying dimensions, specifically focusing on link prediction in networks. For both single-layer and multi-layer networks, we first establish the asymptotic optimality of the proposed averaging prediction in the sense of achieving the lowest possible prediction loss. Then we show that when the candidate models contain some correct models, our method assigns all weights to the correct models. Furthermore, we demonstrate the consistency of the NetMA-based weight estimator tending to the optimal weight vector. Extensive simulation studies show that NetMA performs better than simple averaging and model selection methods, and even outperforms the ``oracle'' method when the real latent space dimension is relatively large. Evaluation on collaboration and virtual event networks further emphasizes the competitiveness of NetMA in link prediction performance.
\end{abstract}

\noindent%
\vfill

\newpage
\spacingset{1.8} 
\section{Introduction}
\label{sec:intro}
Networks are widely used to represent connectivity relationships between entities of a complex system. The entities are often called nodes, and the connections among them are called edges. Network data differ from traditional data in that they not only record features for each node but also provide information on the relationships among all nodes under study, sometimes along with additional pairwise features \citep{ma2020universal}. In an increasingly connected world, network data are ubiquitous. A common example is the social network, where nodes are individuals, groups, and organizations which are linked by various types of inter-dependencies. These include social contacts, familial ties, conflicts, financial transactions, and mutual membership in organizations \citep{serrat2017social}. Exploring the structures of networks can yield valuable insights into the formation of relationships or the mechanisms of interaction between individuals. Network analysis has been found wide applications in various scenarios including economy on understanding international trade patterns \citep{dong2021optimization}, biology such as gene-regulatory network \citep{panni2020non}, public health on studying disease transmission networks \citep{jo2021social} and society on detecting insurance fraud \citep{oskarsdottir2022social}.

In this work, we are concerned with how to estimate the connection probabilities between nodes when their links are missing. The issue of missing data in networks is a common problem due to the time and resource-intensive nature of data collection \citep{10.1093/restud/rdae088}. For instance, in social networks, missing links can arise from several resources, such as (1) the boundary specification problem, where the inclusion rules for actors or relations in a network study are unclear; (2) respondent inaccuracy, where individuals may fail to report actual connections; (3) non-response in network surveys, where participants may not fully engage with the survey process \citep{kossinets2006effects}. In protein-protein interaction networks, biological experiments
are time-consuming, expensive and inefficient, leading to missing connections \citep{li2016predicting}. Additionally, in dynamic networks where connections evolve over time, predicting whether a connection might form in the future is crucial for understanding network dynamics \citep{song2022link}. Link prediction is a fundamental challenge that seeks to estimate the probability of a connection between two nodes, utilizing both observed links and the attributes of those nodes. Estimating these probabilities helps us gain insights into the underlying structure and organization of the network \citep{zhang2017estimating}. One effective approach for estimating connection probabilities from a single realization of network data is to fit a suitable statistical model on them. The existing network models include the random graph model \citep{Erdos:1959:pmd,gilbert1959random}, the $p_1$ model \citep{holland1981exponential}, the stochastic block model (SBM) \citep{holland1983stochastic}, the latent space model \citep{hoff2002latent}, and many variants of them \citep{airoldi2008mixed,karrer2011stochastic,sewell2015latent}. Among them, the latent space model has gained much attention due to its interpretable structure, as well as its capacity to offer meaningful data visualization and provide abundant information for downstream tasks such as community detection, node classification and link prediction \citep{tang2023population,gwee2022latent}. The latent space model assumes that each node can be represented by a latent vector in some low-dimensional metric space. Nodes that are ``close'' in the latent space tend to be connected. Such latent space models account for transitivity, node heterogeneity, homophily for nodal attributes and even node clustering \citep{zhang2022directed} and edge balance \citep{tang2023population}.

Nonetheless, the dimension of the latent space in the latent space model is usually unknown. Typically, the dimension is set to be two for the sake of facilitating visualization and interpretation \citep{zhang2020flexible,sosa2021review,tang2023population}. This method is straightforward and eliminates the process of selecting the dimension. However, we need to acknowledge that this is a somewhat arbitrary decision, potentially resulting in an inadequate or overly intricate depiction of the underlying network structure. Therefore, how to choose the dimension of the latent space is a meaningful problem. \cite{durante2014nonparametric} included a shrinkage hyperprior that can be used to determine the dimension of the latent space autonomously. \cite{gwee2022latent} proposed a latent shrinkage position model to facilitate automatic inference on the dimensionality of the latent space. Additionally, the choice of the dimension of the latent space can be viewed as a model selection issue. Some model selection tools can be used to solve this problem. \cite{oh2001bayesian} proposed a Bayesian dimension choice criterion for multidimensional scaling (MDS), called MDSIC. \cite{oh2007model} extended MDSIC and proposed a new Bayesian selection criterion, named MIC, for choosing both the dimension of the object configuration and the number of clusters. In addition to Bayesian-based criteria, cross-validation is a useful tool for model selection and parameter tuning. \cite{hoff2007modeling} used cross-validation to compare the performance of different latent variable models under a range of dimensions. Furthermore, \cite{li2020network} introduced a general edge cross-validation (ECV) algorithm for networks, primarily focusing on partitioning edge pairs rather than nodes. This algorithm can be applied to a wide range of network model selection tasks with theoretical guarantees. 

However, a potential problem with model selection methods is that when the network size is relatively small but the true dimension of the latent space is relatively large, the latent space model may not be estimated very well. The model and the selection process have higher uncertainty, leading to relatively considerable estimation loss of the connection probabilities. To address this issue, a natural idea is to try different dimensions of the latent variables and combine the estimation results. Indeed, one method to mitigate the negative effects of model instability and uncertainty is through model averaging. This approach makes use of the competing models and safeguards against the selection of a poor model. There are two main types of model averaging methods, i.e., Bayesian model averaging and frequentist model averaging. Bayesian model averaging has been studied for a long time, both in statistics and economics \citep{hoeting1999bayesian,fragoso2018bayesian}. In recent years, there has been a rapid development in frequentist model averaging \citep{buckland1997model,yang2001adaptive,hansen2007least,zhang2011focused,liao2020corrected,liao2021model}. In terms of optimal model averaging, \cite{hansen2012jackknife} introduced a jackknife model averaging (JMA), which determines model averaging weights by minimizing leave-one-out cross-validation for independent data, which is extended to dependent data by \cite{zhang2013model}. \cite{gao2016model} developed a model averaging method based on the leave-subject-out cross-validation for longitudinal data. Additionally, \cite{zhang2022model} proposed an averaging prediction which determines the weights through the $K$-fold cross-validation.

The existing model averaging approaches are mainly designed for structured data, while network data are typically unstructured data.  Unstructured data have no particular format, schema or structure, such as text, images and networks \citep{7359270}. It is worth noting that network data possess special characteristics, such as small sample size, growing parameter space and sparsity, which can cause some difficulties for the methodological design and theoretical analysis. In this paper, we propose a network model averaging (NetMA) method to allow for latent space models with different dimensions. We focus on the optimal model averaging method for link prediction in networks, which has not been explored before. The weights are obtained through the $K$-fold edge cross-validation. Compared to the typical $K$-fold cross-validation designed for the model averaging (e.g., \cite{zhang2022model}), the $K$-fold edge cross-validation here is more adaptive to the network data structure. Furthermore, we demonstrate that when the candidate models are all misspecified, our proposed method exhibits asymptotic optimality. Besides, it assigns all weights to the correct models asymptotically when the candidate models contain the correct models. Additionally, for the NetMA-based weight estimator, we obtain its convergence to the optimal weights. Extensive simulations and empirical applications are conducted to show the advantages of our method on both single-layer and multi-layer networks.

\section{Model Averaging Prediction for Single-Layer Networks}
\label{sec:single}
\subsection{Model framework}
\label{sec:single-model}
Suppose that we have an undirected and unweighted single-layer network, which can be represented by a binary adjacency matrix $A \in \{0,1\}^{N \times N}$, where $A_{ij} = A_{ji} = 1$ if node $i$ and node $j$ are connected, and $A_{ij}= A_{ji} =0$ otherwise. The diagonal elements of $A$ are set to be 0, i.e., $A_{ii}=0$. 
We assume the connectivity between each pair of nodes $i$ and $j$ is conditionally independent Bernoulli random variables, with $E(A)=P$ given certain latent variables and heterogeneity parameters. The objective of general network analysis is to estimate $P$ from $A$. However, we only observe a single realization $A$, which is different from the majority of structured data that we typically encounter. To address this difficulty, we need to make some additional structural assumptions.
We consider the following latent space model proposed by \cite{hoff2002latent}. For any $i,j=1, \dots, N$ and $i<j$, we have
\begin{equation}
    \begin{gathered}
	A_{ij} \stackrel{\text{ind.}}{\sim} \text{Bernoulli}\left(P_{ij}\right), \quad \text{with} \\
	\operatorname{logit}(P_{ij}) = \Theta_{ij} = \alpha_i + \alpha_j + z_i^\top z_j,
    \end{gathered}
    \label{equ: model1}
\end{equation}
where $\alpha_i \in \mR$ reflects the popularity of node $i$, $z_i \in \mathbb{R}^d$ denotes the latent vector of node $i$, and $\operatorname{logit}(x) = \log\{x/(1-x)\}$ for any $x \in (0,1)$. 
In matrix form, we have 
\begin{equation*}
    \Theta = \alpha 1_N^\top + 1_N \alpha^\top + ZZ^\top,
\end{equation*}
where $\alpha=(\alpha_1, \dots, \alpha_N)^\top$, $1_N$ is the all one vector in $\mR^N$ and $Z = (z_1,\dots,z_N)^\top \in \mR^{N \times d}$.  To ensure the identifiability of parameters in Model (\ref{equ: model1}), we assume the latent variables are centred, that is $JZ = Z$, where $J = I_N - 1_N 1_N^\top / N$. This constraint makes $Z$ identifiable up to an orthogonal transformation of its rows.

\noindent
{\bf Remark 1:} Model \eqref{equ: model1} is a basic and classic form of the latent space model. We use this model because we are mainly interested in the weighted average effect of latent spaces of different dimensions. Regarding the functional forms of $\ell(z_i,z_j)$, here we consider the inner product form, i.e., $\ell(z_i,z_j)=z_i^\top z_j$. Apart from this, \cite{hoff2002latent} mentioned the projection form $\ell(z_i,z_j)=z_i^\top z_j/\|z_j\|$ and the distance form $\ell(z_i,z_j) = \|z_i - z_j\|$. Both forms are suitable for our framework. Since most papers on latent space models currently use the inner product form \citep{zhang2022joint,tang2023population,li2023statistical}, we follow this choice as well. Note that the form of $\ell$ is not essential. The theoretical results hold as long as Assumption \ref{assumption1} below is satisfied.

\noindent
{\bf Remark 2:} A common feature in real-world networks is the phenomenon of homophily, which means that nodes sharing common characteristics are more likely to connect with each other \citep{kossinets2009origins}. For example, social scientists have found that school children tend to form friendships and playgroups when they have similar demographic characteristics \citep{mcpherson2001birds}. Therefore, we can incorporate node or edge-specific covariates which represent the similarity between nodes to reflect homophily. Specifically, we introduce the following model which is proposed by \cite{ma2020universal}. Assume that for any $i<j$, 
\begin{equation}
    \operatorname{logit}(P_{ij}) = \alpha_i + \alpha_j + \beta X_{ij} + z_i^\top z_j,
    \label{model with covariate}
\end{equation}
where $X_{ij}$ denotes the covariate of the edge between node $i$ and node $j$. We further require that $X_{ij} = X_{ji}$  to ensure symmetry, and $X_{ii}=0$ to avoid self-loops in the model. The value of $X_{ij}$ can be either binary, indicating whether nodes $i$ and $j$ share a common attribute (e.g., gender, country), or continuous, representing a distance or similarity measure (e.g., differences in age or similarities in hobbies). The estimation of this model has been studied in detail in \cite{ma2020universal}. 

Certainly, if we include covariates in our model, we would need additional assumptions, such as Assumption 4 regarding the stable rank of the covariate matrix in \cite{ma2020universal}. However, this is not fundamentally different from our model. The properties of the estimator obtained from the true candidate model have been extensively studied in the context of different forms of latent space models \citep{ma2020universal,zhang2022directed,zhang2022joint}. We conduct a simulation study on the case with covariates, which can be found in Section \ref{sec:sim-single}. 

As the network size increases, the number of parameters to be estimated also increases. Due to the large number of parameters to be estimated in Model (\ref{equ: model1}), the estimation problem is particularly challenging. In recent years, researchers have proposed estimation methods that can be divided into two categories: one is the Bayesian approach via Markov chain Monte Carlo and the other is the maximum likelihood estimation method \citep{kim2018review}. In the former category, the parameters are treated as random effects \citep{hoff2007modeling, handcock2007model,krivitsky2009representing}, and specific assumptions on the distribution of parameters are required. In the latter category, the parameters are treated as fixed effects. \cite{ma2020universal} was the first to adopt this idea and proposed an efficient projected gradient descent (PGD) algorithm for estimating the single-layer inner-product latent space model. Due to the absence of distributional assumptions and the efficiency and scalability of PGD, many researchers have utilized the algorithm to estimate the latent space model \citep{zhang2020flexible,zhang2022joint,lyu2023latent}. In this paper, we treat the parameters as fixed effects and adopt the PGD algorithm to estimate Model (\ref{equ: model1}).

We estimate the parameters $\alpha$ and $Z$ by minimizing the following conditional negative log-likelihood:
\begin{equation*}
    \mathcal{L}(\alpha,Z) = - \sum_{i,j} \log P(A_{ij}\mid\alpha,Z) = -\sum_{i,j} \left\{ A_{ij} \Theta_{ij} - f(\Theta_{ij})\right\},
\end{equation*}
where $f(x) = \log(1+\exp(x))$. Then we can adopt the PGD algorithm to obtain the estimators of the parameters. 

\subsection{Model averaging criterion}
\label{sec:single_ma_criterion}
Suppose that we have $M$ candidate models, where the dimension of the latent vectors in the $m$th ($m=1,\dots,M$) candidate model is $m$. Specifically, the $m$th candidate model is $\Theta = \alpha 1_N^\top + 1_N \alpha^\top + Z_{(m)}Z_{(m)}^\top$, where $Z_{(m)} \in \mR^{N \times m}$. For the $m$th candidate model, the common practice to estimate its parameters and assess the quality of estimation is to split the data into two parts: one for fitting the model and the other for computing the prediction error. To extend this idea to networks, we should figure out how to split the data in $A$, and how to deal with the resulting partial data which is no longer a complete network. We adopt the data-splitting strategy proposed by \cite{li2020network} that splits nodal pairs instead of nodes into different parts. Specifically, we randomly divide the nodal pairs $\Psi = \{(i,j):\ i,\ j=1,\dots,N\}$ into two parts symmetrically, which are denoted as $\Psi_1$ and $\Psi_2$, respectively. ``Symmetrically'' means that $(i,j)$ and $(j,i)$ are in the same part. Here we assume that the edge information on $\Psi_2$ is missing. We aim to predict the connection probability on $\Psi_2$ using the information from $\Psi_1$. Let $p$ be the proportion of edges in $\Psi_1$, i.e., $p = \lvert \Psi_1 \rvert/\lvert \Psi \rvert$.
Denote $A(\Psi_1)$ as the matrix obtained from $A$ by setting all entries of $A$ with indices in $\Psi_2$ to 0. Following the approaches of \cite{chatterjee2015matrix} and \cite{gao2020discussion}, we apply PGD algorithm to $A(\Psi_1)$ to get estimators of $\alpha$ and $Z_{(m)}$, which are denoted as $\widehat{\alpha}_{(m)} = (\widehat{\alpha}_{(m),1}, \dots, \widehat{\alpha}_{(m),N})^\top \in \mathbb{R}^N$ and $\widehat{Z}_{(m)} = (\widehat{z}_{(m),1}, \dots, \widehat{z}_{(m),N})^\top \in \mathbb{R}^{N \times m}$. According to Model \eqref{equ: model1}, we can obtain the estimators $\widehat{P}^0_{(m)}$ of the probability matrix $E(A(\Psi_1))$, which is a biased estimator of $P=E(A)$. Since we know the sampling probability, we can remove the bias by setting $\widehat{P}_{(m)} = \widehat{P}^0_{(m)}/p$ to be the estimators of $P$. We further constrain the estimators to fall between 0 and 1. 
Let $w=(w_1, \dots, w_M)^\top$ be the weight vector with $w_m \geq 0~(m=1,\dots,M)$ and $\sum_{m=1}^M w_m = 1$. The averaging prediction for $P_{ij},\ (i,j) \in \Psi_2$ is 
$$\widehat{P}_{ij}(w) = \sum_{m=1}^M w_m \widehat{P}_{(m),ij},$$ where $\widehat{P}_{(m),ij}$ is the estimated connection probability between nodes $i$ and 
$j$ for the $m$th candidate model.

To select model weights, our objective is to minimize the squared error, defined as $L(w) = \sum_{(i,j) \in \Psi_2}  \{\widehat{P}_{ij}(w)-P_{ij}\}^2 $ for the single layer network. However, the optimization of $L(w)$ depends on the true probability matrix, which is impractical to obtain. Therefore, rather than directly minimizing $L(w)$, we resort to selecting data-driven weights by the $K$-fold cross-validation criterion. However, it is obvious that nodes in the network are connected by edges, and partitioning nodes would require removing edges, destroying the network structure. Therefore, the traditional $K$-fold cross-validation used for model averaging is no longer applicable.
Here, we propose a $K$-fold edge cross-validation criterion for networks to select the model weights for link prediction. The key idea for the $K$-fold edge cross-validation is to split the nodal pairs into $K$ groups and treat each group as a testing set to evaluate the model performance. Then we describe the calculation of the $K$-fold edge cross-validation criterion and how to conduct link prediction with data-driven weights in detail. The proposed method proceeds as follows.

\begin{enumerate}[Step 1:]
    \item Divide the nodal pairs in $\Psi_1$ into $K$ groups equally. Let $G_k,\ k=1, \dots, K$, denote the set of the nodal pairs in the $k$th group. 
		
    \item For $k=1,\dots,K$,
	\begin{enumerate}[(a)]
		\item Exclude the nodal pairs in the $k$th group from $\Psi_1$ and use the remaining nodal pairs in $\Psi_1$ to calculate the estimators of $Z_{(m)}$ and $\alpha$ in the $m$th model ($m=1,\dots,M$), which are $\widetilde{Z}_{(m)}^{[-k]}$ and  $\widetilde{\alpha}_{(m)}^{[-k]}$, respectively.
			
		\item Calculate the predictions for observations within the $k$th group for each model. That is, we calculate the prediction of $P \circ S^{[k]}$ for the $m$th model by 
		$$\widetilde{P}_{(m)}^{[k]} = f_1 \Big (\widetilde{\alpha}_{(m)}^{[-k]} 1_N^\top +  1_N \widetilde{\alpha}_{(m)}^{[-k]\top} + \widetilde{Z}_{(m)}^{[-k]} \widetilde{Z}_{(m)}^{[-k]\top} \Big) \circ S^{[k]},$$ 
		where $\circ$ denotes the Hadamard product of two matrics, $f_1(x) = K/\{(1+\exp(-x))(K-1)\}$, and $S^{[k]} = \left(S^{[k]}_{ij}\right) = \left(\mathbbm{1}_{(i,j) \in G_k}\right) \in \mR^{N \times N}$, where $\mathbbm{1}_{(i,j) \in G_k}$ is an indicator function that equals 1 if the pair $(i,j)$ is in the set $G_k$ and 0 otherwise.
        \end{enumerate}
		
	\item Construct the $K$-fold edge cross-validation criterion
		$$CV(w) = \frac{1}{\lvert \Psi_1 \rvert} \sum_{k=1}^{K} \left\|A^{[k]}- \widetilde{P}^{[k]}(w) \right\|_F^2,$$
		where $A^{[k]}=A \circ S^{[k]}$ and $\widetilde{P}^{[k]}(w) =  \sum_{m=1}^{M} w_m \widetilde{P}_{(m)}^{[k]} $.
		
	\item Select the model weights by minimizing the $K$-fold edge cross-validation criterion, i.e., $\widehat{w} = \argmin_{w \in \mathcal{W}} CV(w)$, with $w = (w_1, \dots, w_M)^\top$ being a weight vector in the unit simplex in $\mR^M$, i.e., $\mathcal{W} = \left\{w \in [0,1]^M: \sum_{m=1}^{M} w_m = 1\right\}$. Thus, we can construct an averaging prediction for $P_{ij}$, $(i,j) \in \Psi_2$ through $\widehat{P}_{ij}(\widehat{w}) = \sum_{m=1}^{M} \widehat{w}_m \widehat{P}_{(m),ij}$.
\end{enumerate}

Minimizing $CV(w)$ can be transformed into a quadratic programming problem about $w$. Specifically, let $h_{ijk}:= (\widetilde{P}_{(1),ij}^{[k]}, \dots, \widetilde{P}_{(M),ij}^{[k]} )^\top$, $H:= \sum_{k=1}^{K} \sum_{i,j} h_{ijk} h_{ijk}^\top$, and $h:= 2 \sum_{k=1}^K (\langle A^{[k]}, \widetilde{P}_{(1)}^{[k]} \rangle, \dots, \langle A^{[k]}, \widetilde{P}_{(M)}^{[k]} \rangle )^\top$, where $\langle \cdot, \cdot \rangle$ denotes the sum of all elements in the Hadamard product of two matrices. Then we have 
$$\min_{w \in \mathcal{W}} CV(w) \Leftrightarrow \min_{w \in \mathcal{W}} \left(w^\top H w - h^\top w\right),$$
which is a quadratic function of $w$. Therefore, we can use quadratic programming to solve the $K$-fold edge cross-validation weights.

\subsection{Theoretical property}
\label{theory-single}
In this section, we first present an analysis of the asymptotic optimality of the proposed averaging prediction when the candidate models are all misspecified. When some candidate models are correctly specified, as will be shown later, the proposed method assigns all weights to the correct models asymptotically. Then we provide the convergence rate of the weight estimator towards the infeasible optimal weight vector. For the single layer network, Theorem \ref{theorem1} shows that the empirical $K$-fold edge cross-validation weights asymptotically minimize the loss function $L(w)$. The assumptions required for Theorem \ref{theorem1} are discussed as follows. All limiting processes in this section are with respect to $\lvert \Psi_1 \rvert \rightarrow \infty$.
	
\begin{assumption}
    Suppose that $M \leq N$. For $(i,j) \in \Psi_1$, there exist limiting value $P^*_{(m),ij}$ for $\widehat{P}_{(m),ij}$ such that
    $\sum_{(i,j) \in \Psi_1} (\widehat{P}_{(m),ij} - P^*_{(m),ij})^2 = O_p(NM)$ uniformly for $m=1, \dots, M$.
    \label{assumption1}
\end{assumption}

Assumption \ref{assumption1} guarantees that the estimator of $P_{(m),ij}$ in each candidate model has a limit $P^*_{(m),ij}$. Here, $P^*_{(m),ij}$ could be regarded as a pseudo-true value, which is not necessarily equal to the true value. Notice that $\widehat{P}_{(m),ij}$ and $\widetilde{P}^{[-k]}_{(m),ij}$ have the same limiting values $P^*_{(m),ij}$ because $\lvert \Psi_1 \rvert$ and $\lvert \Psi_1 \rvert - \lvert \Psi_1 \rvert/K$ have the same order for any $K \in \{2,\dots, \lvert \Psi_1 \rvert\}$. The existing literature on the latent space model \citep{zhang2022joint,ma2020universal} provided the upper bound on the estimation error of $\widehat{\Theta}_{(m)} = \text{logit} (\widehat{P}_{(m)})$, and showed that $\|\widehat{\Theta}_{(m)} - \Theta\|_F^2 = O_p(Nm)$. When the model is correctly specified, we have $\sum_{(i,j) \in \Psi_1} (\widehat{P}_{(m),ij} - P^*_{(m),ij})^2 \leq \|\widehat{\Theta}_{(m)} - \Theta\|_F^2/16 = O_p(Nm)$, and in this case Assumption \ref{assumption1} holds for fixed $M$.
    
Next, we introduce some notations associated with the limiting value $P^*_{(m),ij}$. The averaging prediction based on the limiting value is $P_{ij}^*(w) = \sum_{m=1}^{M} w_m P^*_{(m),ij}$. Similarly, the loss function based on the limiting value is defined as $L^*(w) =  \sum_{(i,j) \in \Psi_1}  \{P^*_{ij}(w)  -P_{ij}\}^2 $. The minimum loss in the class of averaging estimators based on the limiting value is $\xi^* = \inf_{w \in \mathcal{W}} L^*(w)$. The following assumption is about an upper bound on the expected nodal degree $D$, which is defined to satisfy $\max_{ij} P_{ij} \leq D/N$.

\begin{assumption}
    $N M \xi^{*-1} = o(1)$ and $N\max\{D, \log N\} \xi^{*-1} = O(1)$.
    \label{assumption3}
\end{assumption}

Assumption \ref{assumption3} constraints that $\xi^*$ grows faster than $NM$, and $N\max\{D, \log N\} \xi^{*-1}$ is finite. 
The former is similar to Assumption A3 of \cite{10.1214/17-AOS1538} and Assumption 5 of \cite{zhang2022model}. This assumption rules out the scenario where the model is correctly specified. Here, the correct models refer to candidate models whose dimensions of the latent vectors are greater than or equal to the true dimension. 
In other words, if one of the candidate models is correct, then Assumption \ref{assumption3} does not hold. We will discuss the case where the candidate models contain the correct models later.
	
We then justify that the weights selected by the $K$-fold edge cross-validation criterion are asymptotically optimal. In other words, the $K$-fold edge cross-validation weights asymptotically minimize the prediction loss.

\begin{theorem}
    Under Assumptions \ref{assumption1} and \ref{assumption3}, we have 
    $$\frac{L(\widehat{w})}{\inf _{w \in \mathcal{W}} L(w)} \rightarrow 1$$
    in probability.
    \label{theorem1}
\end{theorem}
	
The optimality statement in Theorem \ref{theorem1} is an important property of the model averaging estimator - that the model averaging estimator based on the $K$-fold edge cross-validation asymptotically achieves the lowest squared loss. The proof of Theorem \ref{theorem1} is presented in the Supplementary Material. 

Next, we discuss the situation where the candidate models contain the correct models. Specifically, denote $\mathcal{T}$ to be the subset of $\{1,\dots,M\}$ which contains the indices of the correct models. For example, when the true dimension is \(d_0\) and \(d_0 \leq M\), then \(\mathcal{T} = \{d_0, \dots, M\}\). Let $\widehat{\zeta} = \sum_{m \in \mathcal{T}} \widehat{w}_m$ be the sum of $K$-fold edge cross-validation weights assigned to the correct models. 
Denote $\mathcal{W}^s = \{w \in \mathcal{W}: \sum_{m \notin \mathcal{T}} w_m = 1\}$ to be the subset of $\mathcal{W}$ which assigns all weights to the misspecified models. We need the following assumption to consider the case where the candidate models contain the correct models.

\begin{assumption}
    $NM\left\{ \inf_{w \in \mathcal{W}^s} L^*(w) \right\}^{-1} = o(1)$ and $N \max\{D, \log N\} \left\{ \inf_{w \in \mathcal{W}^s} L^*(w) \right\}^{-1}$ $= O(1) $.
    \label{assumption4}
\end{assumption}

Assumption \ref{assumption4} gives the restriction on the growth rate of the minimum loss of the averaging estimator over all misspecified models. It is easy to see that Assumption \ref{assumption4} is equivalent to Assumption \ref{assumption3} when all candidate models are misspecified. Next, we show that $\widehat{\zeta} \rightarrow 1$ in probability under some regularity conditions.

\begin{theorem}
    Under Assumptions \ref{assumption1} and \ref{assumption4}, if $\mathcal{T}$ is not empty, we have $\widehat{\zeta} \rightarrow 1$ in probability.
    \label{theorem2}
\end{theorem}

\vspace{-0.4cm}
Theorem \ref{theorem2} implies that when the candidate models contain the correct models, the proposed method asymptotically assigns all weights to the correct models.

Next, we present the convergence rate of the $K$-fold edge cross-validation-based weights. We first introduce some notations. Define the squared risk function as
$R(w)  = \sum_{(i,j) \in \Psi_1} E (\widehat{P}_{ij}(w) - P_{ij})^2$. 
Let $\xi = \inf_{w \in \mathcal{W}}R(w)$, and denote the optimal weight vector $w^0 = \argmin_{w \in \mathcal{W}}R(w)$.  
Let $\lambda_{\min}(B)$ and $\lambda_{\max}(B)$ be the minimum and maximum singular values of a general real matrix $B$, respectively. Arrange $\{A_{ij},\ (i,j) \in \Psi_1\}$, $\{P_{ij},\ (i,j) \in \Psi_1\}$, $\{\widetilde{P}_{ij}(w) = \sum_{k=1}^{K} \widetilde{P}_{ij}^{[k]}(w),\ (i,j) \in \Psi_1\}$, $\{\widetilde{P}_{(m),ij} = \sum_{k=1}^{K} \widetilde{P}_{(m),ij}^{[k]},\ (i,j) \in \Psi_1\}$,  $\{\widehat{P}_{ij}(w),\ (i,j) \in \Psi_1\}$, $\{\widehat{P}_{(m),ij},\ (i,j) \in \Psi_1\}$, $\{P^*_{ij}(w),\ (i,j) \in \Psi_1\}$ and $\{P^*_{(m),ij},\ (i,j) \in \Psi_1\}$ in a same particular order, and denote them as vector $a_1 \in \mR^{\lvert \Psi_1 \rvert}$, $p_1 \in \mR^{\lvert \Psi_1 \rvert}$, $\widetilde{p}_1(w) \in \mR^{\lvert \Psi_1 \rvert}$, $\widetilde{p}_{1(m)} \in \mR^{\lvert \Psi_1 \rvert}$, $\widehat{p}_1(w) \in \mR^{\lvert \Psi_1 \rvert}$, $\widehat{p}_{1(m)} \in \mR^{\lvert \Psi_1 \rvert}$, $p^*_1(w) \in \mR^{\lvert \Psi_1 \rvert}$ and $p^*_{1(m)} \in \mR^{\lvert \Psi_1 \rvert}$, respectively. 
Denote $\widehat{\Lambda}_1 = (\widehat{p}_{1(1)}, \dots, \widehat{p}_{1(M)})$, $\Omega_1 = (p_1-\widehat{p}_{1(1)}, \dots , p_1-\widehat{p}_{1(M)})$, $\widehat{\Lambda} = \widehat{\Lambda}_1^\top \widehat{\Lambda}_1$ and $\Omega = \Omega_1^\top \Omega_1$. Theorem \ref{theorem3} shows the rate of $\widehat{w}$ tending to the infeasible optimal weight vector $w^0$. The following assumptions are needed to show this theorem.
	
\begin{assumption}
    There are two positive constants $\rho_1$ and $\rho_2$, such that $0 < \rho_1 < \lambda_{\min}(\widehat{\Lambda}/ \lvert \Psi_1 \rvert) \leq \lambda_{\max}(\widehat{\Lambda}/ \lvert \Psi_1 \rvert) < \rho_2 < \infty$, in probability tending to 1. 
    \label{assumption5}
\end{assumption}

\begin{assumption}
    $\lambda_{\max}(\Omega/\lvert \Psi_1 \rvert) = O_p(1)$.
    \label{assumption6}
\end{assumption}
	
\begin{assumption}
    $N^{1-4\kappa} \max\{D, \log N\} \xi^{-1} = o(1)$, $N^{1-4\kappa}M \xi^{-1} = o(1)$, and $M = o(N^{1/2})$ where $\kappa \in (0,1/2)$.
    \label{assumption7}
\end{assumption}
	
Similar to the Condition (C.4) in \cite{liao2020corrected}, Assumption \ref{assumption5} puts a constraint on the singular values of $\widehat{\Lambda} / \lvert \Psi_1 \rvert$. It requires that the minimum and maximum singular values of $\widehat{\Lambda} / \lvert \Psi_1 \rvert$ are bounded away from zero and infinity.
Assumption \ref{assumption6} requires that the maximum singular value of $\Omega/\lvert \Psi_1 \rvert$ is bounded in probability. The second part of Assumption \ref{assumption7} further restricts the relationship between $M$, $N$ and $\xi$. The third part of Assumption \ref{assumption7} indicates that the number of candidate models can increase with $N$ but at a rate with a constraint. Additionally, when $M$ is fixed, as long as the first part holds, the second and third parts hold naturally.

\begin{theorem}
    If $w^0$ is an interior point of $\mathcal{W}$, and Assumptions \ref{assumption1} and \ref{assumption5}-\ref{assumption7} are satisfied, then there exists a local minimizer $\widehat{w}$ of $CV(w)$ such that 
    \begin{equation}
	\left\| \widehat{w} - w^0 \right\| = O_p\left(\xi^{1/2} \lvert \Psi_1 \rvert^{-1/2+\kappa}\right),
	\label{equ:theorem3}
    \end{equation}
    where $\kappa$ is defined in Assumption \ref{assumption7}.
    \label{theorem3}
\end{theorem}
	
Theorem \ref{theorem3} gives the convergence rate of the estimated weights $\widehat{w}$ towards the optimal weights $w^0$, which is associated with the sample size $\lvert \Psi_1 \rvert$ and the minimized risk $\xi$. Specifically, the slower the rate of $\xi \rightarrow \infty$, the faster the rate of $\widehat{w} \rightarrow w^0$ as $\lvert \Psi_1 \rvert \rightarrow \infty$. The proof of Theorem \ref{theorem3} is presented in the Supplementary Material.


\section{Model Averaging Prediction for Multi-Layer Networks}
\label{sec:multi}

A multi-layer network is a collection of various networks connecting the same set of nodes. In many applications, some complex relationships can be characterized using multi-layer networks, such as social networks of friendships, work connections, as well as networks that evolve over time.
Here, we propose a model averaging method for link prediction in multi-layer networks, where the weights are determined by considering the performance of candidate models across all layers.

\subsection{Model framework}
Assume that the multi-layer networks are composed of $T$ networks over a common set of $N$ nodes. For $t=1, \dots, T$, the $t$th layer network is represented by an adjacency matrix $A^{(t)} \in \{0,1\}^{N \times N}$, where $A_{ij}^{(t)} = A_{ji}^{(t)} = 1$ if node $i$ and node $j$ are connected and $A_{ij}^{(t)} = A_{ji}^{(t)} = 0$ otherwise. Similarly, we consider using the inner-product latent space model. For any $t=1, \dots, T$, $i,j=1,\dots,N$ and $i<j$, we have
\begin{equation}
    \begin{gathered}
	A_{ij}^{(t)} \sim \text{Bernoulli}\left(P_{ij}^{(t)}\right), \quad \text{with} \\
	\operatorname{logit}(P_{ij}^{(t)}) = \Theta_{ij}^{(t)} = \alpha_i + \alpha_j + z_i^{(t)\top} z_j^{(t)},
    \end{gathered}
    \label{equ: model2}
\end{equation}
where $z_i^{(t)} \in \mR^{d_t}$. For simplicity, we assume all layers share the same $\alpha$, while the latent positions of nodes may vary across different layers. For presentation simplicity, we rewrite the model in matrix form, i.e.,
\begin{equation*}
    \Theta^{(t)} = \alpha 1_N^\top + 1_N \alpha^\top + Z^{(t)}Z^{(t)\top},
\end{equation*}
where $Z^{(t)} = (z_1^{(t)}, \dots, z_N^{(t)})^\top \in \mR^{N \times d_t}$. Denote $\mathbb{Z} = \{Z^{(1)}, \dots, Z^{(T)}\}$. To ensure the identifiability of parameters $\{\alpha, \mathbb{Z}\}$, we assume the latent variables are centered, that is $JZ^{(t)} = Z^{(t)}$. 

Then we develop a PGD algorithm for multi-layer networks to estimate the Model \eqref{equ: model2}. We first define the objective function as the negative conditional log-likelihood of $\{A^{(t)}\}_{t=1}^T$ under Model \eqref{equ: model2}:
\begin{equation*}
    \mathcal{L}_\dagger\left(\alpha, \mathbb{Z}\right) = -\sum_{t=1}^T \sum_{i=1}^N \sum_{j=1}^N \log P\left( A_{ij}^{(t)} \mid \alpha, \mathbb{Z}\right)
    = -\sum_{t=1}^T \sum_{i=1}^N \sum_{j=1}^N \left\{ A_{ij}^{(t)} \Theta_{ij}^{(t)} - f(\Theta_{ij}^{(t)}) \right\}.
\end{equation*}
Then we need to find the estimators of $\alpha$ and $\mathbb{Z}$ that minimize the objective function. The original PGD algorithm is designed for single-layer networks \citep{ma2020universal}. Here, we extend it to multi-layer networks, which is similar to the algorithm proposed by \cite{zhang2020flexible}. The procedure is summarized in Algorithm \ref{alg:algorithm1}. We adopt the initialization method proposed by \cite{ma2020universal} to obtain appropriate initial values. The original projected gradient descent algorithm is extended to the multi-layer network.

\begin{breakablealgorithm} 
\setstretch{1.35} 
    \caption{Projected Gradient Descent Algorithm for Parameter Estimation in Multi-layer Networks} 
    \label{alg:algorithm1} 
    \begin{algorithmic}[1] 
        \REQUIRE network adjacency matrix $\{A^{(t)}\}_{t=1}^T$; latent space dimension $\{d_t\}_{t=1}^T$; initial values $\alpha_0$ and $\mathbb{Z}_0= \{Z_0^{(1)}, \dots, Z_0^{(T)}\}$; step sizes $\eta_{\alpha}$ and $\eta_{Z}$; number of iterations $U$
        \ENSURE $\widehat{\alpha} = \alpha_U$, $\widehat{Z}^{(t)} = Z^{(t)}_U$ for $t=1, \dots, T$ 
        
        \FOR {$ u = 0, 1, \dots,U-1 $ }
            \FOR {$ t = 1, \dots, T $ }
                \STATE {$\Theta_u^{(t)} = \alpha_u 1_N^\top + 1_N \alpha_u^\top + Z_u^{(t)}Z_u^{(t)\top}$}
                \STATE {$Z^{(t)}_{u+1} = Z^{(t)}_{u} + 2 \eta_Z \left\{A^{(t)} - \sigma\left(\Theta^{(t)}_u\right)\right\} Z^{(t)}_u$ \COMMENT{$\sigma(x) = 1/\{1+\exp(-x)\}$}} 
                \STATE {$Z^{(t)}_{u+1} = JZ^{(t)}_{u+1}$}
            \ENDFOR

            \STATE {$\alpha_{u+1} = \alpha_{u} + 2 \eta_{\alpha} \left[\sum_{t=1}^T \left\{A^{(t)} - \sigma\left(\Theta^{(t)}_u\right)\right\}\right] 1_N$}
        \ENDFOR
    \end{algorithmic}
\end{breakablealgorithm}

\subsection{Model averaging criterion}
\label{sec:multilayer-cv}

Similar to the approach in Section \ref{sec:single_ma_criterion}, we consider $M$ candidate models, where the dimension of the latent space in the $m$th ($m=1,\dots,M$) model is $m$. The $m$th candidate model is $\Theta^{(t)} = \alpha 1_{N}^\top + 1_{N} \alpha^\top + Z_{(m)}^{(t)} Z_{(m)}^{(t)\top}$, where $Z_{(m)}^{(t)} \in \mR^{N \times m}$. We propose a $K$-fold edge cross-validation criterion for multi-layer networks to choose the model weights. The objective is to minimize $L_\dagger(w) = \sum_{t=1}^T \sum_{(i,j) \in \Psi_1}  \{\widehat{P}_{ij}^{(t)}(w)-P_{ij}^{(t)}\}^2$ by selecting model weights $w$ from the set $\mathcal{W}$, where $\widehat{P}_{ij}^{(t)}(w) = \sum_{m=1}^M w_m \widehat{P}^{(t)}_{(m),ij}$. In light of the unattainability of the objective function $L_\dagger(w)$, we opt for the determination of data-driven weights through the $K$-fold edge cross-validation criterion. The details of the procedure are given as follows. 
\begin{enumerate}[Step 1:]
    \item Partition the nodal pairs, denoted as $\Psi = \{(i,j):\ i,\ j=1,\dots,N\}$, into two parts symmetrically, namely $\Psi_1$ and $\Psi_2$. The estimator $\widehat{Z}_{(m)}^{(t)} \in \mathbb{R}^{N \times m}$ and $\widehat{\alpha}_{(m)} \in \mathbb{R}^N$ for the $m$th model is obtained by applying Algorithm \ref{alg:algorithm1} to $A^{(t)}(\Psi_1)$. Then we can compute the estimators of $P^{(t)}_{ij}$ for the $m$th model by $\widehat{P}_{(m),ij}^{(t)}=f_2\left(\widehat{\alpha}_{(m),i} +  \widehat{\alpha}_{(m),j} + \widehat{z}^{(t)\top}_{(m),i} \widehat{z}^{(t)}_{(m),j}\right)$, where $f_2(x) = \lvert \Psi \rvert /\{(1+\exp(-x)) \lvert \Psi_1 \rvert\}$, and $\widehat{z}^{(t)}_{(m),i} \in \mR^m$ denotes the $i$th row of $\widehat{Z}_{(m)}^{(t)}$.
    Subsequently, the nodal pairs in $\Psi_1$ are evenly divided into $K$ groups, with $G_k, \ k=1,\dots,K$, representing the set of nodal pairs in the $k$th group.
		
    \item For $k=1,\dots,K$,
	\begin{enumerate}[(a)]
		\item Exclude the nodal pairs in the $k$th group from $\Psi_1$, and utilize the remaining nodal pairs in $\Psi_1$ to compute the estimators of $Z_{(m)}^{(t)}$ and $\alpha$ in the $m$th model ($m=1,\dots,M$), which are $\widetilde{Z}_{(m)}^{(t)[-k]}$ and $\widetilde{\alpha}_{(m)}^{[-k]}$, respectively.
			
		\item Calculate the predictions for observations within the $k$th group for each model. That is, we calculate the prediction of $P^{(t)} \circ S^{[k]}$ by 
		$$\widetilde{P}_{(m)}^{(t)[k]} = f_1 \left (\widetilde{\alpha}_{(m)}^{[-k]} 1_N^\top +  1_N \widetilde{\alpha}_{(m)}^{[-k]\top} + \widetilde{Z}_{(m)}^{(t)[-k]} \widetilde{Z}_{(m)}^{(t)[-k]\top} \right) \circ S^{[k]}.$$
        \end{enumerate}
		
	\item  Construct the $K$-fold edge cross-validation criterion
		$$CV_{\dagger}(w) = \frac{1}{T \lvert \Psi_1 \rvert} \sum_{t=1}^{T} \sum_{k=1}^{K} \left\|A^{(t)[k]}- \widetilde{P}^{(t)[k]}(w) \right\|_F^2,$$
		where $A^{(t)[k]}=A^{(t)} \circ S^{[k]}$ and $\widetilde{P}^{(t)[k]}(w) =  \sum_{m=1}^{M} w_m \widetilde{P}_{(m)}^{(t)[k]} $.
		
	\item Select the model weights by minimizing the $K$-fold edge cross-validation criterion, i.e., $\widehat{w}_\dagger = \argmin_{w \in \mathcal{W}} CV_{\dagger}(w)$, where $\mathcal{W} = \left\{w \in [0,1]^M: \sum_{m=1}^{M} w_m = 1\right\}$. Subsequently, construct an averaging prediction for $P_{ij}^{(t)}$, $(i,j) \in \Psi_2$ through  $\widehat{P}_{ij}^{(t)}(\widehat{w}_\dagger) = \sum_{m=1}^{M} \widehat{w}_{\dagger m} \widehat{P}_{(m),ij}^{(t)}$.
\end{enumerate}

Minimizing $CV_{\dagger}(w)$ can be reformulated as a quadratic programming problem concerning $w$. Therefore, we can use quadratic programming to solve the $K$-fold edge cross-validation weights.

\subsection{Theoretical property}

For multi-layer networks, we also provide three theoretical properties for our method. The objective is to minimize $L_\dagger(w) = \sum_{t=1}^T \sum_{(i,j) \in \Psi_2}  \{\widehat{P}_{ij}^{(t)}(w)-P_{ij}^{(t)}\}^2$ by selecting model weights $w$ from the set $\mathcal{W}$, where $\widehat{P}_{ij}^{(t)}(w) = \sum_{m=1}^M w_m \widehat{P}_{(m),ij}^{(t)}$. In light of the unattainability of the objective function $L_\dagger(w)$, we opt for the determination of data-driven weights through the $K$-fold cross-validation criterion. Theorem \ref{theorem4} shows that the empirical $K$-fold cross-validation weights asymptotically minimize $L_\dagger(w)$. The assumptions required for Theorem \ref{theorem4} are discussed as follows.	

\begin{assumption}
    Suppose that $M \leq N$. For $(i,j) \in \Psi_1$, there exist limiting values $P_{(m),ij}^{(t)*}$ for $\widehat{P}_{(m),ij}^{(t)}$ such that $\sum_{t=1}^T \sum_{(i,j) \in \Psi_1}\left(\widehat{P}_{(m),ij}^{(t)} - P_{(m),ij}^{(t)*}\right)^2 = O_p(NMT)$ uniformly for $m=1, \dots, M$.
    \label{assumptionS1}
\end{assumption}

Assumption \ref{assumptionS1} implies that the estimator of $P_{(m),ij}^{(t)}$ of layer $t$ in the $m$th candidate model has a limit $P_{(m),ij}^{(t)*}$. Assume that $\|\Theta^{(t)}_{(m)}\|_{\max} \leq \mu$ uniformly for $t=1,\dots,T$ and $m=1,\dots,M$, where $\Theta_{(m)}^{(t)} = \text{logit}(P_{(m)}^{(t)})$, and  $\|\cdot\|_{\max}$ represents the maximum absolute value of entries in a matrix. According to the proof of Theorem 1 in \cite{zhang2020flexible}, when the model is correctly specified, we have $\sum_{t=1}^T \|\widehat{\Theta}^{(t)}_{(m)}-\Theta^{(t)}_{(m)}\|_F^2 = O_p(C_2NT + C_3 m^2(2N+T))$, where $C_2$ and $C_3$ depend on $\mu$. Next, we can obtain $\sum_{t=1}^T \sum_{(i,j) \in \Psi_1}(\widehat{P}_{(m),ij}^{(t)} - P_{(m),ij}^{(t)*})^2 \leq \sum_{t=1}^T \|\widehat{\Theta}^{(t)}_{(m)}-\Theta^{(t)}_{(m)}\|_F^2/16 = O_p(C_2^\prime NT + C_3^\prime m^2(2N+T))$, where $C_2^\prime=C_2/16$ and $C_3^\prime = C_3/16$. When we assume $\mu$ and $M$ to be fixed constants, and $M \leq T \leq N$, we have 
$\sum_{t=1}^T \sum_{(i,j) \in \Psi_1}(\widehat{P}_{(m),ij}^{(t)} - P_{(m),ij}^{(t)*})^2 = O_p(NMT)$ uniformly for $m=1,\dots,M$. Additionally, we define an upper bound on the expected nodal degree over the whole layers $D_\dagger$, which satisfies $\max_t \max_{ij} P_{ij}^{(t)} \leq D_\dagger/N$. 

\begin{assumption}
    $T N^{-C_{0_\dagger}} = o(1)$ for some positive constant $C_{0\dagger}$.
    \label{assumptionS2}
\end{assumption}

Assumption \ref{assumptionS2} is imposed to get the uniform order of $\|A^{(t)} - P^{(t)}\|_2$ for $t=1,\dots,T$, where $\|\cdot\|_2$ denotes the spectral norm of a matrix. Next, we denote the loss function for multi-layer networks based on the limiting value as $L_\dagger^*(w) = \sum_{t=1}^T \sum_{(i,j) \in \Psi_1} \{P^{(t)*}_{ij}(w) - P_{ij}^{(t)}\}^2 $, where $P_{ij}^{(t)*} (w)= \sum_{m=1}^M w_m P_{(m),ij}^{(t)*}$ and $P_{(m),ij}^{(t)*} = f_2 (\alpha^*_{(m),i} + \alpha^*_{(m),j} + z^{(t)*\top}_{(m),i} z^{(t)*}_{(m),j})$. The minimum loss in the class of averaging estimators based on the limiting value is denoted as $\xi_\dagger^* = \inf_{w \in \mathcal{W}} L_\dagger^*(w)$. 

\begin{assumption}
    $N M T \xi_\dagger^{*-1} = o(1)$ and $N \max\{D_\dagger, \log N\} T \xi_\dagger^{*-1} = O(1)$.
    \label{assumptionS3}
\end{assumption}

Assumption \ref{assumptionS3} requires that $\xi_\dagger^{*}$ grows faster than $NMT$, and $N \max\{D_\dagger, \log N\} T \xi_\dagger^{*-1}$ is finite. 

\begin{theorem}
    Under Assumptions \ref{assumptionS1}-\ref{assumptionS3}, we have
    $$\frac{L_\dagger(\widehat{w}_\dagger)}{\inf _{w \in \mathcal{W}} L_\dagger(w)} \rightarrow 1$$
    in probability.
    \label{theorem4}
\end{theorem}

Theorem \ref{theorem4} implies that the prediction loss $L_\dagger(w)$ is asymptotically minimized by the $K$-fold edge cross-validation weights $\widehat{w}_\dagger$.
The proof of Theorem \ref{theorem4} is given in the Supplementary Material. Note that when the candidate models contain the correct models, Assumption \ref{assumptionS3} does not hold. Further, we consider the case where there exists correctly specified models. 
We make the following assumption.
\begin{assumption}
    $N M T \{ \inf_{w \in \mathcal{W}^s} L^*_\dagger(w) \}^{-1} = o(1)$ and $NT \max\{D_\dagger, \log N\} \{ \inf_{w \in \mathcal{W}^s} L^*_\dagger(w) \}^{-1}$ $ = O(1) $.
    \label{assumptionS4}
\end{assumption}

Assumption \ref{assumptionS4} is similar to Assumption \ref{assumption4} in single-layer networks. Denote $\widehat{\zeta}_\dagger = \sum_{m \in \mathcal{T}_{\dagger}} \widehat{w}_{\dagger,m}$, where $\widehat{w}_{\dagger,m}$ is the $m$th element of $\widehat{w}_{\dagger}$, and $\mathcal{T}_{\dagger}$ is the index set of the correct models. 

\begin{theorem}
    Under Assumptions \ref{assumptionS1}, \ref{assumptionS2} and \ref{assumptionS4}, if $\mathcal{T}_{\dagger}$ is not empty, we have $\widehat{\zeta}_\dagger \rightarrow 1$ in probability.
    \label{theorem5}
\end{theorem}

Theorem \ref{theorem5} shows that the proposed method for multi-layer networks asymptotically assigns all weights to the correct models if the candidate models include the correct models. In the following content, we provide the convergence rate of the $K$-fold edge cross-validation-based weights. Firstly, we give some notations. For $t=1,\dots,T$, denote the squared risk for multi-layer networks as $R_{\dagger}(w) = \sum_{t=1}^T \sum_{(i,j) \in \Psi_1} E (\widehat{P}_{ij}^{(t)}(w) - P_{ij}^{(t)})$. Let $\xi_{\dagger} = \inf_{w \in \mathcal{W}} R_{\dagger}(w)$ and $w_\dagger^0 = \argmin_{w \in \mathcal{W}} R_{\dagger}(w)$. 
Arrange $\{A_{ij}^{(t)},\ (i,j) \in \Psi_1\}$, $\{P_{ij}^{(t)},\ (i,j) \in \Psi_1\}$, $\{\sum_{k=1}^{K} \widetilde{P}_{ij}^{(t)[k]}(w),\ (i,j) \in \Psi_1\}$, $\{\widetilde{P}_{(m),ij}^{(t)},\ (i,j) \in \Psi_1\}$,  $\{\widehat{P}_{ij}^{(t)}(w),\ (i,j) \in \Psi_1\}$, $\{\widehat{P}_{(m),ij}^{(t)},\ (i,j) \in \Psi_1\}$, $\{P^{(t)*}_{ij}(w),\ (i,j) \in \Psi_1\}$ and $\{P^{(t)*}_{(m),ij},\ (i,j) \in \Psi_1\}$ in same particular order. Then denote them as vector $a_1^{(t)} \in \mR^{\lvert \Psi_1 \rvert}$, $p_1^{(t)} \in \mR^{\lvert \Psi_1 \rvert}$, $\widetilde{p}^{(t)}_1(w) \in \mR^{\lvert \Psi_1 \rvert}$, $\widetilde{p}^{(t)}_{1(m)} \in \mR^{\lvert \Psi_1 \rvert}$, $\widehat{p}^{(t)}_1(w) \in \mR^{\lvert \Psi_1 \rvert}$, $\widehat{p}^{(t)}_{1(m)} \in \mR^{\lvert \Psi_1 \rvert}$, $p^{(t)*}_1(w) \in \mR^{\lvert \Psi_1 \rvert}$ and $p^{(t)*}_{1(m)} \in \mR^{\lvert \Psi_1 \rvert}$, respectively. We next introduce two matrices $A^{(t)\circ} \in \mR^{N \times N}$ and $P^{(t)\circ} \in \mR^{N \times N}$, where the $(i,j)$th element of $A^{(t)\circ}$ is defined as $A^{(t)\circ}_{ij} = A^{(t)}_{ij}$ if $(i,j) \in \Psi_1$, otherwise $A^{(t)\circ}_{ij} = 0$. Similarly, the $(i,j)$th element of $P^{(t)\circ}$ is defined as $P^{(t)\circ}_{ij} = P^{(t)}_{ij}$ if $(i,j) \in \Psi_1$, otherwise $P^{(t)\circ}_{ij} = 0$.
Denote $\widehat{\Lambda}^{(t)}_1 = (\widehat{p}^{(t)}_{1(1)}, \dots, \widehat{p}^{(t)}_{1(M)})$, $\Omega^{(t)}_1 = (p^{(t)}_1-\widehat{p}^{(t)}_{1(1)}, \dots , p^{(t)}_1-\widehat{p}^{(t)}_{1(M)})$, $\widehat{\Lambda}^{(t)} = \widehat{\Lambda}_1^{(t)\top} \widehat{\Lambda}_1^{(t)}$ and $\Omega^{(t)} = \Omega_1^{(t)\top} \Omega_1^{(t)}$. Theorem \ref{theorem6} shows the rate of the $\widehat{w}_\dagger$ tending to the infeasible optimal weight vector $w_\dagger^0$. The following assumptions are needed to show this theorem.

\begin{assumption}
    There are two positive constants $\rho_{1\dagger}$ and $\rho_{2\dagger}$, such that $0 < \rho_{1\dagger} < \lambda_{\min}\{\sum_{t=1}^T\widehat{\Lambda}^{(t)}/ (T \lvert \Psi_1 \rvert)\} \leq \lambda_{\max}\{\sum_{t=1}^T\widehat{\Lambda}^{(t)}/ (T\lvert \Psi_1 \rvert)\} < \rho_{2\dagger} < \infty$, in probability tending to 1.
    \label{assumptionS5}
\end{assumption}

\begin{assumption}
    $\lambda_{\max}\{\sum_{t=1}^T \Omega^{(t)}/(T \lvert \Psi_1 \rvert)\} = O_p(1)$.
    \label{assumptionS7}
\end{assumption}

\begin{assumption}
    $N^{1- 4\kappa_\dagger}\max\{D_\dagger, \log N\} T \xi_{\dagger}^{-1} = o(1)$, $N^{1-4\kappa_\dagger}MT \xi_{\dagger}^{-1} = o(1)$, and $M=o(N^{1/2})$, where $\kappa_\dagger \in (0,1/2)$.
    \label{assumptionS8}
\end{assumption}

Assumptions \ref{assumptionS5} and \ref{assumptionS7} are similar to Assumptions \ref{assumption5} and \ref{assumption6}, respectively. 
Assumption \ref{assumptionS8} involves replacing $\xi^{-1}$ with $T\xi_{\dagger}^{-1}$ in comparison to Assumption \ref{assumption7}.

\begin{theorem}
    If $w_\dagger^0$ is an interior point of $\mathcal{W}$ and Assumptions \ref{assumptionS1}, \ref{assumptionS2} and \ref{assumptionS5}-\ref{assumptionS8} are satisfied, then there exists a local minimizer $\widehat{w}_\dagger$ of $CV_{\dagger}(w)$ such that 
    \begin{equation}
	\| \widehat{w}_\dagger - w_\dagger^0\| = O_p(\xi_{\dagger}^{1/2} T^{-1/2} \lvert \Psi_1 \rvert^{-1/2+\kappa_\dagger}),
	\label{equ:theorem6}
    \end{equation}
    where $\kappa_\dagger$ is defined in Assumption \ref{assumptionS8}.
    \label{theorem6}
\end{theorem}

Theorem \ref{theorem6} shows the convergence rate of $\widehat{w}_\dagger$ towards $w_\dagger^0$. In comparison to the convergence rate in Theorem \ref{theorem3}, the convergence rate in the case of multi-layer networks not only relies on $\xi_{\dagger}$ and $\lvert \Psi_1 \rvert$ but also depends on $T$.
The proof of Theorem \ref{theorem6} is presented in the Supplementary Material.

\section{Simulation Studies}
\label{sec:simulation}

In this section, we evaluate the NetMA methods by using simulation examples. In addition to NetMA, we also consider three other methods. The first one is the oracle method which uses the true model structure and also the true latent dimension. The second one is the ``equal'' version which assigns equal weights to all the candidate models. That is, if there are $M$ candidate models, then the weight of each model is $1/M$. The third one is the best model selected by the edge cross-validation (ECV) procedure of \cite{li2020network} and \cite{gao2020discussion}.  Specifically, we divide the nodal pairs in $\Psi_1$ into $K$ groups, and then use $K-1$ groups to fit the model each time. Select the model that minimizes the loss on the hold-out set as the best model. For example, in a single-layer network, the ECV algorithm simply replaces Step 3 in Section \ref{sec:single_ma_criterion} with 
\begin{equation*}
    m^* = \argmin_m \frac{1}{\lvert \Psi_1 \rvert} \sum_{k=1}^K \left\|A^{[k]} - \widetilde{P}_{(m)}^{[k]}\right\|_F^2.
\end{equation*}
We compare the four methods for both single-layer networks and multi-layer networks. 

In the following simulations, we consider three cases. The first case measures the effect of the dimensions of latent vectors on the prediction performance. The second case evaluates the effect of the size of the network. The third case evaluates the effect of the network density. We set $K=10$ and the ratio of the numbers of nodal pairs in $\Psi_1$ and $\Psi_2$ as 7:3. 

\subsection{Single-layer networks}
\label{sec:sim-single}

 In the following simulation design, we generate a network adjacency matrix $A = (A_{ij}) \in \mR^{N \times N}$, where $A_{ij}$s are generated from the Bernoulli distribution independently. Specifically, $A_{ij}$ takes the value 1 with probability $P_{ij}$, and the value 0 with probability $1-P_{ij}$. The probability $P_{ij}$ is determined by $\operatorname{logit}(P_{ij}) = \alpha_i + \alpha_j + z_i^\top z_j$, where $\alpha_i$ is generated from $\text{Uniform}(-1,1)$ independently for $i=1,\dots,N$. For the latent vectors, we first generate a matrix $Z \in \mR^{N \times d_0}$ such that each entry is generated from $N(0, 1)$ independently. Then we transform $Z$ by setting $Z = JZ$ where $J = I_N-1 _N1_N^\top/N$ and rotate $Z$ such that $Z^\top Z \propto I_{d_0}$. Finally, scale $Z$ such that $Z^\top Z = NI_{d_0}$. In particular, to control the expected average degree of the networks, we consider transforming $P$ by multiplying a constant $\gamma$. For example, if we want to obtain a network with an expected average degree of $\bar{D}$, then $\gamma$ equals $\bar{D}/ARS(P)$, where $ARS(P)$ refers to the average of the row sums of $P$.

We focus on estimating the network connection probability matrix $P$. Given an estimator $\widehat{P}$, the performances of the four methods are measured by the relative empirical risk function, which is calculated as 
\begin{align}
    \widehat{R}(\widehat{w}) = \frac{1}{Q} \sum_{q=1}^{Q} \frac{\sum_{(i,j) \in \Psi_2} \left\{ \widehat{P}_{ij}^{\{q\}}(\widehat{w}^{\{q\}}) - P^{\{q\}}_{ij} \right\}^2}{\sum_{(i,j) \in \Psi_2} \left\{P^{\{q\}}_{ij}\right\}^2},
\end{align}
where $Q$ is the number of simulation replications, and $\widehat{P}_{ij}^{\{q\}}(\widehat{w}^{\{q\}})$ denotes the prediction based on the ``oracle'', the ``equal'', ECV and NetMA weights in the $q$th replication. In the following simulations, we set $Q=100$, and consider a sequence of candidate models, where the dimension of the latent vectors in the $m$th candidate model is $m$, $m=1,\dots,M$. 

\noindent
{\sc Case 1 (Dimension of Latent Space).} In this case, we set the size of the network as $N = 200$, and the expected average nodal degree is 60. We vary $M$ from 2 to 12, and let the true dimensions of the latent vectors $d_0$ be 4, 7 and 10, respectively. 

\begin{figure}[h]
    \centering
    \includegraphics[width=15cm]{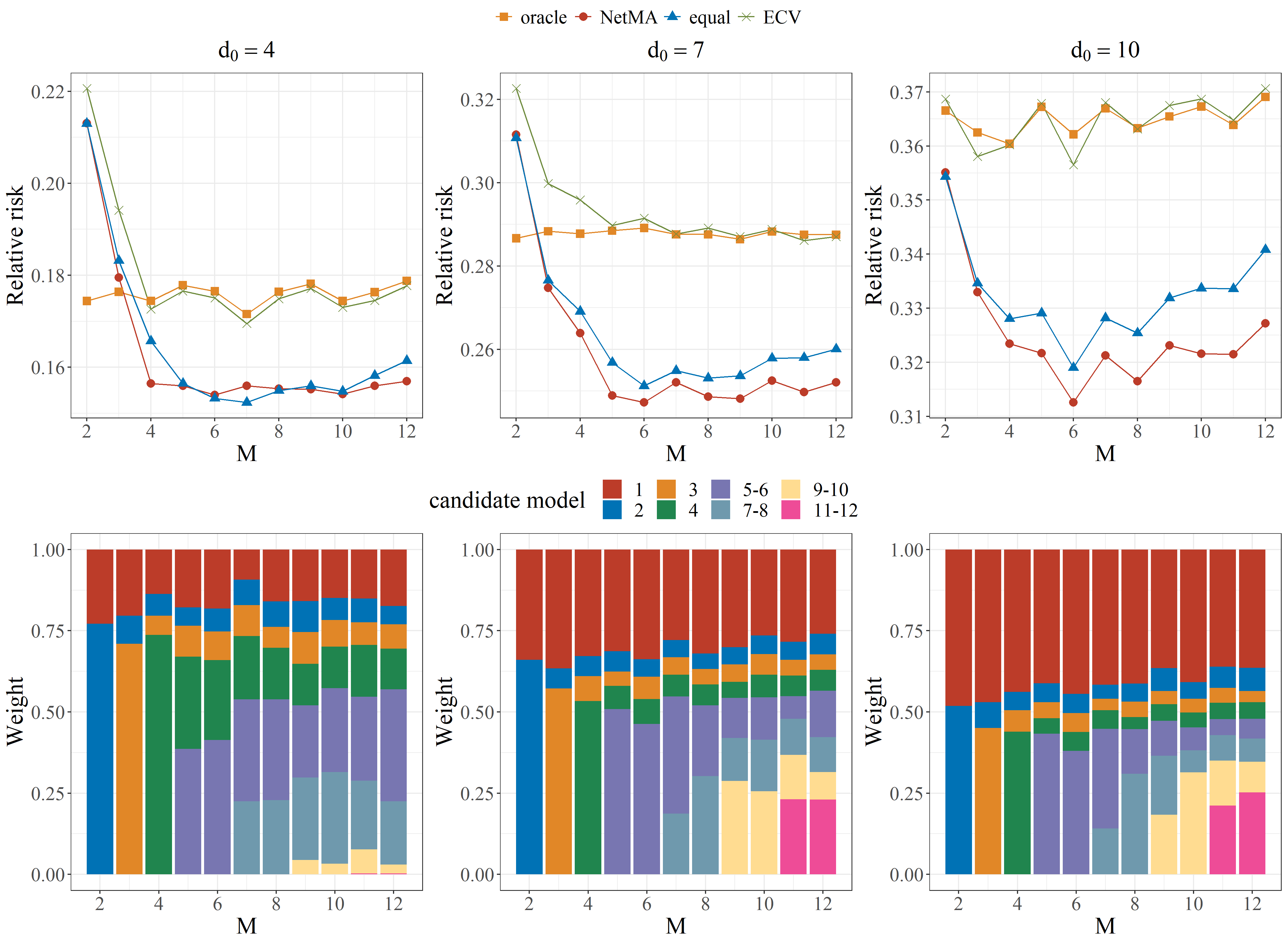}
    \caption{Relative risks and model weights for the four methods with different candidate models in single-layer networks.}
    \label{fig:case_candidate model}
\end{figure}

The resulting relative risks under each $d_0$ are presented in Figure \ref{fig:case_candidate model}. Figure \ref{fig:case_candidate model} shows from left to right the cases where the true dimension is 4, 7 and 10, respectively. The upper panels of Figure \ref{fig:case_candidate model} show the change in relative risk as $M$ changes, and the lower panels of Figure \ref{fig:case_candidate model} show how the weight distribution changes as $M$ varies. It can be seen that, as $d_0$ increases, the advantage of our method becomes more and more obvious, which is even better than the ``oracle''. This is mainly because when the network size and network density are fixed, the increase of $d_0$ introduces more parameters to be estimated. Thus, when $d_0$ is large, the true model may yield poor results. This phenomenon can be verified by observing the relative risk of the ``oracle'' from the upper panels of Figure \ref{fig:case_candidate model} from left to right. To be more specific, in the setting of $d_0=4$, when all the candidate models are misspecified, ``oracle '' performs the best. As the candidate models gradually include the true model, the relative risk of NetMA and the ``equal'' drops dramatically. In the meanwhile, models with higher latent space dimensions are given higher weights.
When the candidate models include the true model, the model selected by ECV performs similarly to the ``oracle'', while NetMA and the ``equal'' perform similarly, and much better than the ``oracle'' and ECV. 
In the setting of $d_0=7$, the ``oracle'' performs the best only when $M=2$. When the dimension of the latent space in the candidate model exceeds 2, NetMA performs the best. As $M$ increases, NetMA's superiority over the ``equal'' also becomes more apparent. 
In the setting of $d_0=10$, regardless of whether the candidate model contains the true model, the performance of ECV is similar to that of the ``oracle'', which is much worse than NetMA and the ``equal''. Additionally, NetMA performs better than the ``equal'', and the gap between NetMA and the ``equal'' is larger than their gap under $d_0=4$ and $d_0=7$.
	
\noindent
{\sc Case 2 (Number of nodes).} In this case, we vary the size of the networks $N$ from 100 to 500. In order to maintain consistent density for networks of different sizes, we set the expected average nodal degree of the network to be $0.3(N-1)$. Then we fix the true dimension of the latent vectors to be 6.

\begin{figure}[h]
    \centering
    \includegraphics[width=15cm]{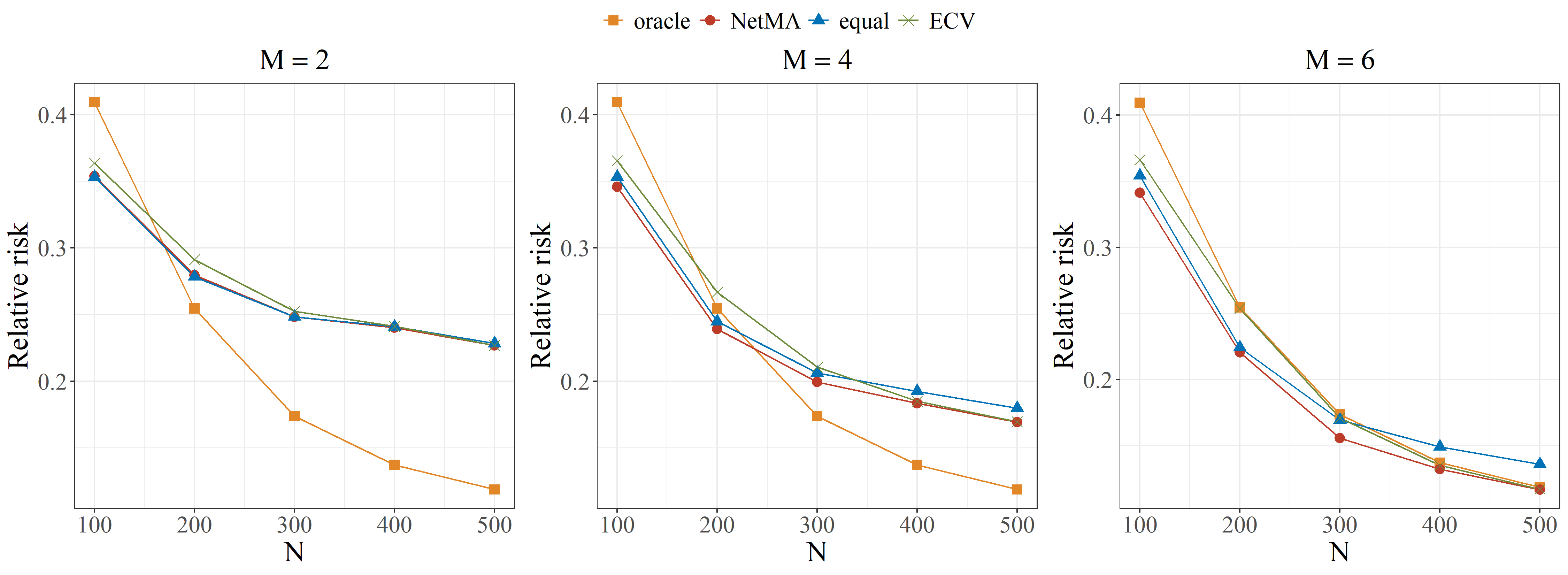}
    \caption{Relative risks of the four methods with different network sizes in single-layer networks.}
    \label{fig:case_N}
\end{figure}

Figure \ref{fig:case_N} displays the results for three different settings of {\sc Case 2} from left to right, namely the number of candidate models $M=2,~4~\text{and}~6$, respectively. When the maximum dimension of the latent space in the candidate models is 2 or 4, which is smaller than the true dimension of latent space, NetMA performs the best under small network sizes. As the network size increases, the advantage of the ``oracle'' becomes more evident. However, apart from the ``oracle'', NetMA performs the best, especially when $M=4$, which is more pronounced. 
When the maximum dimension of the latent space in the candidate models is equal to the true latent space dimension $(M=6)$, NetMA performs the best for most cases. When $N$ is relatively large, the performance of the ``oracle'' and ECV becomes increasingly similar to that of NetMA, while the performance of the ``equal'' shows a certain gap compared to NetMA. 

We further consider two extensions in single-layer networks: the first extension involves considering edge covariates, and the second extension addresses the situation where edges are not missing at random.

\begin{figure}[h]
    \centering
    \includegraphics[width=15cm]{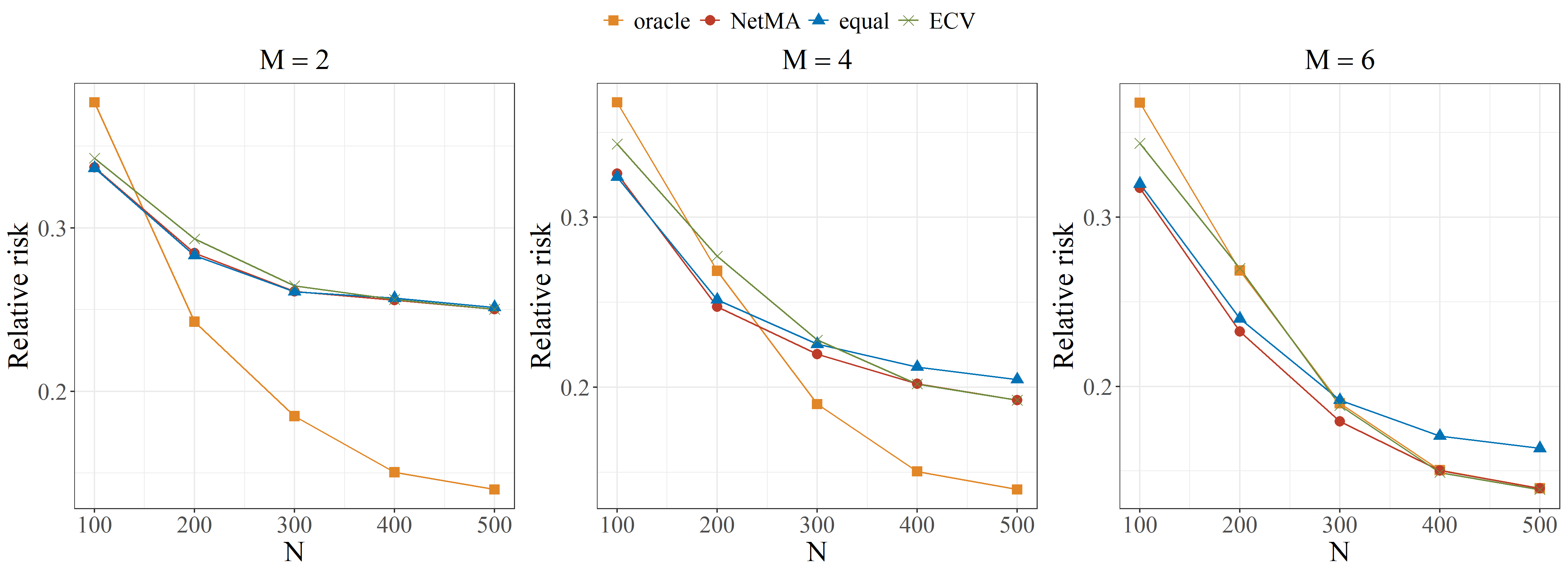}
    \caption{Relative risks of the four methods with different network sizes in single-layer networks with edge covariates.}
    \label{fig:case_singlelayer_covariate_N}
\end{figure}

In the first extension, we consider Model \eqref{model with covariate} with edge covariates. Specifically, we generate $\beta$ and each entry of the covariate matrix $X$ from $\text{Uniform}(0,1)$. The generation for the other parameters remains unchanged. Figure \ref{fig:case_singlelayer_covariate_N} shows the results of the four methods as the number of nodes and the number of candidate models change. It can be observed that Figure \ref{fig:case_singlelayer_covariate_N} is very similar to Figure \ref{fig:case_N}, i.e., our method performs best when the candidate models include the correct models, or when the candidate models do not include the correct models but the number of nodes is relatively small.

\begin{figure}[h]
    \centering
    \includegraphics[width=5cm]{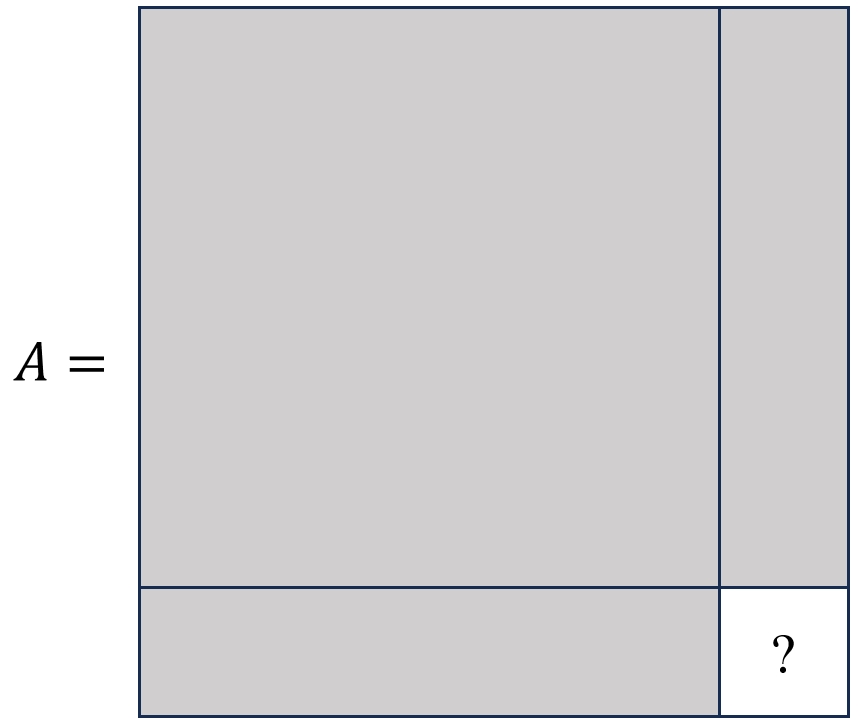}
    \caption{An illustration of the adjacency matrix of an egocentrically sampled network, where grey blocks are observed and the white block is missing.}
    \label{fig:egocentric network}
\end{figure}

\begin{figure}[h]
    \centering
    \includegraphics[width=15cm]{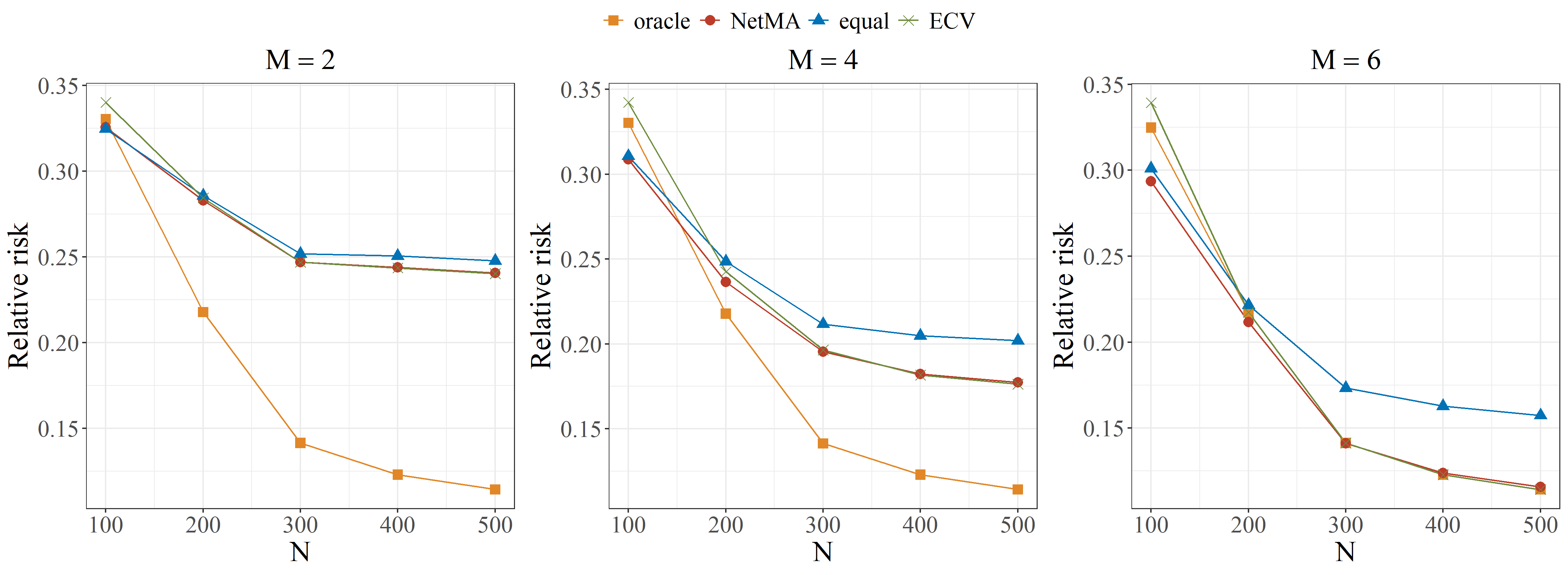}
    \caption{Relative risks of the four methods with different network sizes in single-layer networks with nonrandom missing.}
    \label{fig:case_singlelayer_N_nonrandom}
\end{figure}

In the second extension, we consider a case of non-random missingness, i.e., egocentrically sampled networks. These networks are constructed through egocentric sampling, which is a procedure where a subset of nodes is first sampled, and then the links between these nodes are recorded, while other information remains unknown \citep{Li02102023}. An example of the adjacency matrix of an egocentrically sampled network is shown in Figure \ref{fig:egocentric network}. In Figure \ref{fig:egocentric network}, the grey area represents the observable parts, while the white area indicates the missing parts. Our goal is to predict the missing links based on the information from the observed links. In the simulation, we assume that 90\% of nodes are sampled, meaning that the links between the remaining 10\% of the nodes are missing. All other settings are the same as those in {\sc Case} 2. Figure \ref{fig:case_singlelayer_N_nonrandom} shows the prediction results of different methods under the egocentric missing situation. It can be seen that when the correct models are not included in the candidate models, NetMA performs second only to the oracle, especially when the number of nodes is relatively large. When the correct models are included in the candidate models, NetMA performs the best, particularly when the network size is relatively small.

\noindent
{\sc Case 3 (Network density).} In this case, we examine the influence of network density. Specifically, we fix the network size $N=200$ and the true dimension of latent space $d_0 = 6$. The expected average nodal degree ranges from 10 to 80.

\begin{figure}[ht]
    \centering
    \includegraphics[width=15cm]{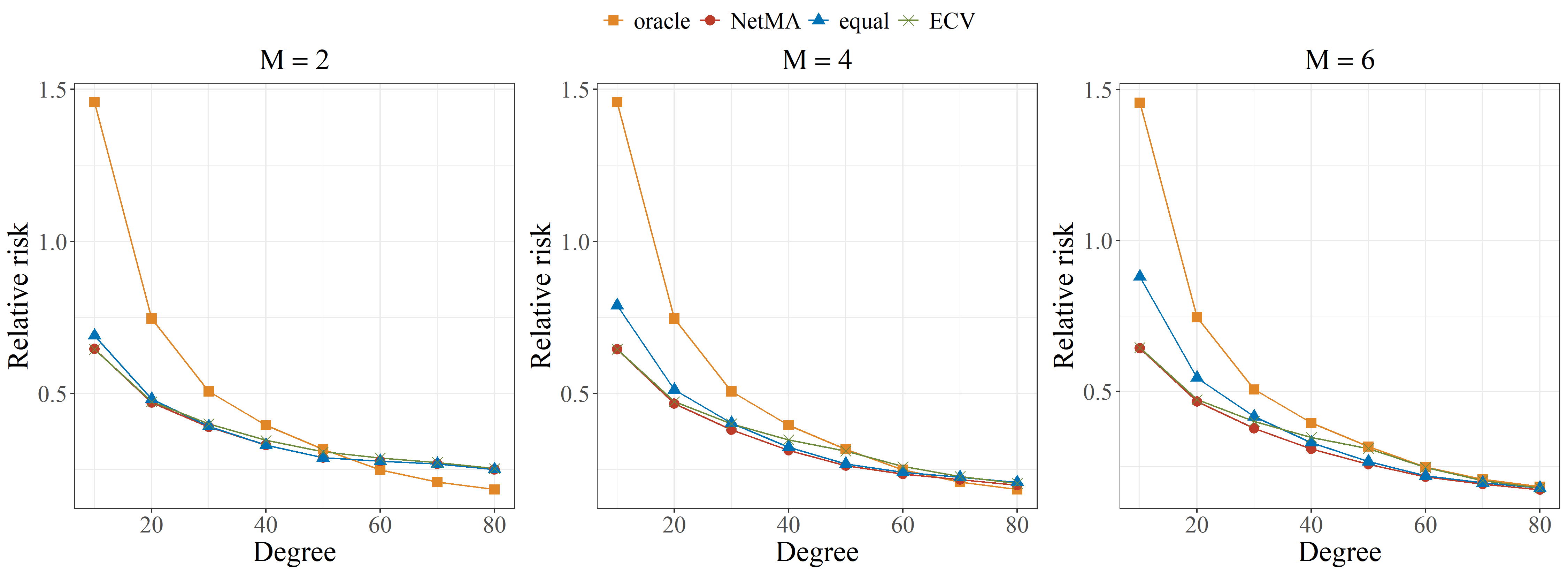}
    \caption{Relative risks of the four methods with different network densities in single-layer networks.}
    \label{fig:density}
\end{figure}

We also design three different settings ($M=2,~4,~6$) to show the performance of our method. Figure \ref{fig:density} presents the results of the three settings for {\sc Case 3}. It can be seen from Figure \ref{fig:density} that  NetMA performs the best for various $M$.  In addition, when the network density is small, the ``oracle'' performs very poorly. This also verifies what we mentioned before, i.e., when the network size and the density are small but the real dimension of the latent space is relatively large, it is challenging to estimate the model based on the true dimension. Under this situation, the advantage of NetMA is very obvious.
	
\subsection{Multi-layer networks}

In this subsection, we conduct some simulations on multi-layer networks, and set $T=5$. The data-generating process is presented as follows. For the latent vectors, we generate $Z^{(t)} \in \mR^{N \times d_{0,t}}$ where $d_{0,t}$ is the true dimension of the latent vector in the $t$th layer by first generating i.i.d $N(0,1)$ entries and then transforming it according to the steps introduced in single-layer networks for $t=1, \dots, 5$. Generate the degree heterogeneity parameters $\alpha = (\alpha_1, \dots, \alpha_N)^\top$ with each entry independent and identically generated from $\text{Uniform}(-1,1)$. Then we can calculate the probability matrix $P^{(t)} \in \mR^{N \times N}$ by $\text{logit}(P_{ij}^{(t)}) = \alpha_i + \alpha_j + z_i^{(t)\top}z_j^{(t)}$. Similarly, we can transform $P^{(t)}$ by multiplying a constant to control the expected average degree. To evaluate the performance of the estimation methods, we introduce the relative empirical risk function for multi-layer networks, which is defined as follows,
\vspace{0cm}
\begin{align}
    \widehat{R}_\dagger(\widehat{w}_\dagger) = \frac{1}{Q} \sum_{q=1}^{Q} \frac{\sum_{t=1}^T \sum_{(i,j) \in \Psi_2} \left\{ \widehat{P}_{ij}^{(t)\{q\}}(\widehat{w}_\dagger^{\{q\}}) - P^{(t)\{q\}}_{ij} \right\}^2}{\sum_{t=1}^T \sum_{(i,j) \in \Psi_2} \left\{P^{(t)\{q\}}_{ij}\right\}^2},
\end{align}
where $Q$ is the number of simulation replications. In the following cases, we set $Q=100$, $K=5$ and $\lvert \Psi_1 \rvert / \lvert \Psi_2 \rvert = 7:3$.
Next, we design three cases and show the results of the simulations.

\noindent
{\sc Case 1 (dimension of latent space).} In this case, we consider a 5-layer network with the number of nodes in each layer $N=200$. The expected average nodal degree is set to be 60. The number of the candidate models ranges from 1 to 10.

\begin{figure}[h]
    \centering
    \includegraphics[width=15cm]{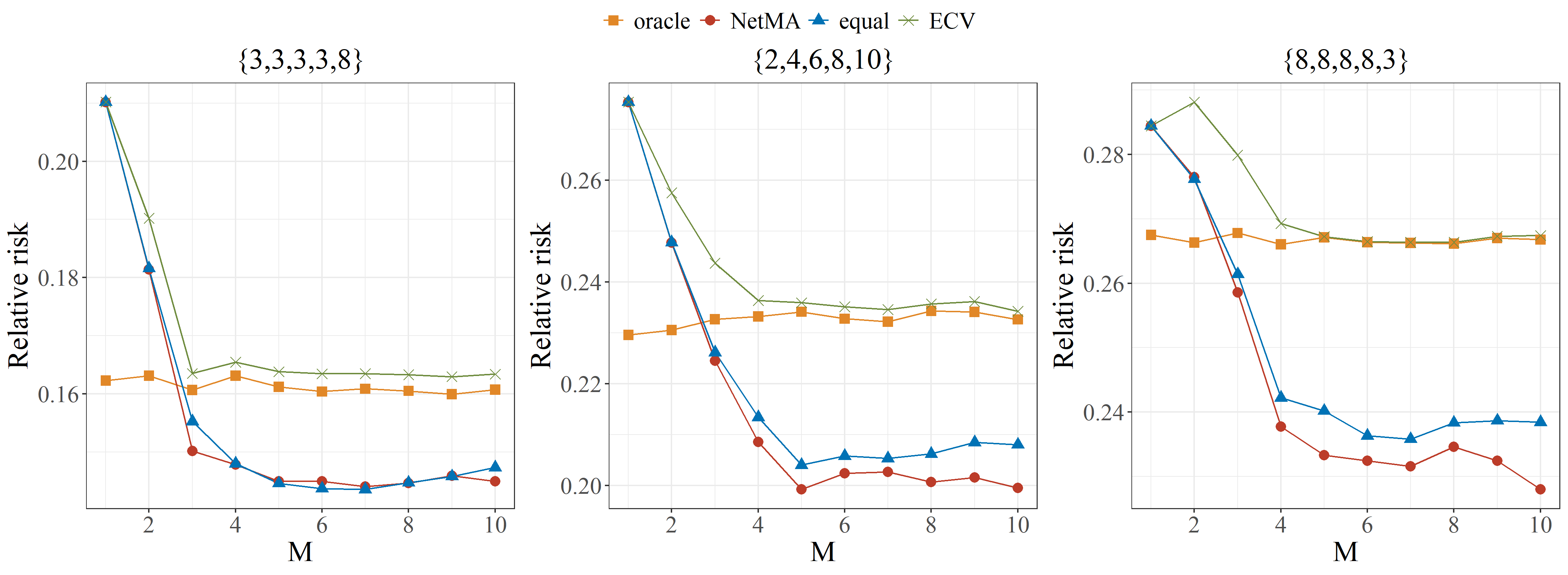}
    \caption{Relative risks of the four methods with different candidate models in multi-layer networks.}
    \label{fig:case_multilayer candidate model}
\end{figure}

For {\sc Case 1}, we design three settings to evaluate the performances of the four methods. The true dimensions of the latent vectors in the five layers for the three settings  are $\{3, 3, 3, 3, 
8\}$, $\{ 2, 4, 6, 8, 10 \}$  and $\{8, 8, 8, 8, 3\}$,  respectively. Figure \ref{fig:case_multilayer candidate model} shows the simulation results of the three different settings from left to right. It's easy to see that as the true dimension increases, the risk of estimating the parameters also increases. Additionally, it is obvious that when $M \geq 3$, NetMA performs the best under different settings, followed by simple averaging. Compared to the ``oracle'' and ECV, NetMA has a substantial advantage, especially when $M$ is large. Except for the ``oracle'', the effectiveness of the other three methods is poor when $M$ is small. This is also easy to understand, as in this situation, the candidate models are not sufficient for modelling the network. However, as $M$ gradually increases, the relative risks of these three methods decrease sharply and eventually stabilize. Additionally, it can be observed that in the first two settings, there always exists a certain gap between the relative risks of the ECV method and those of the ``oracle''.
	
\noindent
{\sc Case 2 (Number of nodes).} In this case, we explore the performance differences of the four methods as the network size increases. Specifically, we construct a 5-layer network and vary the number of nodes in each layer from 100 to 500. The true dimensions of the latent vectors in the five layers are set to be 2, 4, 6, 8, and 10 respectively. In each layer, the expected average nodal degree is $0.3(N-1)$.

\begin{figure}[h]
    \centering
    \includegraphics[width=15cm]{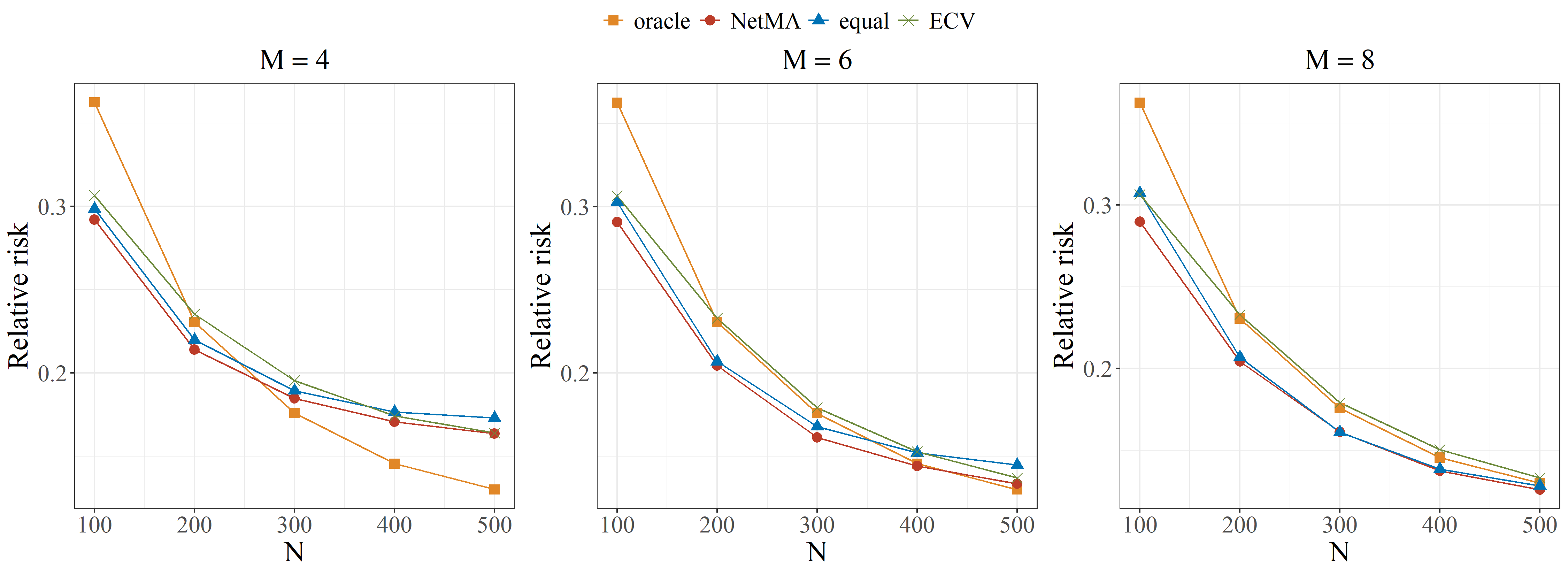}
    \caption{Relative risks of the four methods with different network sizes in multi-layer networks.}
    \label{fig:case_multilayer_N}
\end{figure}

We set $M=4,~6~\text{and}~8$. The results are shown in Figure \ref{fig:case_multilayer_N}. As we can see, when the sample size is small ($N=100, 200$), NetMA almost always outperforms other methods, including the ``oracle'' for all settings. As the sample size increases (e.g., $N=300, 400, 500$), if $M$ is small (such as $M=4$), the ``oracle'' surpasses NetMA. However, when $M$ increases (e.g., $M=6, 8$), NetMA outperforms the ``oracle''. Additionally, NetMA generally outperforms ECV and the ``equal'' methods. 

	
\noindent
{\sc Case 3 (Network density).} In this case, we study the performance of the four methods under different network densities. The network consists of 5 layers, with 200 nodes in each layer. The true dimensions of the latent vectors in the 5 layers are 2, 4, 6, 8, and 10, respectively. We vary the expected average nodal degree from 20 to 80.

\begin{figure}[ht]
    \centering
    \includegraphics[width=\textwidth]{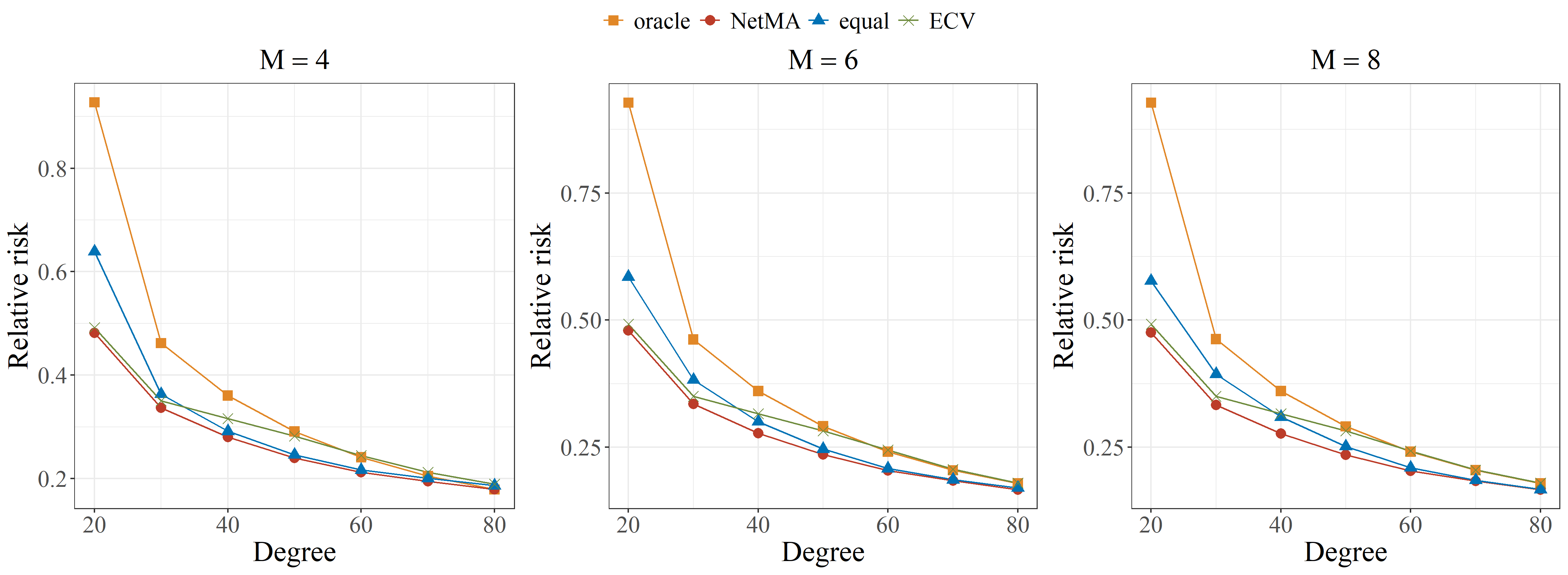}
    \caption{Relative risks of the four methods with different network densities in multi-layer networks.}
    \label{fig:case_multilayer_density}
\end{figure}

Figure \ref{fig:case_multilayer_density} shows the simulation results of Case 3 when $M=4,\ 6,\ 8$ respectively, which are similar to the results of Figure \ref{fig:density}. Specifically, NetMA performs the best and is stable under different $M$ and network densities.
	
\section{Empirical Example}	
\label{sec:empirical}

\subsection{Prediction on ResearchGate data}
\label{sec:empirical-single}
In this section, we apply the proposed method NetMA to real-world data. In practical scenarios, as we do not know the true value of the latent vectors, it is infeasible to assess the model performance in terms of latent space estimation. Instead, we consider the downstream task based on the estimated latent space representation such as link prediction, community detection and node classification. Link prediction is a popular procedure in network science to predict missing or future links \citep{lu2011link,kumar2020link}. Here, we compare the performance of NetMA, ECV and a simple averaging method on link prediction. The three methods are examined by (i) changing the number of candidate models, (ii) changing the ratio of $\lvert \Psi_1 \rvert$ and $\lvert \Psi_2 \rvert$, and (iii) changing the number of folds of cross-validation.

Scientific collaboration refers to the cooperative behaviour between two or more researchers, such as joining together in developing projects or research. One of the key concerns for researchers is how to find suitable collaborators. Scientific and academic social networks, such as LinkedIn\footnote{https://www.linkedin.com/}, ResearchGate\footnote{https://www.researchgate.net/}, and Mendeley\footnote{https://www.mendeley.com/}, provide scholars with an opportunity to understand the current research directions of others. \cite{roozbahani2021presenting} collect a dataset called ResearchGate dataset for Rcommending Systems (RGRS) which contains the information of 266,207 members of ResearchGate. The dataset provides information on followers and following. According to this structural information, we construct a network in which the nodes represent researchers and the edges represent the follower/following relationship. Although there is a directed relationship of following or being followed between two scholars, for convenience, we only consider whether there is a follower or following relationship between them. If such a relationship exists, there is an edge between the two scholars; if it does not exist, there is no edge. Thus, the network is undirected and unweighted. Many real networks have core-periphery structures, where the core contains nodes that are intensively connected and the periphery contains nodes that are weakly connected to the core. The core nodes are the main focus of our research. To abstract core nodes, we first remove nodes whose degrees are no more than 30 and the edges connected with them. Then we abstract the 24-core network according to the idea of \cite{5c16567a-3812-3471-806d-cf443b6d6265}. Specifically, a $c$-core is obtained by removing nodes whose number of neighbours is less than $c$ and the edges connected to them. Keep iterating this step until the nodes in the network no longer change. Finally, the core network has 397 nodes and 8,415 edges. The density of the network is 0.107.

We use the model averaging and model selection methods to analyse the core network. The candidate models are latent space models with different latent space dimensions, which are presented in Model \eqref{equ: model1}. The
dimensions of the latent space for the $m$th ($m=1,\dots,M$) model is $m$ and we let $M=2,~4,~6,~\text{and}~8$. We consider three scenarios for $\pi_{12} = \lvert \Psi_1 \rvert/\lvert \Psi_2 \rvert$: $7/3$, $8/2$, and $9/1$. In addition, when selecting weights for NetMA and models for ECV, we take $K=5$ and $K=10$ for the $K$-fold cross-validation. 
The experiments are replicated 100 times and we report results for link prediction in Table \ref{tab:single-linkprediction}. We use four metrics to compare the prediction performance. Specifically, AUROC is the area under the ROC curve; AUPR is the area under the precision-recall curve; MLogf is the average value of log-likelihood function; MSE is the mean squared error. The bolded numbers in the table indicate the best results under different metrics in a setting.
The results show that the proposed NetMA method performs better than ECV and simple averaging in most scenarios. Besides, the prediction performance of NetMA is quite similar for $K=5$ and $K=10$. 

Next, we record the weights of the candidate models obtained by ECV and NetMA in each replication when $\pi_{12} = 9/1$. Table \ref{tab:single-weight} shows the mean of the weights based on 100 replications for each method under $K=5$ and $K=10$. It can be seen that for this network, NetMA tends to put more weights on the candidate models with larger dimensions of the latent space, and ECV shows a preference for selecting candidate models with higher dimensional latent spaces. 

\begin{table}[h]
\centering
\caption{Average weights estimated by ECV and NetMA on ResearchGate data.}
\label{tab:single-weight}
\begin{tabular}{cccccc}
\hline
                     & Model & \multicolumn{1}{c}{ECV ($K=5$)} & \multicolumn{1}{c}{ECV ($K=10$)} & \multicolumn{1}{c}{NetMA ($K=5$)} & \multicolumn{1}{c}{NetMA ($K=10$)} \\ \hline
\multirow{2}{*}{$M=2$}& 1     & 0.0000                        & 0.0000                         & 0.0449                          & 0.0077                           \\
                     & 2     & 1.0000                        & 1.0000                         & 0.9551                          & 0.9923                           \\ \hline
\multirow{4}{*}{$M=4$}& 1     & 0.0000                        & 0.0000                         & 0.0344                          & 0.0029                           \\
                     & 2     & 0.0000                        & 0.0000                         & 0.1439                          & 0.1094                           \\
                     & 3     & 0.1600                        & 0.0700                         & 0.1530                          & 0.0873                           \\
                     & 4     & 0.8400                        & 0.9300                         & 0.6687                          & 0.8004                           \\ \hline
\multirow{6}{*}{$M=6$}& 1     & 0.0000                        & 0.0000                         & 0.0080                          & 0.0000                           \\
                     & 2     & 0.0000                        & 0.0000                         & 0.1177                          & 0.0587                           \\
                     & 3     & 0.0000                        & 0.0000                         & 0.0536                          & 0.0365                           \\
                     & 4     & 0.0000                        & 0.0000                         & 0.1247                          & 0.1330                           \\
                     & 5     & 0.0000                        & 0.0000                         & 0.0949                          & 0.0959                           \\
                     & 6     & 1.0000                        & 1.0000                         & 0.6010                          & 0.6760    \\ \hline                      
\end{tabular}
\end{table}

\subsection{Prediction on virtual world data}

In this section, we apply the proposed method NetMA to real-world data with a multi-layer network. \cite{jankowski2017multilayer} provides a dataset which contains the record of six types of spreading events that occurred in a virtual world platform. The types of events include campaigns, friends, logins, messages, transactions and visits. Here, we extract a subset of common users present in four events and construct a multi-layer network. The network consists of four layers, representing friend, message, transaction and visit relationships. Each layer has 183 users and thus $N=183$. Considering that some of the relations are directed, we convert the directed networks to undirected networks based on the existence of any single directional edge between nodes. 

We apply the model averaging and model selection methods to analyse the multi-layer network. Here, we consider three scenarios, namely $M=4,~6,~\text{and}~8$. The remaining settings are the same as those in Section \ref{sec:empirical-single}. 
The experiments are replicated 100 times, and the results for link prediction are shown in Table \ref{tab:multi-auc}. As we can see, NetMA with 5-fold cross-validation performs the best in terms of average AUC in all scenarios. As $M$ increases, the advantage of NetMA becomes more pronounced. 
Table \ref{tab:multi-weight} shows the weight allocation results for ECV and NetMA. Interestingly, unlike the results on the weight allocations for single-layer networks, the weight allocations for multi-layer networks are inconsistent for ECV and NetMA. Furthermore, the choice of $K$ seems to have a great impact on the weight allocation of candidate models in ECV, while its influence on NetMA is comparatively small. 

\begin{table}[h]
\caption{Average AUC for link prediction on virtual world data (standard errors are in brackets).}
\label{tab:multi-auc}
\resizebox{\textwidth}{!}{
\begin{tabular}{ccccccc}
\hline 
\multicolumn{1}{l}{} & $\lvert \Psi_1 \rvert / \lvert \Psi_2 \rvert$ & equal                                                                   & ECV ($K=5$)& ECV ($K=10$)& NetMA ($K=5$)& NetMA ($K=10$)\\ \hline
\multirow{5}{*}{$M=4$}& 5/5                                           & \begin{tabular}[c]{@{}c@{}}0.6979\\      (0.0047)\end{tabular} & \begin{tabular}[c]{@{}c@{}}0.6872\\      (0.0055)\end{tabular} & \begin{tabular}[c]{@{}c@{}}0.6859\\      (0.0059)\end{tabular} & \textbf{\begin{tabular}[c]{@{}c@{}}0.7006\\      (0.0047)\end{tabular}} & \begin{tabular}[c]{@{}c@{}}0.7000\\      (0.0046)\end{tabular} \\
                     & 7/3                                           & \begin{tabular}[c]{@{}c@{}}0.7235\\      (0.0050)\end{tabular} & \begin{tabular}[c]{@{}c@{}}0.7117\\      (0.0055)\end{tabular} & \begin{tabular}[c]{@{}c@{}}0.7116\\      (0.0048)\end{tabular} & \textbf{\begin{tabular}[c]{@{}c@{}}0.7246\\      (0.0051)\end{tabular}} & \begin{tabular}[c]{@{}c@{}}0.7235\\      (0.005)\end{tabular}  \\
                     & 9/1                                           & \begin{tabular}[c]{@{}c@{}}0.7388\\      (0.0087)\end{tabular} & \begin{tabular}[c]{@{}c@{}}0.7283\\      (0.0085)\end{tabular} & \begin{tabular}[c]{@{}c@{}}0.7270\\      (0.0088)\end{tabular} & \textbf{\begin{tabular}[c]{@{}c@{}}0.7394\\      (0.0087)\end{tabular}} & \begin{tabular}[c]{@{}c@{}}0.7385\\      (0.0088)\end{tabular} \\ \hline
\multirow{5}{*}{$M=6$}& 5/5                                           & \begin{tabular}[c]{@{}c@{}}0.6962\\      (0.0050)\end{tabular} & \begin{tabular}[c]{@{}c@{}}0.6870\\      (0.0055)\end{tabular} & \begin{tabular}[c]{@{}c@{}}0.6848\\      (0.0057)\end{tabular} & \textbf{\begin{tabular}[c]{@{}c@{}}0.7029\\      (0.0046)\end{tabular}} & \begin{tabular}[c]{@{}c@{}}0.7021\\      (0.0044)\end{tabular} \\
                     & 7/3                                           & \begin{tabular}[c]{@{}c@{}}0.7220\\      (0.0049)\end{tabular} & \begin{tabular}[c]{@{}c@{}}0.7111\\      (0.0055)\end{tabular} & \begin{tabular}[c]{@{}c@{}}0.7106\\      (0.0045)\end{tabular} & \textbf{\begin{tabular}[c]{@{}c@{}}0.7258\\      (0.0048)\end{tabular}} & \begin{tabular}[c]{@{}c@{}}0.7247\\      (0.0048)\end{tabular} \\
                     & 9/1                                           & \begin{tabular}[c]{@{}c@{}}0.7387\\      (0.0085)\end{tabular} & \begin{tabular}[c]{@{}c@{}}0.7273\\      (0.0086)\end{tabular} & \begin{tabular}[c]{@{}c@{}}0.7259\\      (0.0086)\end{tabular} & \textbf{\begin{tabular}[c]{@{}c@{}}0.7417\\      (0.0088)\end{tabular}} & \begin{tabular}[c]{@{}c@{}}0.7408\\      (0.0086)\end{tabular} \\ \hline
\multirow{5}{*}{$M=8$}& 5/5                                           & \begin{tabular}[c]{@{}c@{}}0.6986\\      (0.0050)\end{tabular} & \begin{tabular}[c]{@{}c@{}}0.6871\\      (0.0054)\end{tabular} & \begin{tabular}[c]{@{}c@{}}0.6842\\      (0.0059)\end{tabular} & \textbf{\begin{tabular}[c]{@{}c@{}}0.7063\\      (0.0046)\end{tabular}} & \begin{tabular}[c]{@{}c@{}}0.7055\\      (0.0042)\end{tabular} \\
                     & 7/3                                           & \begin{tabular}[c]{@{}c@{}}0.7240\\      (0.0048)\end{tabular} & \begin{tabular}[c]{@{}c@{}}0.7110\\      (0.0052)\end{tabular} & \begin{tabular}[c]{@{}c@{}}0.7097\\      (0.0054)\end{tabular} & \textbf{\begin{tabular}[c]{@{}c@{}}0.7297\\      (0.0046)\end{tabular}} & \begin{tabular}[c]{@{}c@{}}0.7285\\      (0.0046)\end{tabular} \\
                     & 9/1                                           & \begin{tabular}[c]{@{}c@{}}0.7404\\      (0.0083)\end{tabular} & \begin{tabular}[c]{@{}c@{}}0.7262\\      (0.0085)\end{tabular} & \begin{tabular}[c]{@{}c@{}}0.7257\\      (0.0085)\end{tabular} & \textbf{\begin{tabular}[c]{@{}c@{}}0.7445\\      (0.0083)\end{tabular}} & \begin{tabular}[c]{@{}c@{}}0.7434\\      (0.0083)\end{tabular} \\ \hline
\end{tabular}}
\end{table}

\begin{table}[ht]
\caption{Average weights estimated by ECV and NetMA on virtual world data.}
\label{tab:multi-weight}
\centering
\resizebox{\textwidth}{!}{
\begin{tabular}{cccccc}
\hline
                     & Model & \multicolumn{1}{c}{ECV ($K=5$)} & \multicolumn{1}{c}{ECV ($K=10$)} & \multicolumn{1}{c}{NetMA ($K=5$)} & \multicolumn{1}{c}{NetMA ($K=10$)} \\ \hline
\multirow{4}{*}{$M=4$}& 1     & 0.0000    & 0.0000     & 0.2856      & 0.1905       \\
                     & 2     & 0.8200    & 0.5900     & 0.2794      & 0.3311       \\
                     & 3     & 0.1100    & 0.0700     & 0.0400      & 0.0143       \\
                     & 4     & 0.0700    & 0.3400     & 0.3950      & 0.4641       \\ \hline
\multirow{6}{*}{$M=6$}& 1     & 0.0000    & 0.0000     & 0.2793      & 0.1838       \\
                     & 2     & 0.6600    & 0.2700     & 0.2594      & 0.3060       \\
                     & 3     & 0.0800    & 0.0100     & 0.0173      & 0.0041       \\
                     & 4     & 0.0100    & 0.0300     & 0.0212      & 0.0068       \\
                     & 5     & 0.0100    & 0.0100     & 0.0227      & 0.0012       \\
                     & 6     & 0.2400    & 0.6800     & 0.4001      & 0.4981       \\ \hline
\multirow{8}{*}{$M=8$}& 1     & 0.0000    & 0.0000     & 0.2584      & 0.1651       \\
                     & 2     & 0.3800    & 0.0400     & 0.2375      & 0.2848       \\
                     & 3     & 0.0600    & 0.0000     & 0.0063      & 0.0008       \\
                     & 4     & 0.0000    & 0.0100     & 0.0020      & 0.0020       \\
                     & 5     & 0.0000    & 0.0000     & 0.0007      & 0.0013       \\
                     & 6     & 0.0300    & 0.0200     & 0.0085      & 0.0071       \\
                     & 7     & 0.4300    & 0.7500     & 0.1353      & 0.1640       \\
                     & 8     & 0.1000    & 0.1800     & 0.3512      & 0.3748   \\ \hline                         
\end{tabular}}
\end{table}

\section{Discussion}
\label{sec:conclusion}
In this paper, we introduce a model averaging strategy for link prediction. Specifically, we focus on latent space models and allow for different dimensions of latent space. We propose the $K$-fold edge cross-validation procedures to select the data-driven optimal weights for candidate models in the single-layer and multi-layer networks. The proposed method fully leverages information from multiple models and thus leads to the desirable prediction performance. Specifically, when the candidate models are misspecified, NetMA is proved to be asymptotically optimal in terms of squared errors; when the candidate model set includes the correct models, it assigns all weights to the correct models asymptotically. Besides, we derive the rate of the NetMA-based empirical weights converging to the theoretically optimal weights. Simulation studies show the promise of the NetMA method in both single-layer networks and multi-layer networks. We also evaluate its competitive performance empirically in link prediction problems. 

Although we focus on latent space models in this paper, the NetMA procedure, along with its theoretical properties, can be applied to models for any undirected and unweighted network. As \cite{doi:10.1080/01621459.2023.2252137} demonstrates, we can consider integrating various models for networks, such as SBM, the degree-corrected stochastic block model of \cite{karrer2011stochastic}, and the universal singular value thresholding of \cite{chatterjee2015matrix} in the future. For other types of networks, such as directed networks, we can consider candidate models designed for directed networks. For example, \cite{zhang2022directed} proposed a latent space model for directed networks, where $\text{logit}(P_{ij}) = \nu_i^\top \omega_j$, with $\nu_i \in \mR^d$ and $\omega_j \in \mR^d$ being the latent vectors of out-node $i$ and in-node $j$. It makes sense to extend the model averaging method to this situation.
Furthermore, in the study of networks, there are not only issues related to link prediction but also several other problems such as community detection \citep{lancichinetti2009community} and predicting the response observed for each node in social network \citep{zhu2017network}. 
It would be interesting to extend model averaging to these research problems.

\bigskip
\begin{center}
{\large\bf SUPPLEMENTARY MATERIAL}
\end{center}

Supplementary material includes the proofs of Theorems \ref{theorem1}-\ref{theorem6}, as well as the verifications of Assumptions \ref{assumption1}-\ref{assumption4}.
  
\begin{landscape}
\begin{table}[ht]
\label{tab:single-linkprediction}
\centering
\caption{Evaluation for link prediction on ResearchGate data (standard errors are in brackets).}
\begin{adjustbox}{width=24cm}
\centering
\begin{tabular}{cccccccccccccc}
\toprule
\multirow{2}{*}{Metric} & \multirow{2}{*}{Method} & \multicolumn{3}{c}{M=2}                                                                                                                                                                                                                    & \multicolumn{3}{c}{M=4}                                                                                                                                                                                                                    & \multicolumn{3}{c}{M=6}                                                                                                                                                                                                                    & \multicolumn{3}{c}{M=8}                                                                                                                                                                                                                     \\
\cmidrule(lr){3-5} \cmidrule(lr){6-8} \cmidrule(lr){9-11} \cmidrule(lr){12-14}
                        &                         & $\pi_{12}=7/3$                                                          & $8/2$                                                          & $9/1$                                                          & $\pi_{12}=7/3$                                                          & $8/2$                                                          & $9/1$                                                          & $\pi_{12}=7/3$                                                          & $8/2$                                                          & $9/1$                                                          & $\pi_{12} = 7/3$                                                          & $8/2$                                                          & $9/1$                                                           \\
                        \hline
\multirow{10}{*}{AUROC}  & equal                   & \begin{tabular}[c]{@{}c@{}}0.7983\\ (0.0053)\end{tabular}                    & \begin{tabular}[c]{@{}c@{}}0.8022\\ (0.0065)\end{tabular}                    & \begin{tabular}[c]{@{}c@{}}0.8069\\ (0.0082)\end{tabular}                    & \begin{tabular}[c]{@{}c@{}}0.8194\\ (0.0047)\end{tabular}                    & \begin{tabular}[c]{@{}c@{}}0.8235\\ (0.0060)\end{tabular}                    & \begin{tabular}[c]{@{}c@{}}0.8279\\ (0.0081)\end{tabular}                    & \begin{tabular}[c]{@{}c@{}}0.8363\\ (0.0038)\end{tabular}                    & \begin{tabular}[c]{@{}c@{}}0.8415\\ (0.0050)\end{tabular}                    & \begin{tabular}[c]{@{}c@{}}0.8464\\ (0.0072)\end{tabular}                    & \begin{tabular}[c]{@{}c@{}}0.8414\\ (0.0036)\end{tabular}                    & \begin{tabular}[c]{@{}c@{}}0.8476\\ (0.0046)\end{tabular}                    & \begin{tabular}[c]{@{}c@{}}0.8531\\ (0.0070)\end{tabular}                     \\
                        & ECV (K=5)               & \begin{tabular}[c]{@{}c@{}}0.8009\\ (0.0053)\end{tabular}                    & \begin{tabular}[c]{@{}c@{}}0.8062\\ (0.0067)\end{tabular}                    & \begin{tabular}[c]{@{}c@{}}\textbf{0.8118}\\\textbf{(0.0081)}\end{tabular}  & \begin{tabular}[c]{@{}c@{}}0.8217\\ (0.0055)\end{tabular}                    & \begin{tabular}[c]{@{}c@{}}0.8269\\ (0.0079)\end{tabular}                    & \begin{tabular}[c]{@{}c@{}}0.8331\\ (0.0082)\end{tabular}                    & \begin{tabular}[c]{@{}c@{}}0.8373\\ (0.0033)\end{tabular}                    & \begin{tabular}[c]{@{}c@{}}0.8451\\ (0.0040)\end{tabular}                    & \begin{tabular}[c]{@{}c@{}}0.8516\\ (0.0067)\end{tabular}                    & \begin{tabular}[c]{@{}c@{}}0.8375\\ (0.0033)\end{tabular}                    & \begin{tabular}[c]{@{}c@{}}0.8454\\ (0.0039)\end{tabular}                    & \begin{tabular}[c]{@{}c@{}}0.8522\\ (0.0067)\end{tabular}                     \\
                        & ECV (K=10)              & \begin{tabular}[c]{@{}c@{}}0.8009\\ (0.0053)\end{tabular}                    & \begin{tabular}[c]{@{}c@{}}0.8062\\ (0.0067)\end{tabular}                    & \begin{tabular}[c]{@{}c@{}}\textbf{0.8118}\\\textbf{(0.0081)}\end{tabular}  & \begin{tabular}[c]{@{}c@{}}0.8224\\ (0.0039)\end{tabular}                    & \begin{tabular}[c]{@{}c@{}}0.8284\\ (0.0061)\end{tabular}                    & \begin{tabular}[c]{@{}c@{}}\textbf{0.8341}\\\textbf{(0.0073)}\end{tabular}  & \begin{tabular}[c]{@{}c@{}}0.8375\\ (0.0033)\end{tabular}                    & \begin{tabular}[c]{@{}c@{}}0.8451\\ (0.0040)\end{tabular}                    & \begin{tabular}[c]{@{}c@{}}0.8516\\ (0.0067)\end{tabular}                    & \begin{tabular}[c]{@{}c@{}}0.8376\\ (0.0033)\end{tabular}                    & \begin{tabular}[c]{@{}c@{}}0.8455\\ (0.0039)\end{tabular}                    & \begin{tabular}[c]{@{}c@{}}0.8526\\ (0.0065)\end{tabular}                     \\
                        & NetMA (K=5)             & \begin{tabular}[c]{@{}c@{}}0.8018\\ (0.0054)\end{tabular}                    & \begin{tabular}[c]{@{}c@{}}0.8063\\ (0.0067)\end{tabular}                    & \begin{tabular}[c]{@{}c@{}}0.8111\\ (0.0081)\end{tabular}                    & \begin{tabular}[c]{@{}c@{}}0.8226\\ (0.0056)\end{tabular}                    & \begin{tabular}[c]{@{}c@{}}0.8265\\ (0.0079)\end{tabular}                    & \begin{tabular}[c]{@{}c@{}}0.8317\\ (0.0081)\end{tabular}                    & \begin{tabular}[c]{@{}c@{}}0.8419\\ (0.0034)\end{tabular}                    & \begin{tabular}[c]{@{}c@{}}0.8478\\ (0.0044)\end{tabular}                    & \begin{tabular}[c]{@{}c@{}}0.8526\\ (0.0070)\end{tabular}                    & \begin{tabular}[c]{@{}c@{}}0.8428\\ (0.0033)\end{tabular}                    & \begin{tabular}[c]{@{}c@{}}0.8492\\ (0.0043)\end{tabular}                    & \begin{tabular}[c]{@{}c@{}}0.8544\\ (0.0069)\end{tabular}                     \\
                        & NetMA (K=10)            & \begin{tabular}[c]{@{}c@{}}\textbf{0.8020}\\\textbf{(0.0054)}\end{tabular}  & \begin{tabular}[c]{@{}c@{}}\textbf{0.8066}\\\textbf{(0.0067)}\end{tabular}  & \begin{tabular}[c]{@{}c@{}}0.8116\\ (0.0081)\end{tabular}                    & \begin{tabular}[c]{@{}c@{}}\textbf{0.8237}\\\textbf{(0.0040)}\end{tabular}  & \begin{tabular}[c]{@{}c@{}}\textbf{0.8285}\\\textbf{(0.0063)}\end{tabular}  & \begin{tabular}[c]{@{}c@{}}0.8334\\ (0.0076)\end{tabular}                    & \begin{tabular}[c]{@{}c@{}}\textbf{0.8421}\\\textbf{(0.0033)}\end{tabular}  & \begin{tabular}[c]{@{}c@{}}\textbf{0.8484}\\\textbf{(0.0043)}\end{tabular}  & \begin{tabular}[c]{@{}c@{}}\textbf{0.8535}\\\textbf{(0.0069)}\end{tabular}  & \begin{tabular}[c]{@{}c@{}}\textbf{0.8431}\\\textbf{(0.0034)}\end{tabular}  & \begin{tabular}[c]{@{}c@{}}\textbf{0.8498}\\\textbf{(0.0043)}\end{tabular}  & \begin{tabular}[c]{@{}c@{}}\textbf{0.8554}\\\textbf{(0.007)}\end{tabular}    \\
                        \hline
\multirow{10}{*}{AUPR}   & equal                   & \begin{tabular}[c]{@{}c@{}}0.3812\\ (0.0098)\end{tabular}                    & \begin{tabular}[c]{@{}c@{}}0.3890\\ (0.0139)\end{tabular}                    & \begin{tabular}[c]{@{}c@{}}0.3965\\ (0.0170)\end{tabular}                    & \begin{tabular}[c]{@{}c@{}}0.4140\\ (0.0100)\end{tabular}                    & \begin{tabular}[c]{@{}c@{}}0.4230\\ (0.0143)\end{tabular}                    & \begin{tabular}[c]{@{}c@{}}0.4320\\ (0.0170)\end{tabular}                    & \begin{tabular}[c]{@{}c@{}}0.4423\\ (0.0088)\end{tabular}                    & \begin{tabular}[c]{@{}c@{}}0.4548\\ (0.0131)\end{tabular}                    & \begin{tabular}[c]{@{}c@{}}0.4658\\ (0.0167)\end{tabular}                    & \begin{tabular}[c]{@{}c@{}}0.4530\\ (0.0083)\end{tabular}                    & \begin{tabular}[c]{@{}c@{}}0.4686\\ (0.0127)\end{tabular}                    & \begin{tabular}[c]{@{}c@{}}0.4820\\ (0.0169)\end{tabular}                     \\
                        & ECV (K=5)               & \begin{tabular}[c]{@{}c@{}}0.3816\\ (0.0100)\end{tabular}                    & \begin{tabular}[c]{@{}c@{}}0.3930\\ (0.0144)\end{tabular}                    & \begin{tabular}[c]{@{}c@{}}0.4039\\ (0.0166)\end{tabular}                    & \begin{tabular}[c]{@{}c@{}}0.4092\\ (0.0104)\end{tabular}                    & \begin{tabular}[c]{@{}c@{}}0.4236\\ (0.0153)\end{tabular}                    & \begin{tabular}[c]{@{}c@{}}0.4383\\ (0.0180)\end{tabular}                    & \begin{tabular}[c]{@{}c@{}}0.4353\\ (0.0098)\end{tabular}                    & \begin{tabular}[c]{@{}c@{}}0.4564\\ (0.0129)\end{tabular}                    & \begin{tabular}[c]{@{}c@{}}0.4754\\ (0.0180)\end{tabular}                    & \begin{tabular}[c]{@{}c@{}}0.4359\\ (0.0096)\end{tabular}                    & \begin{tabular}[c]{@{}c@{}}0.4575\\ (0.0125)\end{tabular}                    & \begin{tabular}[c]{@{}c@{}}0.4773\\ (0.0183)\end{tabular}                     \\
                        & ECV (K=10)              & \begin{tabular}[c]{@{}c@{}}0.3816\\ (0.0100)\end{tabular}                    & \begin{tabular}[c]{@{}c@{}}0.3930\\ (0.0144)\end{tabular}                    & \begin{tabular}[c]{@{}c@{}}0.4039\\ (0.0166)\end{tabular}                    & \begin{tabular}[c]{@{}c@{}}0.4104\\ (0.0088)\end{tabular}                    & \begin{tabular}[c]{@{}c@{}}0.4256\\ (0.0138)\end{tabular}                    & \begin{tabular}[c]{@{}c@{}}0.4397\\ (0.0171)\end{tabular}                    & \begin{tabular}[c]{@{}c@{}}0.4356\\ (0.0096)\end{tabular}                    & \begin{tabular}[c]{@{}c@{}}0.4564\\ (0.0129)\end{tabular}                    & \begin{tabular}[c]{@{}c@{}}0.4754\\ (0.0180)\end{tabular}                    & \begin{tabular}[c]{@{}c@{}}0.4360\\ (0.0094)\end{tabular}                    & \begin{tabular}[c]{@{}c@{}}0.4579\\ (0.0120)\end{tabular}                    & \begin{tabular}[c]{@{}c@{}}0.4787\\ (0.0173)\end{tabular}                     \\
                        & NetMA (K=5)             & \begin{tabular}[c]{@{}c@{}}\textbf{0.3857}\\\textbf{(0.0098)}\end{tabular}  & \begin{tabular}[c]{@{}c@{}}0.3953\\ (0.0142)\end{tabular}                    & \begin{tabular}[c]{@{}c@{}}0.4041\\ (0.0169)\end{tabular}                    & \begin{tabular}[c]{@{}c@{}}0.4174\\ (0.0110)\end{tabular}                    & \begin{tabular}[c]{@{}c@{}}0.4287\\ (0.0157)\end{tabular}                    & \begin{tabular}[c]{@{}c@{}}0.4402\\ (0.0171)\end{tabular}                    & \begin{tabular}[c]{@{}c@{}}\textbf{0.4530}\\\textbf{(0.0088)}\end{tabular}  & \begin{tabular}[c]{@{}c@{}}\textbf{0.4696}\\\textbf{(0.0127)}\end{tabular}  & \begin{tabular}[c]{@{}c@{}}0.4831\\ (0.0169)\end{tabular}                    & \begin{tabular}[c]{@{}c@{}}\textbf{0.4549}\\\textbf{(0.0086)}\end{tabular}  & \begin{tabular}[c]{@{}c@{}}\textbf{0.4731}\\\textbf{(0.0122)}\end{tabular}  & \begin{tabular}[c]{@{}c@{}}0.4880\\ (0.0174)\end{tabular}                     \\
                        & NetMA (K=10)            & \begin{tabular}[c]{@{}c@{}}0.3854\\ (0.0098)\end{tabular}                    & \begin{tabular}[c]{@{}c@{}}\textbf{0.3955}\\\textbf{(0.0142)}\end{tabular}  & \begin{tabular}[c]{@{}c@{}}\textbf{0.4047}\\\textbf{(0.0169)}\end{tabular}  & \begin{tabular}[c]{@{}c@{}}\textbf{0.4179}\\\textbf{(0.0087)}\end{tabular}  & \begin{tabular}[c]{@{}c@{}}\textbf{0.4309}\\\textbf{(0.0140)}\end{tabular}  & \begin{tabular}[c]{@{}c@{}}\textbf{0.4423}\\\textbf{(0.0168)}\end{tabular}  & \begin{tabular}[c]{@{}c@{}}0.4515\\ (0.0089)\end{tabular}                    & \begin{tabular}[c]{@{}c@{}}0.4693\\ (0.0126)\end{tabular}                    & \begin{tabular}[c]{@{}c@{}}\textbf{0.4838}\\\textbf{(0.0171)}\end{tabular}  & \begin{tabular}[c]{@{}c@{}}0.4536\\ (0.0084)\end{tabular}                    & \begin{tabular}[c]{@{}c@{}}0.4730\\ (0.0123)\end{tabular}                    & \begin{tabular}[c]{@{}c@{}}\textbf{0.4889}\\\textbf{(0.0175)}\end{tabular}   \\
                        \hline
\multirow{10}{*}{Mlogf}  & equal                   & \begin{tabular}[c]{@{}c@{}}\textbf{-0.2951}\\\textbf{(0.0054)}\end{tabular} & \begin{tabular}[c]{@{}c@{}}-0.2902\\ (0.0063)\end{tabular}                   & \begin{tabular}[c]{@{}c@{}}-0.2855\\ (0.0087)\end{tabular}                   & \begin{tabular}[c]{@{}c@{}}\textbf{-0.2821}\\\textbf{(0.0048)}\end{tabular} & \begin{tabular}[c]{@{}c@{}}-0.2772\\ (0.0055)\end{tabular}                   & \begin{tabular}[c]{@{}c@{}}-0.2727\\ (0.0079)\end{tabular}                   & \begin{tabular}[c]{@{}c@{}}-0.2700\\ (0.0044)\end{tabular}                   & \begin{tabular}[c]{@{}c@{}}-0.2646\\ (0.0050)\end{tabular}                   & \begin{tabular}[c]{@{}c@{}}-0.2599\\ (0.0075)\end{tabular}                   & \begin{tabular}[c]{@{}c@{}}\textbf{-0.2669}\\\textbf{(0.0046)}\end{tabular} & \begin{tabular}[c]{@{}c@{}}-0.2607\\ (0.0052)\end{tabular}                   & \begin{tabular}[c]{@{}c@{}}-0.2554\\ (0.0078)\end{tabular}                    \\
                        & ECV (K=5)               & \begin{tabular}[c]{@{}c@{}}-0.3048\\ (0.0059)\end{tabular}                   & \begin{tabular}[c]{@{}c@{}}-0.2968\\ (0.0072)\end{tabular}                   & \begin{tabular}[c]{@{}c@{}}-0.2896\\ (0.0095)\end{tabular}                   & \begin{tabular}[c]{@{}c@{}}-0.2993\\ (0.0053)\end{tabular}                   & \begin{tabular}[c]{@{}c@{}}-0.2890\\ (0.0065)\end{tabular}                   & \begin{tabular}[c]{@{}c@{}}-0.2804\\ (0.0089)\end{tabular}                   & \begin{tabular}[c]{@{}c@{}}-0.3017\\ (0.0053)\end{tabular}                   & \begin{tabular}[c]{@{}c@{}}-0.2862\\ (0.0059)\end{tabular}                   & \begin{tabular}[c]{@{}c@{}}-0.2734\\ (0.0090)\end{tabular}                   & \begin{tabular}[c]{@{}c@{}}-0.3021\\ (0.0054)\end{tabular}                   & \begin{tabular}[c]{@{}c@{}}-0.2869\\ (0.0063)\end{tabular}                   & \begin{tabular}[c]{@{}c@{}}-0.2737\\ (0.0094)\end{tabular}                    \\
                        & ECV (K=10)              & \begin{tabular}[c]{@{}c@{}}-0.3048\\ (0.0059)\end{tabular}                   & \begin{tabular}[c]{@{}c@{}}-0.2968\\ (0.0072)\end{tabular}                   & \begin{tabular}[c]{@{}c@{}}-0.2896\\ (0.0095)\end{tabular}                   & \begin{tabular}[c]{@{}c@{}}-0.2992\\ (0.0053)\end{tabular}                   & \begin{tabular}[c]{@{}c@{}}-0.2886\\ (0.0061)\end{tabular}                   & \begin{tabular}[c]{@{}c@{}}-0.2801\\ (0.0090)\end{tabular}                   & \begin{tabular}[c]{@{}c@{}}-0.3017\\ (0.0053)\end{tabular}                   & \begin{tabular}[c]{@{}c@{}}-0.2862\\ (0.0059)\end{tabular}                   & \begin{tabular}[c]{@{}c@{}}-0.2734\\ (0.0090)\end{tabular}                   & \begin{tabular}[c]{@{}c@{}}-0.3023\\ (0.0056)\end{tabular}                   & \begin{tabular}[c]{@{}c@{}}-0.2873\\ (0.0062)\end{tabular}                   & \begin{tabular}[c]{@{}c@{}}-0.2737\\ (0.0093)\end{tabular}                    \\
                        & NetMA (K=5)             & \begin{tabular}[c]{@{}c@{}}-0.2962\\ (0.0055)\end{tabular}                   & \begin{tabular}[c]{@{}c@{}}\textbf{-0.2900}\\\textbf{(0.0066)}\end{tabular} & \begin{tabular}[c]{@{}c@{}}\textbf{-0.2846}\\\textbf{(0.0089)}\end{tabular} & \begin{tabular}[c]{@{}c@{}}-0.2830\\ (0.0051)\end{tabular}                   & \begin{tabular}[c]{@{}c@{}}\textbf{-0.2770}\\\textbf{(0.0067)}\end{tabular} & \begin{tabular}[c]{@{}c@{}}-0.2713\\ (0.0081)\end{tabular}                   & \begin{tabular}[c]{@{}c@{}}\textbf{-0.2680}\\\textbf{(0.0043)}\end{tabular} & \begin{tabular}[c]{@{}c@{}}\textbf{-0.2609}\\\textbf{(0.0046)}\end{tabular} & \begin{tabular}[c]{@{}c@{}}\textbf{-0.2554}\\\textbf{(0.0075)}\end{tabular} & \begin{tabular}[c]{@{}c@{}}-0.2673\\ (0.0045)\end{tabular}                   & \begin{tabular}[c]{@{}c@{}}\textbf{-0.2598}\\\textbf{(0.0049)}\end{tabular} & \begin{tabular}[c]{@{}c@{}}\textbf{-0.2539}\\\textbf{(0.0078)}\end{tabular}  \\
                        & NetMA (K=10)            & \begin{tabular}[c]{@{}c@{}}-0.2972\\ (0.0055)\end{tabular}                   & \begin{tabular}[c]{@{}c@{}}-0.2905\\ (0.0066)\end{tabular}                   & \begin{tabular}[c]{@{}c@{}}-0.2848\\ (0.0089)\end{tabular}                   & \begin{tabular}[c]{@{}c@{}}-0.2842\\ (0.0048)\end{tabular}                   & \begin{tabular}[c]{@{}c@{}}\textbf{-0.2770}\\\textbf{(0.0058)}\end{tabular} & \begin{tabular}[c]{@{}c@{}}\textbf{-0.2711}\\\textbf{(0.0082)}\end{tabular} & \begin{tabular}[c]{@{}c@{}}-0.2703\\ (0.0043)\end{tabular}                   & \begin{tabular}[c]{@{}c@{}}-0.2619\\ (0.0047)\end{tabular}                   & \begin{tabular}[c]{@{}c@{}}-0.2557\\ (0.0076)\end{tabular}                   & \begin{tabular}[c]{@{}c@{}}-0.2695\\ (0.0046)\end{tabular}                   & \begin{tabular}[c]{@{}c@{}}-0.2607\\ (0.0050)\end{tabular}                   & \begin{tabular}[c]{@{}c@{}}-0.2541\\ (0.0079)\end{tabular}                    \\
                        \hline
\multirow{10}{*}{MSE}    & equal                   & \begin{tabular}[c]{@{}c@{}}0.0817\\ (0.0012)\end{tabular}                    & \begin{tabular}[c]{@{}c@{}}0.0810\\ (0.0015)\end{tabular}                    & \begin{tabular}[c]{@{}c@{}}0.0802\\ (0.0023)\end{tabular}                    & \begin{tabular}[c]{@{}c@{}}0.0790\\ (0.0012)\end{tabular}                    & \begin{tabular}[c]{@{}c@{}}0.0782\\ (0.0015)\end{tabular}                    & \begin{tabular}[c]{@{}c@{}}0.0774\\ (0.0022)\end{tabular}                    & \begin{tabular}[c]{@{}c@{}}0.0766\\ (0.0012)\end{tabular}                    & \begin{tabular}[c]{@{}c@{}}0.0755\\ (0.0014)\end{tabular}                    & \begin{tabular}[c]{@{}c@{}}0.0746\\(0.0021)\end{tabular}  & \begin{tabular}[c]{@{}c@{}}0.0757\\ (0.0012)\end{tabular}                    & \begin{tabular}[c]{@{}c@{}}0.0744\\ (0.0014)\end{tabular}                    & \begin{tabular}[c]{@{}c@{}}0.0732\\ (0.0022)\end{tabular}                     \\
                        & ECV (K=5)               & \begin{tabular}[c]{@{}c@{}}0.0822\\ (0.0013)\end{tabular}                    & \begin{tabular}[c]{@{}c@{}}0.0812\\ (0.0016)\end{tabular}                    & \begin{tabular}[c]{@{}c@{}}0.0801\\ (0.0024)\end{tabular}                    & \begin{tabular}[c]{@{}c@{}}0.0802\\ (0.0012)\end{tabular}                    & \begin{tabular}[c]{@{}c@{}}0.0788\\ (0.0016)\end{tabular}                    & \begin{tabular}[c]{@{}c@{}}0.0775\\ (0.0023)\end{tabular}                    & \begin{tabular}[c]{@{}c@{}}0.0788\\ (0.0012)\end{tabular}                    & \begin{tabular}[c]{@{}c@{}}0.0767\\ (0.0013)\end{tabular}                    & \begin{tabular}[c]{@{}c@{}}0.0748\\ (0.0022)\end{tabular}                    & \begin{tabular}[c]{@{}c@{}}0.0787\\ (0.0012)\end{tabular}                    & \begin{tabular}[c]{@{}c@{}}0.0766\\ (0.0013)\end{tabular}                    & \begin{tabular}[c]{@{}c@{}}0.0747\\ (0.0022)\end{tabular}                     \\
                        & ECV (K=10)              & \begin{tabular}[c]{@{}c@{}}0.0822\\ (0.0013)\end{tabular}                    & \begin{tabular}[c]{@{}c@{}}0.0812\\ (0.0016)\end{tabular}                    & \begin{tabular}[c]{@{}c@{}}0.0801\\ (0.0024)\end{tabular}                    & \begin{tabular}[c]{@{}c@{}}0.0801\\ (0.0012)\end{tabular}                    & \begin{tabular}[c]{@{}c@{}}0.0787\\ (0.0015)\end{tabular}                    & \begin{tabular}[c]{@{}c@{}}0.0774\\ (0.0023)\end{tabular}                    & \begin{tabular}[c]{@{}c@{}}0.0788\\ (0.0012)\end{tabular}                    & \begin{tabular}[c]{@{}c@{}}0.0767\\ (0.0013)\end{tabular}                    & \begin{tabular}[c]{@{}c@{}}0.0748\\ (0.0022)\end{tabular}                    & \begin{tabular}[c]{@{}c@{}}0.0788\\ (0.0012)\end{tabular}                    & \begin{tabular}[c]{@{}c@{}}0.0766\\ (0.0013)\end{tabular}                    & \begin{tabular}[c]{@{}c@{}}0.0746\\ (0.0022)\end{tabular}                     \\
                        & NetMA (K=5)             & \begin{tabular}[c]{@{}c@{}}\textbf{0.0816}\\\textbf{(0.0012)}\end{tabular}  & \begin{tabular}[c]{@{}c@{}}\textbf{0.0807}\\\textbf{(0.0016)}\end{tabular}  & \begin{tabular}[c]{@{}c@{}}\textbf{0.0798}\\\textbf{(0.0024)}\end{tabular}  & \begin{tabular}[c]{@{}c@{}}\textbf{0.0789}\\\textbf{(0.0013)}\end{tabular}  & \begin{tabular}[c]{@{}c@{}}\textbf{0.0778}\\\textbf{(0.0016)}\end{tabular}  & \begin{tabular}[c]{@{}c@{}}0.0767\\ (0.0022)\end{tabular}                    & \begin{tabular}[c]{@{}c@{}}\textbf{0.0758}\\\textbf{(0.0011)}\end{tabular}  & \begin{tabular}[c]{@{}c@{}}\textbf{0.0743}\\\textbf{(0.0013)}\end{tabular}  & \begin{tabular}[c]{@{}c@{}}\textbf{0.0731}\\\textbf{(0.0021)}\end{tabular}  & \begin{tabular}[c]{@{}c@{}}\textbf{0.0756}\\\textbf{(0.0012)}\end{tabular}  & \begin{tabular}[c]{@{}c@{}}\textbf{0.0740}\\\textbf{(0.0013)}\end{tabular}  & \begin{tabular}[c]{@{}c@{}}\textbf{0.0727}\\\textbf{(0.0022)}\end{tabular}   \\
                        & NetMA (K=10)            & \begin{tabular}[c]{@{}c@{}}0.0817\\ (0.0012)\end{tabular}                    & \begin{tabular}[c]{@{}c@{}}\textbf{0.0807}\\\textbf{(0.0016)}\end{tabular}  & \begin{tabular}[c]{@{}c@{}}\textbf{0.0798}\\\textbf{(0.0024)}\end{tabular}  & \begin{tabular}[c]{@{}c@{}}0.0790\\ (0.0012)\end{tabular}                    & \begin{tabular}[c]{@{}c@{}}\textbf{0.0778}\\\textbf{(0.0015)}\end{tabular}  & \begin{tabular}[c]{@{}c@{}}\textbf{0.0766}\\\textbf{(0.0022)}\end{tabular}  & \begin{tabular}[c]{@{}c@{}}0.0762\\ (0.0011)\end{tabular}                    & \begin{tabular}[c]{@{}c@{}}0.0745\\ (0.0013)\end{tabular}                    & \begin{tabular}[c]{@{}c@{}}\textbf{0.0731}\\\textbf{(0.0021)}\end{tabular}  & \begin{tabular}[c]{@{}c@{}}0.0759\\ (0.0012)\end{tabular}                    & \begin{tabular}[c]{@{}c@{}}0.0741\\ (0.0013)\end{tabular}                    & \begin{tabular}[c]{@{}c@{}}\textbf{0.0727}\\\textbf{(0.0022)}\end{tabular} 
                        \\
                        \bottomrule
\end{tabular}
\end{adjustbox}
\end{table}
\end{landscape}

\bibliographystyle{agsm}

\bibliography{ref}
\end{document}


\def\spacingset#1{\renewcommand{\baselinestretch}%
{#1}\small\normalsize} \spacingset{1}


\if1\blind
{
  \title{\bf Supplementary Material for ``Network Model Averaging Prediction for Latent Space Models by $K$-Fold Edge Cross-Validation''}
  \author{Yan Zhang\thanks{Yan Zhang, Jun Liao and Xinyan Fan are co-first authors and contributed equally to this work.} \hspace{.2cm}\\
    School of Economics, Xiamen University\\
    Jun Liao\hspace{.2cm}\\
    School of Statistics, Renmin University of China\\
    Xinyan Fan\hspace{.2cm}\\
    School of Statistics, Renmin University of China\\
    Kuangnan Fang\thanks{Corresponding author. E-mail address: xmufkn@xmu.edu.cn.
    }\hspace{.2cm}\\
    School of Economics, Xiamen University\\
    and \\
    Yuhong Yang \\
    Yau Mathematical Sciences Center, Tsinghua University}
  \maketitle
} \fi

\if0\blind
{
  \bigskip
  \bigskip
  \bigskip
  \begin{center}
    {\LARGE\bf Supplementary Material for ``Network Model Averaging Prediction for Latent Space Models by $K$-Fold Edge Cross-Validation''}
\end{center}
  \medskip
} \fi

\bigskip
\begin{abstract}
The supplementary material contains the proofs for the theoretical results and the verifications of the assumptions. Specifically, Section \ref{s4} gives the proofs of Theorems 1-6. Section \ref{s7} gives verifications of Assumptions 1-3.
\end{abstract}

\spacingset{1.8} 

\section{Proof of Theoretical Results}
\label{s4}
\subsection{Proof of Theorem 1}
\label{appendix1}
This subsection gives the proof of Theorem 1. We need the following lemmas.
	
\begin{lemma}[\cite{gao2019frequentist,zhang2022model}]
    \label{lemma1}
    Let
    $$\widetilde{w}=\underset{w \in \mathcal{W}}{\operatorname{argmin}}\left\{L(w)+a_n(w)+b_n\right\},$$
    where $a_n(w)$ is a term related to $w$, and $b_n$ is a term unrelated to $w$. If
    $$\sup _{w \in \mathcal{W}}\left|a_n(w)\right| / L^*(w)=o_p(1),  \quad \sup _{w \in \mathcal{W}}\left|L^*(w)-L(w)\right| / L^*(w)=o_p(1),$$
    and there exists a constant $c$ and a positive integer $N^*$ so that when $n \geq N^*, \inf _{w \in \mathcal{W}} L^*(w) \geq c>0$ almost surely, then $L(\widetilde{w}) / \inf _{w \in \mathcal{W}} L(w) \rightarrow 1$ in probability.
\end{lemma}

\begin{lemma}[\cite{10.1214/14-AOS1274,ma2020universal}]
    \label{lemma2-new}
    Let $A$ be the symmetric adjacency matrix of a random graph on $N$ nodes with independent edges. Set $E(A_{ij}) = P_{ij}$ for all $i \neq j$ and $P_{ii} \in [0,1]$. Assume that $N \max_{i,j} P_{ij} \leq D$. Then for any $C_0$, there exists a constant $C=C(C_0)$ such that 
    $$
    \|A-P\|_2 \leq C \sqrt{D+\log N}
    $$
    with probability at least $1-N^{-C_0}$.
\end{lemma}

Let $CV^*(w) = CV(w) -  \lvert \Psi_1 \rvert^{-1}\sum_{k=1}^{K} ( \|A^{[k]}\|_F^2 - \|P^{[k]}\|_F^2 )$, where $P^{[k]} = P \circ S^{[k]}$, and the second term is unrelated to $w$. Therefore, 
\begin{equation*}
    \widehat{w} = \argmin_{w \in \mathcal{W}} CV(w) = \argmin_{w \in \mathcal{W}} \lvert \Psi_1 \rvert CV^*(w).
\end{equation*}
According to Lemma \ref{lemma1}, Theorem 1 is valid if the following conditions hold:
\begin{equation}
    \label{equ:condition1}
    \sup_{w \in \mathcal{W}} \frac{\lvert L(w) - L^*(w) \rvert}{ L^*(w)} = o_p(1),
\end{equation}
and
\begin{equation}
    \label{equ:condition2}
    \sup_{w \in \mathcal{W}} \frac{\lvert \lvert \Psi_1 \rvert CV^*(w) - L^*(w) \rvert}{ L^*(w)} = o_p(1).
\end{equation}
We first consider (\ref{equ:condition1}). It can be shown that 
\begin{align}
    &\sup_{w \in \mathcal{W}} \frac{\lvert L(w) - L^*(w) \rvert}{ L^*(w)} 
    = \sup_{w \in \mathcal{W}} \left[ \bigg \lvert  \sum_{(i,j) \in \Psi_1} \left\{\left(\widehat{P}_{ij}(w) - P_{ij}\right)^2 - \Big(P^*_{ij}(w) - P_{ij}\Big)^2 \right\} \bigg \rvert / L^*(w) \right]\nonumber \\
    \leq & \sup_{w \in \mathcal{W}} \left[ \left\lvert \sum_{(i,j) \in \Psi_1} \left\{\left(\widehat{P}_{ij}(w) - P^*_{ij}(w)\right)^2 + 2 \left(\widehat{P}_{ij}(w) - P^*_{ij}(w)\right) \Big( P^*_{ij}(w) - P_{ij}\Big) \right\} \right\rvert / L^*(w) \right] \nonumber \\
    \leq & \xi^{*-1} \sup_{w \in \mathcal{W}} \sum_{(i,j) \in \Psi_1} \left\{ \sum_{m=1}^M w_m \left(\widehat{P}_{(m),ij} - P^*_{(m),ij}\right) \right\}^2  \nonumber \\
    & + 2 \sup_{w \in \mathcal{W}} \left[ \frac{1}{L^*(w)} \Bigg\{ \sum_{(i,j) \in \Psi_1} \left(\widehat{P}_{ij}(w) - P^*_{ij}(w)\right)^2 \Bigg\}^{1/2} \Bigg\{ \sum_{(i,j) \in \Psi_1} \Big( P^*_{ij}(w) - P_{ij}\Big)^2 \Bigg\}^{1/2} \right] \nonumber \\
    \leq & \xi^{*-1} \sup_{w \in \mathcal{W}} \sum_{(i,j) \in \Psi_1} \sum_{m=1}^M w_m \left(\widehat{P}_{(m),ij} - P^*_{(m),ij}\right)^2 
    + 2 \xi^{*-1/2}\sup_{w \in \mathcal{W}} \Bigg\{ \sum_{(i,j) \in \Psi_1} \left(\widehat{P}_{ij}(w) - P^*_{ij}(w)\right)^2 \Bigg\}^{1/2} \nonumber \\
    = & \xi^{*-1} O_p\left(N M \right) + \xi^{*-1/2} O_p\left(N^{1/2} M^{1/2} \right) = o_p(1), \label{L-L*/L*}
\end{align}
where the fifth step uses Assumption 1, and the sixth step uses Assumption 2. Thus, we obtain (\ref{equ:condition1}).

We next verify (\ref{equ:condition2}). Observe that
\begin{align}
    \label{equ:5}
    &\sup_{w \in \mathcal{W}} \left\{\big\lvert \lvert \Psi_1 \rvert CV^*(w) - L^*(w) \big\rvert / L^*(w) \right\} \nonumber  \\
    =& \sup_{w \in \mathcal{W}}\left\{ \left \lvert   \|a_1 - \widetilde{p}_1(w) \|^2 - (a_1 - p_1)^\top (a_1 + p_1)  - \| p_1^*(w) - p_1\|^2 \right \rvert /L^*(w) \right\}\nonumber  \\
    \leq& \sup_{w \in \mathcal{W}} \left\{ \big \lvert  \|a_1 - p_1^*(w) \|^2 - (a_1 - p_1)^\top (a_1 + p_1)  - \| p_1^*(w) - p_1\|^2  \big \rvert /L^*(w)\right\} \nonumber  \\
    & + \sup_{w \in \mathcal{W}} \left\{ \big \lvert \|a_1 - \widetilde{p}_1(w) \|^2 - \|a_1 - p_1^*(w) \|^2 \big \rvert /L^*(w) \right\} \nonumber  \\
    =&  \sup_{w \in \mathcal{W}} \left\{ \big \lvert  \| p_1^*(w) - p_1 - a_1 + p_1 \|^2 - (a_1 - p_1)^\top (a_1 + p_1) - \| p_1^*(w) - p_1\|^2 \big \rvert /L^*(w)  \right\} \nonumber \\
    & +  \sup_{w \in \mathcal{W}} \left\{ \big \lvert \|a_1 - \widetilde{p}_1(w) \|^2 - \|a_1 - p_1^*(w) \|^2 \big \rvert /L^*(w) \right\} \nonumber \\
    \leq &  2 \xi^{*-1} \sup_{w \in \mathcal{W}} \big \lvert p_1^*(w)^{\top}(a_1 - p_1) \big \rvert + \sup_{w \in \mathcal{W}} \left\{\big \lvert \|a_1 - \widetilde{p}_1(w) \|^2 - \|a_1 - p_1^*(w) \|^2 \big \rvert/L^*(w) \right\}. 
\end{align}	 
Similar to the derivation in \eqref{L-L*/L*}, based on Assumptions 1 and 2, we can obtain	
\begin{align}
    & \sup_{w \in \mathcal{W}} \left\{ \big \lvert \|a_1 - \widetilde{p}_1(w) \|^2 - \|a_1 - p_1^*(w) \|^2 \big \rvert /L^*(w) \right\} \nonumber \\
    = & \sup_{w \in \mathcal{W}} \left\{\left\lvert \| p^*_{1}(w)-\widetilde{p}_{1}(w)\|^2 + 2 \big(a_1-p^*_{1}(w)\big)^\top \big( p^*_{1}(w)-\widetilde{p}_{1}(w) \big) \right\rvert /L^*(w)\right\}\nonumber  \\
    \leq & \xi^{*-1} \sup_{w \in \mathcal{W}} \left\{ \sum_{(i,j)\in \Psi_1} \left( P_{ij}^*(w) - \widetilde{P}_{ij}(w) \right)^2\right\} + \nonumber  \\
    & 2 \sup_{w \in \mathcal{W}} \left\{ \left\lvert \sum_{(i,j)\in \Psi_1} \left(A_{ij} - P_{ij}^*(w)\right)\left(P_{ij}^*(w)-\widetilde{P}_{ij}(w)\right) \right\rvert /L^*(w) \right\} \nonumber  \\
    \leq & \xi^{*-1} \sup_{w \in \mathcal{W}} \left[  \sum_{(i,j) \in \Psi_1} \left\{ \sum_{m=1}^M w_m \left(P^*_{(m),ij} - \widetilde{P}_{(m),ij}\right) \right\}^2 \right]  \nonumber \\
    & + 2 \sup_{w \in \mathcal{W}} \left[ \frac{1}{L^*(w)} \Bigg\{ \sum_{(i,j) \in \Psi_1} \left(A_{ij} - P_{ij}^*(w)\right)^2 \Bigg\}^{1/2} \Bigg\{ \sum_{(i,j) \in \Psi_1} \Big( P_{ij}^*(w)-\widetilde{P}_{ij}(w)\Big)^2 \Bigg\}^{1/2} \right] \nonumber \\
    \leq & \xi^{*-1} \sup_{w \in \mathcal{W}} \left\{ \sum_{(i,j) \in \Psi_1} \sum_{m=1}^M w_m \left(P^*_{(m),ij} - \widetilde{P}_{(m),ij}\right)^2 \right\} \nonumber \\
    & + 2 \sup_{w \in \mathcal{W}} \left[ \Bigg\{ \frac{1}{L^*(w)} \sum_{(i,j) \in \Psi_1} \left(A_{ij} - P_{ij}^*(w)\right)^2 \Bigg\}^{1/2} \Bigg\{ \frac{1}{L^*(w)} \sum_{(i,j) \in \Psi_1} \Big( P_{ij}^*(w)-\widetilde{P}_{ij}(w)\Big)^2 \Bigg\}^{1/2} \right] \nonumber \\
    \leq & \xi^{*-1} O_p(NM) + 2 \left\{\xi^{*-1} O_p\left(NM\right)\right\}^{1/2} \sup_{w \in \mathcal{W}} \Bigg\{ \frac{1}{L^*(w)} \sum_{(i,j) \in \Psi_1} \left(A_{ij} - P_{ij}^*(w)\right)^2 \Bigg\}^{1/2} \nonumber\\
    = & o_p(1) + 2o_p(1)\sup_{w \in \mathcal{W}} \Bigg\{ \frac{1}{L^*(w)} \sum_{(i,j) \in \Psi_1} \left(A_{ij} - P_{ij}^*(w)\right)^2 \Bigg\}^{1/2}. \label{equ:6.1} 
\end{align}
From Lemma \ref{lemma2-new}, we have $P\big(\|A - P\|_2 > C \sqrt{D+\log N}\big) \leq N^{-C_0}$ for any $C_0 > 0$. Then $\|A - P\|_2 = O_p(\max\{D, \log N\}^{1/2})$. Subsequently, we can obtain 
\begin{equation}
    \|p_1 - a_1\|^2  \leq \| A - P\|_F^2 \leq  \rank(A - P) \|A - P\|_2^2 = O_p\left(N\max\{D, \log N\}\right). 
\label{p1-a1}
\end{equation}
Thus, together with Assumption 2, we have
\begin{align}
    & \sup_{w \in \mathcal{W}} \Bigg\{ \frac{1}{L^*(w)} \sum_{(i,j) \in \Psi_1} \left(A_{ij} - P_{ij}^*(w)\right)^2 \Bigg\}^{1/2} \nonumber\\
    \leq & \sup_{w \in \mathcal{W}} \left[ \frac{1}{L^*(w)} \sum_{(i,j) \in \Psi_1} \left\{ 2 \left(A_{ij} - P_{ij}\right)^2 + 2 \left( P_{ij} - P_{ij}^*(w) \right)^2 \right\} \right]^{1/2} \nonumber\\
    \leq & \sqrt{2} \xi^{*-1/2} O_p\left(N^{1/2}\max\{D, \log N\}^{1/2}\right) + \sqrt{2} = O_p(1). \label{equ:6.2}
\end{align}
According to \eqref{equ:6.1} and \eqref{equ:6.2}, we have 
\begin{equation}
\label{equ:6}
    \sup_{w \in \mathcal{W}} \left\{ \big \lvert \|a_1 - \widetilde{p}_1(w) \|^2 - \|a_1 - p_1^*(w) \|^2 \big \rvert /L^*(w) \right\} = o_p(1).
\end{equation}
For the first summand in \eqref{equ:5}, we first denote the $l$th element of $a_1,~p_1$ and $p_{1(m)}^*$ to be $a_{1,l},~p_{1,l}$ and $p_{1(m),l}^*$, respectively. Then
it follows that for any $\delta > 0$,
\begin{align}
    \label{equ:8}
    &P \left\{\xi^{*-1} \sup_{w \in \mathcal{W}} \left \lvert p_1^{*\top}(w)(a_1 - p_1) \right \rvert > \delta \right \} \nonumber \\
    =&P \bigg \{\xi^{*-1} \sup_{w \in \mathcal{W}} \bigg \lvert \sum_{l=1}^{ \lvert \Psi_1 \rvert} \left\{p_{1,l}^*(w) (a_{1,l} - p_{1,l})\right\} \bigg \rvert > \delta \bigg \}  \nonumber  \\  
    =& P \bigg \{ \xi^{*-1} \sup_{w \in \mathcal{W}} \bigg \lvert \sum_{l=1}^{ \lvert \Psi_1 \rvert} \bigg \{ \sum_{m=1}^{M} w_m p_{1(m),l}^* (a_{1,l} - p_{1,l}) \bigg\} \bigg \rvert > \delta \bigg \}  \nonumber  \\  
    \leq & \sum_{m=1}^{M} P \bigg\{ \bigg\lvert  \sum_{l=1}^{ \lvert \Psi_1 \rvert} \left(p_{1(m),l}^*a_{1,l} - p_{1(m),l}^*p_{1,l}\right) \bigg \rvert > \xi^*\delta \bigg\} \nonumber  \\  
    \leq & \xi^{*-2} \delta^{-2} \sum_{l=1}^{ \lvert \Psi_1 \rvert} \sum_{m=1}^{M} \var \left(p_{1(m),l}^*a_{1,l} - p_{1(m),l}^*p_{1,l}\right) \nonumber  \\  
    \leq & \xi^{*-2} \delta^{-2} M \lvert \Psi_1 \rvert = o(1), 
\end{align}
where the third step uses Boole’s inequality and the fourth step uses Chebyshev’s Inequality. For the fifth step, since $p_{1(m),l}^*,\ p_{1,l} \in [0,1]$ and $a_{1,l} \in \{0,1\}$ for any $m = 1,\dots,M$ and $l=1,\dots,\lvert \Psi_1 \rvert$, we have $\big(p_{1(m),l}^*a_{1,l} - p_{1(m),l}^*p_{1,l}\big)^2 \leq 1$. Therefore, we can get $ \var \big(p_{1(m),l}^*a_{1,l} - p_{1(m),l}^*p_{1,l}\big) = E \big(p_{1(m),l}^*a_{1,l} - p_{1(m),l}^*p_{1,l}\big)^2 \leq 1$. The sixth step uses the fact that $\lvert \Psi_1 \rvert$ has the same order of $N^2$ and Assumption 2.
By (\ref{equ:5}), \eqref{equ:6} and (\ref{equ:8}), we obtain (\ref{equ:condition2}). This completes the proof of Theorem 1.	

\subsection{Proof of Theorem 2}
\label{theorem2}
This subsection gives the proof of Theorem 2. 
Similar to \eqref{equ:5}, we have 
\begin{equation}
    \label{CV-L}
    \left\lvert \lvert \Psi_1 \rvert CV^*(w) - L^*(w) \right\rvert \leq \lvert p_1^*(w)^\top (a_1-p_1) \rvert + \left\lvert \|a_1-\widetilde{p}_1(w)\|^2 - \|a_1 - p_1^*(w) \|^2 \right\rvert.
\end{equation}
From \eqref{equ:6.1}, \eqref{equ:6.2}, and Assumption 1, we can get
\begin{align}
    \label{CV-L_1}
    &\left\lvert \|a_1-\widetilde{p}_1(w)\|^2 - \|a_1 - p_1^*(w) \|^2 \right\rvert \nonumber \\
    = &O_p(NM) + O_p(N M^{1/2} \max\{D, \log N\}^{1/2}) +(L^*(w))^{1/2} O_p(N^{1/2}M^{1/2}).
\end{align}
In addition, similar to \eqref{equ:8}, it is seen that  
\begin{equation}
    \label{CV-L_2}
    \sup_{w \in \mathcal{W}} \left \lvert p_1^{*\top}(w)(a_1 - p_1) \right \rvert = O_p(NM^{1/2}).
\end{equation}
According to \eqref{CV-L}, \eqref{CV-L_1} and \eqref{CV-L_2}, we have 
\begin{align}
    \label{CV-L_3}
    \lvert \Psi_1 \rvert CV^*(w) = &L^*(w) + O_p(NM) + O_p(N M^{1/2} \max\{D, \log N\}^{1/2}) \nonumber \\
    &+(L^*(w))^{1/2} O_p(N^{1/2}M^{1/2}).
\end{align}
Denote $\nu \in \mR^M$ to be a weight vector with $\nu_m=0$ if $m \in \mathcal{T}$ and $\nu_m=w_m/(1-\zeta)$ if $m \notin \mathcal{T}$. It is obvious that if $m \in \mathcal{T}$, we have $P^*_{(m),ij} = P_{(m),ij}$. In view of \cite{zhang2022model}, we can obtain 
\begin{align}
    L^*(w) =& \sum_{(i,j) \in \Psi_1} \left\{ \sum_{m=1}^M w_m \left( P^*_{(m),ij} - P_{ij} \right) \right\}^2 \nonumber \\
    =& \sum_{(i,j) \in \Psi_1} \left\{ \sum_{m \notin \mathcal{T}} w_m \left( P^*_{(m),ij} - P_{ij} \right) \right\}^2 \nonumber \\
    = & (1-\zeta)^2 \sum_{(i,j) \in \Psi_1} \left\{ \sum_{m \notin \mathcal{T}} \frac{w_m}{1-\zeta} \left( P^*_{(m),ij} - P_{ij} \right) \right\}^2 \nonumber \\
    =& (1-\zeta)^2 \sum_{(i,j) \in \Psi_1} \left\{ \sum_{m=1}^M \nu_m \left( P^*_{(m),ij} - P_{ij} \right) \right\}^2 = (1-\zeta)^2 L^*(\nu). \label{L*w}
\end{align}
Combining \eqref{CV-L_3} and \eqref{L*w}, we have
\begin{align}
    \label{hatw}
    \lvert \Psi_1 \rvert CV^*(\widehat{w}) = &(1-\widehat{\zeta})^2 L^*(\widehat{\nu}) + (1-\widehat{\zeta}) (L^*(\widehat{\nu}))^{1/2} O_p(N^{1/2}M^{1/2}) \nonumber \\
    &+ O_p(NM) + O_p(N M^{1/2} \max\{D, \log N\}^{1/2}),
\end{align}
where $\widehat{\nu} = (\widehat{\nu}_1, \dots, \widehat{\nu}_M)^\top \in \mR^M$ and $\widehat{\nu}_m = \widehat{w}_m/(1-\widehat{\zeta})$ for $m=1,\dots,M$. Let $\overline{w} \in \mR^M$ be a weight vector with $\sum_{m \in \mathcal{T}} \overline{w}_m = 1$. By $\eqref{L*w}$, we have $L^*(\overline{w}) = 0$. Therefore, we have 
\begin{equation}
    \label{overlinew}
    \lvert \Psi_1 \rvert CV^*(\overline{w}) = O_p(NM) + O_p(N M^{1/2} \max\{D, \log N\}^{1/2}).
\end{equation}
Since $\widehat{w}$ minimizes $CV^*(w)$, by \eqref{hatw} and \eqref{overlinew}, we have
\begin{align*}
    &(1-\widehat{\zeta})^2 L^*(\widehat{\nu}) + (1-\widehat{\zeta}) (L^*(\widehat{\nu}))^{1/2} O_p(N^{1/2}M^{1/2}) + O_p(NM) \\
    &+ O_p(N M^{1/2} \max\{D, \log N\}^{1/2}) 
    \leq O_p(NM) + O_p(N M^{1/2} \max\{D, \log N\}^{1/2}).
\end{align*}
It follows that 
\begin{align}
\label{1-zeta}
    &(1-\widehat{\zeta})^2 \left(\inf_{w \in \mathcal{W}^s} L^*(w)\right) + (1-\widehat{\zeta}) \left(\inf_{w \in \mathcal{W}^s} L^*(w)\right)^{1/2} O_p(N^{1/2}M^{1/2}) + O_p(NM) \nonumber \\
    & + O_p(N M^{1/2} \max\{D, \log N\}^{1/2}) 
    \leq O_p(NM) + O_p(N M^{1/2} \max\{D, \log N\}^{1/2}). 
\end{align}
Dividing both sides of the \eqref{1-zeta} by $\inf_{w \in \mathcal{W}^s} L^*(w)$, we obtain
\begin{align}
    &(1-\widehat{\zeta})^2 + (1-\widehat{\zeta}) \left(\inf_{w \in \mathcal{W}^s} L^*(w)\right)^{-1/2}O_p(N^{1/2}M^{1/2}) 
    + \left(\inf_{w \in \mathcal{W}^s} L^*(w)\right)^{-1} O_p(NM) \nonumber \\
    &+ \left(\inf_{w \in \mathcal{W}^s} L^*(w)\right)^{-1/2}O_p(N^{1/2}M^{1/2}) \left(\inf_{w \in \mathcal{W}^s} L^*(w)\right)^{-1/2} O_p(N^{1/2} \max\{D, \log N\}^{1/2}) \nonumber \\
    \leq &\left(\inf_{w \in \mathcal{W}^s} L^*(w)\right)^{-1} \left\{ O_p(NM) + O_p(N M^{1/2} \max\{D, \log N\}^{1/2})\right\}. \label{infw}
\end{align}
Together with Assumption 3, we obtain $\widehat{\zeta} \rightarrow 1$ in probability. This completes the proof of Theorem 2.

\subsection{Proof of Theorem 3}
\label{theorem3}
This subsection gives the proof of Theorem 3.
Denote $\epsilon = \xi^{1/2} \lvert \Psi_1 \rvert^{-1/2+\kappa}$. To verify Theorem 3, following \cite{liao2020corrected} and \cite{liao2021model}, it suffices to show that there exists a constant $C_1$ such that, for the $M \times 1$ vector $r = (r_1, \dots, r_M)^\top$, 
\begin{equation}
    \lim\limits_{\lvert \Psi_1 \rvert \rightarrow \infty} P \left\{\inf_{\|r_0 \| = C_1, (w^0 + \epsilon r) \in \mathcal{W}} CV(w^0 + \epsilon r) > CV(w^0) \right\} = 1,
    \label{equ:limcv}
\end{equation}
which indicates that there exists a minimizer $\widehat{w}$ in the bounded closed domain $\{w^0 + \epsilon r: \| r\| \leq C_1, (w^0 + \epsilon r) \in \mathcal{W}\}$ such that $\| \widehat{w} - w^0\| = O_p(\epsilon)$.
	
Denote $\Phi_1 = (\widehat{p}_{1(1)}-\widetilde{p}_{1(1)}, \dots , \widehat{p}_{1(M)}-\widetilde{p}_{1(M)})$. Then we can write $\| \widehat{p}_1(w) - \widetilde{p}_1(w) \|^2 = w^\top \Phi_1^\top \Phi_1 w$, and $(a_1 - \widehat{p}_1(w))^\top (\widehat{p}_1(w) - \widetilde{p}_1(w)) = w^\top \Phi_1^\top e_1 + w^\top \Phi_1^\top \Omega_1 w$, where $e_1 = a_1 - p_1$. Thus, we can decompose $CV(w)$ as
\begin{equation*}
\begin{aligned}
    CV(w) &= \frac{1}{\lvert \Psi_1 \rvert} \sum_{k=1}^{K} \left\| A^{[k]} - \widetilde{P}^{[k]}(w) \right\|_F^2 = \| a_1 - \widetilde{p}_1(w) \|^2 /  \lvert \Psi_1 \rvert \\
    & = \left\{ \| a_1 - \widehat{p}_1(w) \|^2 + \|\widehat{p}_1(w) -               \widetilde{p}_1(w) \|^2 + 2 \left(a_1 - \widehat{p}_1(w)\right)^\top                  \left(\widehat{p}_1(w) - \widetilde{p}_1(w)\right) \right\} / \lvert \Psi_1 \rvert \\
    & = \left\{ \| a_1 - \widehat{p}_1(w) \|^2 + w^\top \Phi_1^\top \Phi_1      w + 2w^\top \Phi_1^\top e_1 + 2w^\top \Phi_1^\top \Omega_1 w\right\} / \lvert \Psi_1 \rvert.
\end{aligned}
\end{equation*}
Note that 
\begin{equation*}
\begin{aligned}
    &\| a_1 - \widehat{p}_1(w^0 + \epsilon r) \|^2 - \| a_1 -                  \widehat{p}_1(w^0) \|^2 \\
    =& \widehat{p}_1(w^0 + \epsilon r)^\top \widehat{p}_1(w^0 + \epsilon       r) - \widehat{p}_1(w^0)^\top \widehat{p}_1(w^0) - 2a_1^\top \left\{\widehat{p}_1(w^0 + \epsilon r) - \widehat{p}_1(w^0)\right\} \\
    =& (w^0 + \epsilon r)^\top \widehat{\Lambda}_1^\top  \widehat{\Lambda}_1(w^0 + \epsilon r) -     w^{0\top}\widehat{\Lambda}_1^\top \widehat{\Lambda}_1w^0 - 2a_1^\top \left\{\widehat{\Lambda}_1(w^0 + \epsilon r) - \widehat{\Lambda}_1w^0\right\} \\
    =& \epsilon^2 r^\top \widehat{\Lambda} r + 2(\widehat{\Lambda}_1 w^0 - a_1)^\top \widehat{\Lambda}_1 \epsilon r.
\end{aligned}
\end{equation*}
As a result, 
\begin{align}
    \label{equ:cv-cv}
    &CV(w^0 + \epsilon r) - CV(w^0)  \nonumber  \\
    =& \left\{\| a_1 - \widehat{p}_1(w^0 + \epsilon r) \|^2 - \| a_1 - \widehat{p}_1(w^0) \|^2  + \epsilon^2 r^\top \Phi_1^\top \Phi_1 r + 2 \epsilon r^\top \Phi_1^\top e_1  \right. \nonumber \\
    & \left. + 2 \epsilon^2 r^\top \Phi_1^\top \Omega_1 r + 2 \epsilon r^\top \Phi_1^\top \Omega_1 w^0 + 2 \epsilon w^{0\top} \Phi_1^\top \Omega_1 r + 2 \epsilon r^\top \Phi_1^\top \Phi_1 w^0 \right\} / \lvert \Psi_1 \rvert   \nonumber  \\
    =& \left\{ \epsilon^2 r^\top \widehat{\Lambda} r + 2\epsilon (\widehat{\Lambda}_1 w^0 - a_1)^\top \widehat{\Lambda}_1 r + \epsilon^2 r^\top \Phi_1^\top \Phi_1 r + 2 \epsilon r^\top \Phi_1^\top e_1 + 2 \epsilon^2 r^\top \Phi_1^\top \Omega_1 r  \right. \nonumber  \\
    & \left. + 2 \epsilon r^\top \Phi_1^\top \Omega_1 w^0 + 2 \epsilon w^{0\top} \Phi_1^\top \Omega_1 r + 2 \epsilon r^\top \Phi_1^\top \Phi_1 w^0 \right \} / \lvert \Psi_1 \rvert.
\end{align}
It is obvious that $\epsilon^2 r^\top \Phi_1^\top \Phi_1 r \geq 0$ for any $r$. Besides, under Assumption 4, we have 
\begin{equation}
    \epsilon^2 r^\top \widehat{\Lambda} r > \rho_1 \epsilon^2 \lvert \Psi_1 \rvert \|r \|^2 >0
    \label{equ:positive}
\end{equation}
in probability tending to 1. In the following, we show that the remaining six terms of \eqref{equ:cv-cv} are asymptotically dominated by $\epsilon^2 r^\top \widehat{\Lambda} r$.
	
We first consider the item $\lvert \epsilon  (\widehat{\Lambda}_1 w^0 - a_1)^\top \widehat{\Lambda}_1 r \rvert$. According to the definition of $\xi$ and $w^0$, we have $E (\|\widehat{\Lambda}_1 w^0 - p_1\|^2) = \xi$. Therefore, 
$\|\widehat{\Lambda}_1 w^0 - p_1\| = O_p(\xi^{1/2})$. 
From Assumption 4, it is clear that $\| \widehat{\Lambda}_1 \|_2 = \lambda_{\max}^{1/2}(\widehat{\Lambda}) = O_p(\lvert \Psi_1 \rvert^{1/2})$. Besides, from (\ref{p1-a1}), we have
\begin{equation}
    \label{e1}
    \|e_1\|^2 = \|a_1 - p_1\|^2 = O_p\left(N\max\{D, \log N\}\right).
\end{equation}
Thus,	
\begin{align}
    \label{equ:item1}
    \lvert \epsilon  (\widehat{\Lambda}_1 w^0 - a_1)^\top \widehat{\Lambda}_1 r \rvert = &\lvert \epsilon  (\widehat{\Lambda}_1 w^0 -p_1 +p_1- a_1)^\top \widehat{\Lambda}_1 r \rvert \nonumber \\
    \leq & \lvert \epsilon  (\widehat{\Lambda}_1 w^0 -p_1)^\top \widehat{\Lambda}_1 r \rvert + \lvert \epsilon  (p_1- a_1)^\top \widehat{\Lambda}_1 r \rvert \nonumber \\
    \leq & \epsilon \|\widehat{\Lambda}_1 w^0 -p_1\| \| \widehat{\Lambda}_1 \|_2 \| r \| + \epsilon \|p_1- a_1\|  \| \widehat{\Lambda}_1 \|_2 \| r \| \nonumber \\
    = & O_p\left(\epsilon \xi^{1/2} \lvert \Psi_1 \rvert^{1/2}\right)\|r\| +
    O_p\left(\epsilon N^{1/2} \max\{D, \log N\}^{1/2} \lvert \Psi_1 \rvert^{1/2}\right)\|r\|.
\end{align}
Notice that
\begin{align}
    \label{equ:gamma1}
    & \| \Phi_1 \|_2^2 \leq \text{tr}(\Phi_1^\top \Phi_1)
    = \sum_{m=1}^{M} \left\| \widetilde{p}_{1(m)} - \widehat{p}_{1(m)}\right\|^2 \nonumber \\ 
    \leq & \sum_{m=1}^{M} 2\left(\big\| \widetilde{p}_{1(m)} - p^*_{1(m)}\big\|^2 + \big\|p^*_{1(m)} -\widehat{p}_{1(m)}\big\|^2 \right)=O_p\left(NM^2\right),
\end{align}
where the last step is due to Assumption 1. 
Based on (\ref{equ:gamma1}) and (\ref{e1}), we can obtain
\begin{equation}
    \lvert \epsilon r^\top \Phi_1^\top e_1 \rvert \leq \epsilon \|e_1\| \|\Phi_1\|_2 \|r\| = O_p\left(\epsilon N \max\{D, \log N\}^{1/2} M\right)\|r\|.
    \label{equ:item2}
\end{equation}
Observe from Assumption 5 that $\lambda_{\max}^2(\Omega_1) = \lambda_{\max}(\Omega_1^\top \Omega_1) = \lambda_{\max}(\Omega) = O_p(\lvert \Psi_1 \rvert)$, and hence,
\begin{equation}
    \lvert \epsilon^2 r^\top \Phi_1^\top \Omega_1 r \rvert \leq \epsilon^2  \lambda_{\max}(\Omega_1) \|\Phi_1\|_2 \|r\|^2 = O_p\left(\epsilon^2 \lvert \Psi_1 \rvert^{1/2} N^{1/2} M \right)\|r\|^2.
    \label{equ:item3}
\end{equation}
Noting that $E(\| \Omega_1 w^0 \|^2) = E (\| p_1 - \widehat{p}_1(w^0) \|^2) = \xi$, we have $\| \Omega_1 w^0 \| = O_p\left(\xi^{1/2}\right)$, which implies that
\begin{equation}
    \lvert \epsilon r^\top \Phi_1^\top \Omega_1 w^0 \rvert \leq \epsilon \|\Phi_1\|_2 \| \Omega_1 w^0 \| \|r\| = O_p\left(\epsilon N^{1/2} M \xi^{1/2}\right)\|r\|.
    \label{equ:item4}
\end{equation}
Similar to the process of getting (\ref{equ:gamma1}), we can obtain $\| \widehat{p}_{1(m)} - \widetilde{p}_{1(m)} \|^2 = O_p(NM)$, uniformly in $m$ $(m=1,...,M)$. Thus, 
\begin{align}
    \label{equ:gamma1w0}
    \| \Phi_1 w^0 \|^2 = \left\| \sum_{m=1}^{M} w^0_m (\widehat{p}_{1(m)} - \widetilde{p}_{1(m)}) \right\|^2  \leq \sum_{m=1}^{M} w^0_m \| \widehat{p}_{1(m)} - \widetilde{p}_{1(m)} \|^2 = O_p(NM),
\end{align}
which indicates that 
\begin{equation}
    \lvert \epsilon w^{0\top} \Phi_1^\top \Omega_1 r \rvert \leq \epsilon \lambda_{\max}(\Omega_1) \|\Phi_1 w^0\| \|r\| = O_p\left(\epsilon \lvert \Psi_1 \rvert^{1/2} N^{1/2} M^{1/2}\right)\|r\| .
    \label{equ:item5}
\end{equation}
Similarly, utilizing (\ref{equ:gamma1}) and (\ref{equ:gamma1w0}) again, we obtain
\begin{equation}
    \lvert \epsilon r^\top \Phi_1^\top \Phi_1 w^0 \rvert \leq \epsilon \|\Phi_1 w^0\| \| \Phi_1 \|_2 \|r\| = O_p\left(\epsilon N M^{3/2}\right)\|r\|.
    \label{equ:item6}
\end{equation}
	
Finally,  recalling $\epsilon = \xi^{1/2} \lvert \Psi_1 \rvert^{-1/2+\kappa}$, we see that 
\begin{equation}
    \frac{\epsilon \xi^{1/2} \lvert \Psi_1 \rvert^{1/2}}{\epsilon^2 \lvert \Psi_1 \rvert} =  \frac{1}{ \lvert \Psi_1 \rvert^{\kappa}} = o(1).
    \label{equ:conc0}
\end{equation}
Since $\lvert \Psi_1 \rvert$ has the order of $N^2$, by Assumption 6, we have 
\begin{gather}
    \label{equ:conc1}
    \frac{\epsilon N^{1/2} \max\{D, \log N\}^{1/2} \lvert \Psi_1 \rvert^{1/2}}{\epsilon^2 \lvert \Psi_1 \rvert} = \frac{N^{1/2} \max\{D, \log N\}^{1/2}}{\xi^{1/2} \lvert \Psi_1 \rvert^{\kappa}} = o(1), \\
    \label{equ:conc2}
    \frac{\epsilon^2 \lvert \Psi_1 \rvert^{1/2} N^{1/2} M}{\epsilon^2 \lvert \Psi_1 \rvert} = \frac{N^{1/2} M}{ \lvert \Psi_1 \rvert^{1/2}} = o(1), \\
    \label{equ:conc3}
    \frac{\epsilon N^{1/2} M \xi^{1/2}}{\epsilon^2 \lvert \Psi_1 \rvert} = \frac{N^{1/2} M}{\lvert \Psi_1 \rvert^{1/2+\kappa}} = o(1), \\
    \label{equ:conc4}
    \frac{\epsilon \lvert \Psi_1 \rvert^{1/2} N^{1/2} M^{1/2}}{\epsilon^2 \lvert \Psi_1 \rvert} = \frac{N^{1/2 - 2\kappa} M^{1/2}}{\xi^{1/2}} = o(1),\\
    \label{equ:conc5}
    \frac{\epsilon N \max\{D, \log N\}^{1/2} M}{\epsilon N^{1/2} \max\{D, \log N\}^{1/2} \lvert \Psi_1 \rvert^{1/2}} = \frac{N^{1/2} M}{\lvert \Psi_1 \rvert^{1/2}} = o(1),
\end{gather}
and 
\begin{equation}
    \label{equ:conc6}
    \frac{\epsilon N M^{3/2}}{\epsilon \lvert \Psi_1 \rvert^{1/2} N^{1/2} M^{1/2}} = \frac{N^{1/2} M}{\lvert \Psi_1 \rvert^{1/2}} = o(1).
\end{equation}
Therefore, from (\ref{equ:item1}),  (\ref{equ:item2}), (\ref{equ:item3}), (\ref{equ:item4}), (\ref{equ:item5}) and (\ref{equ:item6}), and using (\ref{equ:positive}), (\ref{equ:conc0})- (\ref{equ:conc6}) we can prove that the remaining six terms of  (\ref{equ:cv-cv}) are asymptotically dominated by $\epsilon^2 r^\top \widehat{\Lambda} r$. This completes the proof of Theorem 3.

\subsection{Proof of Theorem 4}
\label{appendix3}
This subsection gives the proof of Theorem 4.
Let $$CV_{\dagger}^*(w) = CV_{\dagger}(w) -  \sum_{t=1}^T \sum_{k=1}^{K} \left( \|A^{(t)[k]}\|_F^2 - \|P^{(t)[k]}\|_F^2 \right)/(T \lvert \Psi_1 \rvert),$$ where $P^{(t)[k]} = P^{(t)} \circ S^{[k]}$ and the second term is unrelated to $w$. Therefore, 
\begin{equation*}
    \widehat{w}_\dagger = \argmin_{w \in \mathcal{W}} CV_{\dagger}(w) = \argmin_{w \in \mathcal{W}} T \lvert \Psi_1 \rvert CV_{\dagger}^*(w).
\end{equation*}
According to Lemma \ref{lemma1}, Theorem 4 is valid if the following conditions hold:
\begin{equation}
    \label{equ:condition3}
    \sup_{w \in \mathcal{W}} \frac{\lvert L_\dagger(w) - L_\dagger^*(w) \rvert}{ L_\dagger^*(w)} = o_p(1),
\end{equation}
and
\begin{equation}
    \label{equ:condition4}
    \sup_{w \in \mathcal{W}} \frac {\lvert T \lvert \Psi_1 \rvert CV_{\dagger}^*(w) - L_\dagger^*(w) \rvert}{ L_\dagger^*(w)} = o_p(1).
\end{equation}
	
We first consider (\ref{equ:condition3}). It can be shown that 
\begin{align}
    &\sup_{w \in \mathcal{W}} \frac{\lvert L_\dagger(w) - L_\dagger^*(w) \rvert}{ L_\dagger^*(w)} \nonumber \\
    = & \sup_{w \in \mathcal{W}} \left[ \left\lvert \sum_{t=1}^T \sum_{(i,j) \in \Psi_1} \left\{\left(\widehat{P}_{ij}^{(t)}(w) - P_{ij}^{(t)}\right)^2 - \left(P^{(t)*}_{ij}(w) - P_{ij}^{(t)}\right)^2 \right\}\right\rvert/ L_\dagger^*(w) \right] \nonumber \\
    \leq & \sup_{w \in \mathcal{W}} \left[ \left\lvert\sum_{t=1}^T \sum_{(i,j) \in \Psi_1} \left\{\left(\widehat{P}_{ij}^{(t)}(w) - P_{ij}^{(t)*}(w)\right)^2 + 2 \left(\widehat{P}_{ij}^{(t)}(w) - P_{ij}^{(t)*}(w)\right) \left(P_{ij}^{(t)*}(w)-P_{ij}^{(t)}\right)  \right\} \right\rvert / L_\dagger^*(w) \right]\nonumber \\
    \leq & \xi_\dagger^{*-1} \sup_{w \in \mathcal{W}}  \sum_{t=1}^T \sum_{(i,j) \in \Psi_1} \left\{ \sum_{m=1}^M w_m \left( \widehat{P}_{(m),ij}^{(t)} - P_{(m),ij}^{(t)*} \right) \right\}^2 \nonumber \\
    & + 2 \sup_{w \in \mathcal{W}} \left[ \frac{1}{L_\dagger^*(w)} \left\{ \sum_{t=1}^T \sum_{(i,j) \in \Psi_1} \left(\widehat{P}_{ij}^{(t)}(w) - P_{ij}^{(t)*}(w)\right)^2 \right\}^{1/2} \left\{ \sum_{t=1}^T \sum_{(i,j) \in \Psi_1} \left(P_{ij}^{(t)*}(w)-P_{ij}^{(t)}\right)^2  \right\}^{1/2} \right] \nonumber \\
    \leq & \xi_\dagger^{*-1} \sup_{w \in \mathcal{W}}  \sum_{t=1}^T \sum_{(i,j) \in \Psi_1} \sum_{m=1}^M w_m \left( \widehat{P}_{(m),ij}^{(t)} - P_{(m),ij}^{(t)*} \right)^2 \nonumber \\
    & + 2\xi_\dagger^{*-1/2} \sup_{w \in \mathcal{W}} \left\{ \sum_{t=1}^T \sum_{(i,j) \in \Psi_1} \left(\widehat{P}_{ij}^{(t)}(w) - P_{ij}^{(t)*}(w)\right)^2 \right\}^{1/2} \nonumber \\
    = & \xi_\dagger^{*-1} O_p(NMT) + 2 \xi_\dagger^{*-1/2} O_p(N^{1/2}M^{1/2}T^{1/2}) = o_p(1),
\end{align}
where the fifth step uses Assumption 7, and the sixth step uses Assumption 9. Thus, we obtain (\ref{equ:condition3}).
We next verify (\ref{equ:condition4}). Observe that
\begin{align}
    \label{equ:28}
    &\sup_{w \in \mathcal{W}} \left\{\left \lvert T \lvert \Psi_1 \rvert CV_{\dagger}^*(w) - L_\dagger^*(w) \right \rvert / L_\dagger^*(w) \right\}\nonumber  \\
    =& \sup_{w \in \mathcal{W}} \left[ \frac{1}{L_\dagger^*(w)} \left \lvert \sum_{t=1}^T \left\{ \left\|a_1^{(t)} - \widetilde{p}^{(t)}_1(w) \right\|^2 - \left(a_1^{(t)} - p_1^{(t)}\right)^\top \left(a_1^{(t)} + p_1^{(t)}\right)  - \left\| p_1^{(t)*}(w) - p_1^{(t)}\right\|^2 \right\} \right \rvert \right] \nonumber  \\
    \leq& \sup_{w \in \mathcal{W}} \left[ \frac{1}{L_\dagger^*(w)} \left \lvert \sum_{t=1}^T \left\{ \left\|a_1^{(t)} - p_1^{(t)*}(w) \right\|^2 - \left(a_1^{(t)} - p_1^{(t)}\right)^\top \left(a_1^{(t)} + p_1^{(t)}\right) - \left\| p_1^{(t)*}(w) - p_1^{(t)}\right\|^2  \right\} \right \rvert \right] \nonumber \\ 
    & + \sup_{w \in \mathcal{W}} \left[ \frac{1}{L_\dagger^*(w)} \left \lvert \sum_{t=1}^T \left\{ \left\|a_1^{(t)} - \widetilde{p}_1^{(t)}(w) \right\|^2 - \left\|a_1^{(t)} - p_1^{(t)*}(w) \right\|^2 \right\} \right \rvert \right] \nonumber  \\
    \leq & 2 \xi_\dagger^{*-1} \sup_{w \in \mathcal{W}} \left \lvert \sum_{t=1}^T p_1^{(t)*}(w)^\top \left(a_1^{(t)} - p_1^{(t)}\right) \right \rvert \nonumber  \\
    & + \sup_{w \in \mathcal{W}} \left[ \frac{1}{L_\dagger^*(w)} \left \lvert \sum_{t=1}^T \left\{ \left\|a_1^{(t)} - \widetilde{p}^{(t)}_1(w) \right\|^2 - \left\|a_1^{(t)} - p_1^{(t)*}(w) \right\|^2 \right\} \right \rvert \right].
\end{align}	 
By Assumptions 7 and 9, we have
\begin{align}
    \label{equ:29}
    & \sup_{w \in \mathcal{W}} \left[ \frac{1}{L_\dagger^*(w)} \left \lvert \sum_{t=1}^T \left\{ \left\|a_1^{(t)} - \widetilde{p}^{(t)}_1(w) \right\|^2 - \left\|a_1^{(t)} - p_1^{(t)*}(w) \right\|^2 \right\} \right \rvert \right] \nonumber  \\
    = & \sup_{w \in \mathcal{W}} \left[ \frac{1}{L_\dagger^*(w)} \left\lvert \sum_{t=1}^T \left\{ \| p^{(t)*}_{1}(w)-\widetilde{p}^{(t)}_{1}(w)\|^2 + 2 \big(a_1^{(t)}-p^{(t)*}_{1}(w)\big)^\top \big( p^{(t)*}_{1}(w)-\widetilde{p}^{(t)}_{1}(w) \big) \right\} \right\rvert \right]\nonumber  \\
    \leq & \xi_\dagger^{*-1} \sup_{w \in \mathcal{W}} \left\{ \sum_{t=1}^T \sum_{(i,j)\in \Psi_1} \left( P_{ij}^{(t)*}(w) - \widetilde{P}^{(t)}_{ij}(w) \right)^2 \right\} \nonumber  \\
&+ 2 \sup_{w \in \mathcal{W}} \left\lvert \frac{1}{L_\dagger^*(w)} \sum_{t=1}^T \sum_{(i,j)\in \Psi_1} \left(A^{(t)}_{ij} - P_{ij}^{(t)*}(w)\right)\left(P_{ij}^{(t)*}(w)-\widetilde{P}^{(t)}_{ij}(w)\right) \right\rvert  \nonumber  \\
    \leq & \xi_\dagger^{*-1} \sup_{w \in \mathcal{W}} \left[  \sum_{t=1}^T \sum_{(i,j) \in \Psi_1} \left\{ \sum_{m=1}^M w_m \left(P^{(t)*}_{(m),ij} - \widetilde{P}^{(t)}_{(m),ij}\right) \right\}^2 \right]  \nonumber \\
    & + 2 \sup_{w \in \mathcal{W}} \left[ \frac{1}{L_\dagger^*(w)} \Bigg\{ \sum_{t=1}^T \sum_{(i,j) \in \Psi_1} \left(A^{(t)}_{ij} - P_{ij}^{(t)*}(w)\right)^2 \Bigg\}^{1/2} \Bigg\{ \sum_{t=1}^T \sum_{(i,j) \in \Psi_1} \Big( P_{ij}^{(t)*}(w)-\widetilde{P}^{(t)}_{ij}(w)\Big)^2 \Bigg\}^{1/2} \right] \nonumber \\
    \leq & \xi_\dagger^{*-1} \sup_{w \in \mathcal{W}} \left[ \sum_{t=1}^T \sum_{(i,j) \in \Psi_1} \sum_{m=1}^M w_m \left(P^{(t)*}_{(m),ij} - \widetilde{P}^{(t)}_{(m),ij}\right)^2 \right] \nonumber \\
    & + 2 \sup_{w \in \mathcal{W}} \left[ \Bigg\{ \frac{1}{L_\dagger^*(w)} \sum_{t=1}^T\sum_{(i,j) \in \Psi_1} \left(A^{(t)}_{ij} - P_{ij}^{(t)*}(w)\right)^2 \Bigg\}^{1/2} \Bigg\{ \frac{1}{L_\dagger^*(w)} \sum_{t=1}^T \sum_{(i,j) \in \Psi_1} \Big( P_{ij}^{(t)*}(w)-\widetilde{P}^{(t)}_{ij}(w)\Big)^2 \Bigg\}^{1/2} \right] \nonumber \\
    \leq & \xi_\dagger^{*-1} O_p(NMT) + 2 \left\{\xi_\dagger^{*-1} O_p\left(NMT\right)\right\}^{1/2} \sup_{w \in \mathcal{W}} \Bigg\{ \frac{1}{L_\dagger^*(w)} \sum_{t=1}^T\sum_{(i,j) \in \Psi_1} \left(A^{(t)}_{ij} - P_{ij}^{(t)*}(w)\right)^2 \Bigg\}^{1/2} \nonumber\\
    = & o_p(1) + 2o_p(1)\sup_{w \in \mathcal{W}} \Bigg\{ \frac{1}{L_\dagger^*(w)} \sum_{t=1}^T \sum_{(i,j) \in \Psi_1} \left(A^{(t)}_{ij} - P_{ij}^{(t)*}(w)\right)^2 \Bigg\}^{1/2}.
\end{align}
From Lemma \ref{lemma2-new} and Assumption 8, we have $P(\|A^{(t)} - P^{(t)}\|_2 > C_\dagger \sqrt{D_\dagger+\log N}) \leq N^{-C_{0\dagger}}$ for any $C_{0\dagger}>0$.  Under Assumption 8, there exists a constant $C_{0\dagger}>0$ such that 
\begin{align}
    &P\left(\sup_t \left\|A^{(t)} - P^{(t)}\right\|_2 > C_\dagger \sqrt{D_\dagger+\log N}\right) \nonumber \\
    \leq & \sum_{t=1}^T P \left( \left\|A^{(t)} - P^{(t)}\right\|_2 > C_\dagger \sqrt{D_\dagger+\log N} \right) 
    \leq TN^{-C_{0\dagger}} = o(1).
\end{align}
Therefore, we have $\sup_t \left\|A^{(t)} - P^{(t)}\right\|_2 = O_p(\max\{D_\dagger, \log N\}^{1/2})$.
Then, we can get
\begin{align*}
    &\sup_t \big\|p_1^{(t)} - a_1^{(t)}\big\|^2 \leq \sup_t\big\|A^{(t)} - P^{(t)}\big\|_F^2 \nonumber \\
    \leq & \sup_t \left\{ \operatorname{rank} \big(A^{(t)} - P^{(t)}\big) \big\|A^{(t)} - P^{(t)}\big\|_2^2 \right\} = O_p\left(N\max\{D_\dagger, \log N\}\right).
\end{align*}
Then we have 
\begin{equation}
    \sum_{t=1}^T \|p_1^{(t)}-a_1^{(t)}\|^2 \leq T \sup_t \big\|p_1^{(t)} - a_1^{(t)}\big\|^2 = O_p\left(N\max\{D_\dagger, \log N\} T\right).\label{equ:p1t-a1t}
\end{equation}
By Assumption 9, we have
\begin{align}
    & \sup_{w \in \mathcal{W}} \Bigg\{ \frac{1}{L_\dagger^*(w)} \sum_{t=1}^T \sum_{(i,j) \in \Psi_1} \left(A^{(t)}_{ij} - P_{ij}^{(t)*}(w)\right)^2 \Bigg\}^{1/2} \nonumber\\
    \leq & \sup_{w \in \mathcal{W}} \left[ \frac{1}{L_\dagger^*(w)} \sum_{t=1}^T \sum_{(i,j) \in \Psi_1} \left\{ 2 \left(A^{(t)}_{ij} - P^{(t)}_{ij}\right)^2 + 2 \left( P^{(t)}_{ij} - P_{ij}^{(t)*}(w) \right)^2 \right\} \right]^{1/2} \nonumber\\
    \leq & \sqrt{2} \xi_\dagger^{*-1/2} O_p\left(N^{1/2} \max\{D_\dagger, \log N\}^{1/2} T^{1/2}\right) + \sqrt{2} = O(1). \label{equ:35}
\end{align}
According to \eqref{equ:29} and \eqref{equ:35}, we have 
\begin{equation}
\label{equ:36}
    \sup_{w \in \mathcal{W}} \left[ \frac{1}{L_\dagger^*(w)} \left \lvert \sum_{t=1}^T \left\{ \left\|a_1^{(t)} - \widetilde{p}^{(t)}_1(w) \right\|^2 - \left\|a_1^{(t)} - p_1^{(t)*}(w) \right\|^2 \right\} \right \rvert \right] = o_p(1).
\end{equation}
Denote the $l$th element of $a_1^{(t)}$, $p_1^{(t)}$ and $p_1^{(t)*}$ to be  $a_{1,l}^{(t)}$, $p_{1,l}^{(t)}$ and $p_{1,l}^{(t)*}$. Then, by Assumption 9 and considering that $\lvert \Psi_1 \rvert$ has the order of $N^2$, for any $\delta > 0$,
\begin{align}
    \label{equ:37}
    &P \left\{\xi_\dagger^{*-1} \sup_{w \in \mathcal{W}} \left \lvert \sum_{t=1}^T p_1^{(t)*}(w)^\top \left(a_1^{(t)} - p_1^{(t)}\right) \right \rvert > \delta \right \} \nonumber \\
    =& P \left \{\xi_\dagger^{*-1} \sup_{w \in \mathcal{W}} \left \lvert \sum_{t=1}^T \sum_{l=1}^{ \lvert \Psi_1 \rvert} \sum_{m=1}^{M} w_m p_{1(m),l}^{(t)*} \left(a_{1,l}^{(t)} - p_{1,l}^{(t)}\right)  \right \rvert > \delta \right\}  \nonumber  \\  
    \leq & \sum_{m=1}^{M} P \left\{ \left\lvert \sum_{t=1}^T  \sum_{l=1}^{ \lvert \Psi_1 \rvert} \left(p_{1(m),l}^{(t)*} a_{1,l}^{(t)} - p_{1(m),l}^{(t)*}p_{1,l}^{(t)}\right) \right \rvert > \xi_\dagger^*\delta \right\} \nonumber  \\  
    \leq & \xi_\dagger^{*-2} \delta^{-2} \sum_{t=1}^T \sum_{l=1}^{ \lvert \Psi_1 \rvert} \sum_{m=1}^{M} \var \left(p_{1(m),l}^{(t)*}a_{1,l}^{(t)} - p_{1(m),l}^{(t)*}p_{1,l}^{(t)}\right) \nonumber  \\  
    \leq & \xi_\dagger^{*-2} \delta^{-2} T M \lvert \Psi_1 \rvert = o(1).
\end{align}
By (\ref{equ:28}) ,  (\ref{equ:36}) and \eqref{equ:37}, we obtain (\ref{equ:condition4}). This completes the proof of Theorem 4.	

\subsection{Proof of Theorem 5}
\label{app:proof-theorem 5}
This subsection gives the proof of Theorem 5.
According to \eqref{equ:28}, we can get 
\begin{align}
    \left \lvert T \lvert \Psi_1 \rvert CV_{\dagger}^*(w) - L_\dagger^*(w) \right \rvert \leq & \left\lvert \sum_{t=1}^T p_1^{(t)*}(w)^\top \left(a_1^{(t)} - p_1^{(t)}\right) \right\rvert \nonumber \\
    & + \left\lvert \sum_{t=1}^T \left\{\left\|a_1^{(t)} - \widetilde{p}_1^{(t)}(w)\right\|^2 - \left\|a_1^{(t)} - p_1^{(t)*}(w)\right\|^2 \right\} \right\rvert.
    \label{CV-L-T}
\end{align}
Under Assumptions 7 and 8, we have
\begin{align}
    &\left\lvert \sum_{t=1}^T \left\{\left\|a_1^{(t)} - \widetilde{p}_1^{(t)}(w)\right\|^2 - \left\|a_1^{(t)} - p_1^{(t)*}(w)\right\|^2 \right\} \right\rvert \nonumber \\
    = & O_p(NMT) + O_p(NM^{1/2}T \max\{D_\dagger, \log N\}^{1/2}) +(L^*_\dagger(w))^{1/2} O_p(N^{1/2}M^{1/2}T^{1/2}).
    \label{CV-L-T-1}
\end{align}
Similar to \eqref{equ:37}, we have
\begin{equation}
    \sup_{w \in \mathcal{W}} \left \lvert \sum_{t=1}^T p_1^{(t)*}(w)^\top \left(a_1^{(t)} - p_1^{(t)}\right) \right \rvert = O_p(NM^{1/2}T^{1/2}).
    \label{CV-L-T-2}
\end{equation}
Based on \eqref{CV-L-T}, \eqref{CV-L-T-1} and \eqref{CV-L-T-2}, we have 
\begin{align}
     T \lvert \Psi_1 \rvert CV_{\dagger}^*(w) = &L_\dagger^*(w) + O_p(NMT) + O_p(NM^{1/2}T \max\{D_\dagger, \log N\}^{1/2}) \nonumber \\
     &+(L^*_\dagger(w))^{1/2} O_p(N^{1/2}M^{1/2}T^{1/2}).
     \label{CV=L}
\end{align}
It is easy to see that $P_{(m),ij}^{(t)*} = P_{(m),ij}^{(t)}$ if $m \in \mathcal{T}$. Hence, we have
\begin{align}
    L_\dagger^*(w) =& \sum_{t=1}^T \sum_{(i,j) \in \Psi_1} \left\{ \sum_{m=1}^M w_m \left( P_{(m),ij}^{(t)*} - P_{ij}^{(t)}\right) \right\}^2 \nonumber \\
    =& \sum_{t=1}^T \sum_{(i,j) \in \Psi_1} \left\{ \sum_{m \notin \mathcal{T}} w_m \left( P_{(m),ij}^{(t)*} - P_{ij}^{(t)}\right) \right\}^2 \nonumber \\
    = & (1-\zeta)^2 \sum_{t=1}^T \sum_{(i,j) \in \Psi_1} \left\{ \sum_{m \notin \mathcal{T}} \frac{w_m}{1-\zeta} \left( P_{(m),ij}^{(t)*} - P_{ij}^{(t)}\right) \right\}^2 \nonumber \\
    =& (1-\zeta)^2 \sum_{t=1}^T \sum_{(i,j) \in \Psi_1} \left\{ \sum_{m=1}^M \nu_m \left( P_{(m),ij}^{(t)*} - P_{ij}^{(t)}\right) \right\}^2 = (1-\zeta)^2 L^*_\dagger(\nu).
    \label{L-T*}
\end{align}
Together with \eqref{CV=L}, we have
\begin{align}
    T \lvert \Psi_1 \rvert CV_{\dagger}^*(\widehat{w}_\dagger) = &(1-\widehat{\zeta}_\dagger)^2 L^*_\dagger(\widehat{\nu}_\dagger) +(1-\widehat{\zeta}_\dagger)(L^*_\dagger(\widehat{\nu}_\dagger))^{1/2} O_p(N^{1/2}M^{1/2}T^{1/2}) \nonumber \\
    &+ O_p(NMT) + O_p(NM^{1/2}T \max\{D_\dagger, \log N\}^{1/2}),
    \label{hat-w-T}
\end{align}
where $\widehat{\nu}_\dagger = \widehat{w}_\dagger/(1-\widehat{\zeta}_\dagger)$. It is obvious that $L^*_\dagger(\overline{w}) = 0$. By \eqref{CV=L}, we have
\begin{equation}
    T \lvert \Psi_1 \rvert CV_{\dagger}^*(\overline{w})= O_p(NMT) + O_p(NM^{1/2}T \max\{D_\dagger, \log N\}^{1/2}).
    \label{bar-w-T}
\end{equation}
Because $\widehat{w}_\dagger$ minimizes $CV_{\dagger}^*(w)$, according to \eqref{hat-w-T} and \eqref{bar-w-T}, we have 
\begin{align*}
    &(1-\widehat{\zeta}_\dagger)^2 L^*_\dagger(\widehat{\nu}_\dagger) +(1-\widehat{\zeta}_\dagger)(L^*_\dagger(\widehat{\nu}_\dagger))^{1/2} O_p(N^{1/2}M^{1/2}T^{1/2}) \\
    & + O_p(NMT) + O_p(NM^{1/2}T \max\{D_\dagger, \log N\}^{1/2}) \nonumber\\
    \leq & O_p(NMT) + O_p(NM^{1/2}T \max\{D_\dagger, \log N\}^{1/2}).
\end{align*}
It is easy to see that 
\begin{align*}
    &(1-\widehat{\zeta}_\dagger)^2 \left( \inf_{w \in \mathcal{W}^s}L^*_\dagger(w) \right) + (1-\widehat{\zeta}_\dagger) \left( \inf_{w \in \mathcal{W}^s}L^*_\dagger(w) \right)^{1/2}O_p(N^{1/2}M^{1/2}T^{1/2}) \nonumber \\
    &+ O_p(NMT) + O_p(NM^{1/2}T \max\{D_\dagger, \log N\}^{1/2}) \\
    \leq & O_p(NMT) + O_p(NM^{1/2}T \max\{D_\dagger, \log N\}^{1/2}).
\end{align*}
Similar to \eqref{infw}, and by Assumption 10, we can obtain $\widehat{\zeta}_\dagger \rightarrow 1$ in probability. This completes the proof of Theorem 5.

\subsection{Proof of Theorem 6}
\label{appendix4}
This subsection gives the proof of Theorem 6.
Denote $\epsilon_\dagger = \xi_{\dagger}^{1/2} T^{-1/2} \lvert \Psi_1 \rvert^{-1/2+\kappa_\dagger}$. To verify Theorem 6, following \cite{liao2020corrected} and \cite{liao2021model}, we only need to show that there exists a constant $C_{0\dagger}$ such that, for the $M \times 1$ vector $r = (r_1, \dots, r_M)^\top$, 
\begin{equation}
    \lim\limits_{\lvert \Psi_1 \rvert \rightarrow \infty} P \left\{ \inf_{\|r_0 \| = C_{0\dagger}, (w_\dagger^0 + \epsilon_\dagger r) \in \mathcal{W}} CV_{\dagger}(w_\dagger^0 + \epsilon_\dagger r) > CV_{\dagger}(w_\dagger^0) \right\} = 1,
    \label{equ:limcv2}
\end{equation}
which indicates that there exists a minimizer $\widehat{w}_\dagger$ in the bounded closed domain $\big\{w_\dagger^0 + \epsilon_\dagger r: \| r\| \leq C_{0\dagger}, (w_\dagger^0 + \epsilon_\dagger r) \in \mathcal{W}\big\}$ such that $\big\| \widehat{w}_\dagger - w_\dagger^0\big\| = O_p(\epsilon_\dagger)$.
Denote $\Phi_1^{(t)} = \big(\widehat{p}_{1(1)}^{(t)}-\widetilde{p}_{1(1)}^{(t)}, \dots , \widehat{p}_{1(M)}^{(t)}-\widetilde{p}_{1(M)}^{(t)}\big)$. Then we can write $\big\| \widehat{p}_1^{(t)}(w) - \widetilde{p}_1^{(t)}(w) \big\|^2 = w^\top \Phi_1^{(t)\top} \Phi_1^{(t)} w$, and $\big(a_1^{(t)} - \widehat{p}_1^{(t)}(w)\big)^\top \big(\widehat{p}_1^{(t)}(w) - \widetilde{p}_1^{(t)}(w)\big) = w^\top \Phi_1^{(t)\top} e_1^{(t)} + w^\top \Phi_1^{(t)\top} \Omega_1^{(t)} w$, where $e_1^{(t)} = a_1^{(t)} - p_1^{(t)}$. Thus, we can decompose $CV_{\dagger}(w)$ as
\begin{equation*}
    \begin{aligned}
        CV_{\dagger}(w) =& \frac{1}{T \lvert \Psi_1 \rvert} \sum_{t=1}^{T} \sum_{k=1}^{K} \left\| A^{(t)[k]} - \widetilde{P}^{(t)[k]}(w) \right\|_F^2 \nonumber \\
        =& \frac{1}{T \lvert \Psi_1 \rvert} \sum_{t=1}^{T} \left\| a_1^{(t)} - \widetilde{p}_1^{(t)}(w) \right\|^2 \\
        = & \frac{1}{T \lvert \Psi_1 \rvert} \sum_{t=1}^{T} \left\{ \left\| a_1^{(t)} - \widehat{p}_1^{(t)}(w) \right\|^2 + \left\|\widehat{p}_1^{(t)}(w) -  \widetilde{p}_1^{(t)}(w) \right\|^2 \right. \nonumber \\
        &\left. + 2 \left(a_1^{(t)} - \widehat{p}_1^{(t)}(w)\right)^\top \left(\widehat{p}_1^{(t)}(w) - \widetilde{p}_1^{(t)}(w)\right) \right\} \\
        = & \frac{1}{T \lvert \Psi_1 \rvert} \sum_{t=1}^{T} \left\{ \left\| a_1^{(t)} - \widehat{p}_1^{(t)}(w) \right\|^2 + w^\top \Phi_1^{(t)\top} \Phi_1^{(t)} w + 2w^\top \Phi_1^{(t)\top} e_1 + 2w^\top \Phi_1^{(t)\top} \Omega_1^{(t)} w \right \}.
    \end{aligned}
\end{equation*}
Note that 
\begin{equation*}
\begin{aligned}
    &\left\| a_1^{(t)} - \widehat{p}_1^{(t)}(w_\dagger^0 + \epsilon_\dagger r) \right\|^2 - \left\| a_1^{(t)} - \widehat{p}_1(w_\dagger^0) \right\|^2 \\
    =& \widehat{p}_1^{(t)}(w_\dagger^0 + \epsilon_\dagger r)^\top \widehat{p}_1^{(t)}(w_\dagger^0 + \epsilon_\dagger r) - \widehat{p}_1^{(t)}(w_\dagger^0)^{\top} \widehat{p}_1^{(t)}(w_\dagger^0) - 2a_1^{(t)\top}\left\{\widehat{p}_1^{(t)}(w_\dagger^0 + \epsilon_\dagger r) - \widehat{p}_1^{(t)}(w_\dagger^0)\right\} \\
    =& (w_\dagger^0 + \epsilon_\dagger r)^\top \widehat{\Lambda}_1^{(t)\top}                        \widehat{\Lambda}_1^{(t)}(w_\dagger^0 + \epsilon_\dagger r) -      w_\dagger^{0\top}\widehat{\Lambda}_1^{(t)\top} \widehat{\Lambda}_1^{(t)} w_\dagger^0 - 2a_1^{(t)\top} \left\{\widehat{\Lambda}_1^{(t)}(w_\dagger^0 + \epsilon_\dagger r) - \widehat{\Lambda}_1^{(t)} w_\dagger^0\right\} \\
    =& \epsilon_\dagger^2 r^\top \widehat{\Lambda}^{(t)} r + 2\left(\widehat{\Lambda}_1^{(t)} w_\dagger^0 - a_1^{(t)}\right)^\top \widehat{\Lambda}_1^{(t)} \epsilon_\dagger r.
\end{aligned}
\end{equation*}
As a result, 
\begin{align}
    \label{equ:cv-cv2}
    &CV_{\dagger}(w_\dagger^0 + \epsilon_\dagger r) - CV_{\dagger}(w_\dagger^0)  \nonumber  \\
    =& \frac{1}{T \lvert \Psi_1 \rvert} \sum_{t=1}^{T} \left\{\| a_1^{(t)} - \widehat{p}_1^{(t)}(w_\dagger^0 + \epsilon_\dagger r) \|^2 - \| a_1^{(t)} - \widehat{p}_1^{(t)}(w_\dagger^0) \|^2  + \epsilon_\dagger^2 r^\top \Phi_1^{(t)\top} \Phi_1^{(t)} r + 2 \epsilon_\dagger r^\top \Phi_1^{(t)\top} e_1^{(t)}  \right. \nonumber \\
    & \left. + 2 \epsilon_\dagger^2 r^\top \Phi_1^{(t)\top} \Omega_1^{(t)} r + 2 \epsilon_\dagger r^{(t)\top} \Phi_1^{(t)\top} \Omega_1^{(t)} w_\dagger^0 + 2 \epsilon_\dagger w_\dagger^{0\top} \Phi_1^{(t)\top} \Omega_1^{(t)} r + 2 \epsilon_\dagger r^\top \Phi_1^{(t)\top} \Phi_1^{(t)} w_\dagger^0 \right\} \nonumber  \\
    =& \frac{1}{T \lvert \Psi_1 \rvert} \sum_{t=1}^{T} \left\{ \epsilon_\dagger^2 r^\top \widehat{\Lambda}^{(t)} r + 2\epsilon_\dagger \left(\widehat{\Lambda}_1^{(t)} w_\dagger^0 - a_1^{(t)}\right)^\top \widehat{\Lambda}_1^{(t)} r + \epsilon_\dagger^2 r^\top \Phi_1^{(t)\top} \Phi_1^{(t)} r + 2 \epsilon_\dagger r^\top \Phi_1^{(t)\top} e_1^{(t)}  \right. \nonumber  \\
    & \left. + 2 \epsilon_\dagger^2 r^\top \Phi_1^{(t)\top} \Omega_1^{(t)} r  + 2 \epsilon_\dagger r^\top \Phi_1^{(t)\top} \Omega_1^{(t)} w_\dagger^0 + 2 \epsilon_\dagger w_\dagger^{0\top} \Phi_1^{(t)\top} \Omega_1^{(t)} r + 2 \epsilon_\dagger r^\top \Phi_1^{(t)\top} \Phi_1^{(t)} w_\dagger^0 \right\}.
\end{align}
It is obvious that $\epsilon_\dagger^2 r^\top \Phi_1^{(t)\top} \Phi_1^{(t)} r \geq 0$ for any $r$. Besides, under Assumption 11, we have 
\begin{equation}
    \sum_{t=1}^T \epsilon_\dagger^2 r^\top \widehat{\Lambda}^{(t)} r > \rho_{1\dagger} \epsilon_\dagger^2 T \lvert \Psi_1 \rvert \|r \|^2 >0
    \label{equ:positive2}
\end{equation}
in probability tending to 1. In the following, we show that the remaining six terms of \eqref{equ:cv-cv2} are asymptotically dominated by $\sum_{t=1}^T \epsilon_\dagger^2 r^\top \widehat{\Lambda}^{(t)} r$.
	
We first consider the item $\sum_{t=1}^T \big\lvert \epsilon_\dagger (\widehat{\Lambda}_1^{(t)} w_\dagger^0 - a_1^{(t)})^\top \widehat{\Lambda}_1^{(t)} r \big\rvert$. According to the definitions of $\xi_{\dagger}$ and $w_\dagger^0$, we have $E \big( \sum_{t=1}^T \|\widehat{\Lambda}^{(t)}_1 w_\dagger^0 - p_1^{(t)}\|^2\big) = \xi_{\dagger}$. 
Further, by Assumption 11, we have	
\begin{align}
\label{equ:item1-2}
    & \sum_{t=1}^T \left\lvert \epsilon_\dagger \left(\widehat{\Lambda}_1^{(t)} w_\dagger^0 - a_1^{(t)}\right)^\top \widehat{\Lambda}_1^{(t)} r \right\rvert \nonumber \\
    \leq & \sum_{t=1}^T \left\lvert \epsilon_\dagger \left(\widehat{\Lambda}_1^{(t)} w_\dagger^0 -p_1^{(t)}\right)^\top \widehat{\Lambda}^{(t)}_1 r\right\rvert + \sum_{t=1}^T \left\lvert \epsilon_\dagger \left(p_1^{(t)}- a_1^{(t)}\right)^\top \widehat{\Lambda}^{(t)}_1 r\right\rvert \nonumber \\
    \leq & \epsilon_\dagger \left( \sum_{t=1}^T \left\|\widehat{\Lambda}_1^{(t)} w_\dagger^0 -p_1^{(t)}\right\|^2 \right)^{1/2} \left( \sum_{t=1}^T \left\| \widehat{\Lambda}^{(t)}_1 r \right\|^2 \right)^{1/2} + \epsilon_\dagger \left( \sum_{t=1}^T \left\|p_1^{(t)}- a_1^{(t)}\right\|^2 \right)^{1/2} \left( \sum_{t=1}^T \left\| \widehat{\Lambda}^{(t)}_1 r \right\|^2 \right)^{1/2} \nonumber \\
    \leq & \epsilon_\dagger \left\{ \left( \sum_{t=1}^T \left\|\widehat{\Lambda}_1^{(t)} w_\dagger^0 -p_1^{(t)}\right\|^2 \right)^{1/2} + \left( \sum_{t=1}^T \left\|p_1^{(t)}- a_1^{(t)}\right\|^2 \right)^{1/2} \right\} \left\{ r^\top \left( \sum_{t=1}^T \widehat{\Lambda}^{(t)} \right) r \right\}^{1/2}  \nonumber \\
    = & O_p\left(\epsilon_\dagger \xi_{\dagger}^{1/2} T^{1/2} \lvert \Psi_1 \rvert^{1/2}\right) \|r\| + O_p\left(\epsilon_\dagger T N^{1/2} \max\{D_\dagger, \log N\}^{1/2} \lvert \Psi_1 \rvert^{1/2} \right)\|r\|. 
\end{align}
Notice that
\begin{align}
    \label{equ:gamma1-2}
    &\sum_{t=1}^T \left\| \Phi^{(t)}_1 \right\|_2^2 \leq \sum_{t=1}^T \text{tr}\left(\Phi^{(t)\top}_1 \Phi^{(t)}_1 \right) 
    = \sum_{t=1}^T \sum_{m=1}^{M} \left\| \widetilde{p}^{(t)}_{1(m)} - \widehat{p}^{(t)}_{1(m)}\right\|^2 \nonumber \\
    \leq & \sum_{m=1}^{M} 2\left( \sum_{t=1}^T \left\| \widetilde{p}^{(t)}_{1(m)} - p^{(t)*}_{1(m)}\right\|^2 + \sum_{t=1}^T \left\|p^{(t)*}_{1(m)} -\widehat{p}^{(t)}_{1(m)}\right\|^2 \right) 
    = O_p\left(NM^2T\right),
\end{align}
where the last step is due to Assumption 7. 
Based on (\ref{equ:gamma1-2}) and (\ref{equ:p1t-a1t}), we can obtain
\begin{align}
    \sum_{t=1}^T \left \lvert \epsilon_\dagger r^\top \Phi_1^{(t)\top} e_1^{(t)} \right\rvert \leq & \epsilon_\dagger \|r\| \left( \sum_{t=1}^T \left\| \Phi^{(t)}_1 \right\|_2^2 \right)^{1/2} \left( \sum_{t=1}^T \left\| e_1^{(t)} \right\|^2 \right)^{1/2} \nonumber \\
    \leq& O_p\left(\epsilon_\dagger T M N \max\{D_\dagger, \log N\}^{1/2}\right)\|r\|.
    \label{equ:item2-2}
\end{align}
Observe from Assumption 12 that $\lambda_{\max}(\sum_{t=1}^T \Omega^{(t)}) = O_p(T\lvert \Psi_1 \rvert)$, and hence,
\begin{align}
    &\sum_{t=1}^T \left \lvert \epsilon_\dagger^2 r^\top \Phi_1^{(t)\top} \Omega_1^{(t)} r \right\rvert \leq \epsilon_\dagger^2 \left(\sum_{t=1}^T \left\| \Phi_1^{(t)} r \right\|^2 \right)^{1/2} \left( \sum_{t=1}^T \left\| \Omega_1^{(t)} r  \right\|^2 \right)^{1/2} \nonumber \\
    \leq &  \epsilon_\dagger^2 \|r\|^2 \left( \sum_{t=1}^T \big\|\Phi_1^{(t)}\big\|_2^2 \right)^{1/2} \lambda_{\max}^{1/2} \left(\sum_{t=1}^T \Omega^{(t)}\right)
=   O_p\left(\epsilon_\dagger^2 T M N^{1/2} \lvert \Psi_1 \rvert^{1/2}\right)\|r\|^2.
    \label{equ:item3-2}
\end{align}
Noting that $ E \big( \sum_{t=1}^T \big\| \Omega_1^{(t)} w_\dagger^0 \big\|^2 \big) = E \big(\sum_{t=1}^T \big\| p_1^{(t)} - \widehat{p}^{(t)}_1(w_\dagger^0) \big\|^2\big) = \xi_{\dagger}$, we have $\sum_{t=1}^T \big\| \Omega_1^{(t)} w_\dagger^0 \big\|^2 = O_p\left(\xi_{\dagger}\right)$, which implies that
\begin{align}
\label{equ:item4-2}
    &\sum_{t=1}^T \left \lvert \epsilon_\dagger r^\top \Phi_1^{(t)\top} \Omega_1^{(t)} w_\dagger^0 \right \rvert 
    \leq \epsilon_\dagger \left( \sum_{t=1}^T \left\|\Phi_1^{(t)}r \right\|^2 \right)^{1/2}  \left( \sum_{t=1}^T \left\|\Omega_1^{(t)} w_\dagger^0 \right\|^2 \right)^{1/2}\nonumber \\
    \leq & \epsilon_\dagger \|r\| \left( \sum_{t=1}^T \left\|\Phi_1^{(t)} \right\|_2^2 \right)^{1/2} \left( \sum_{t=1}^T \left\|\Omega_1^{(t)} w_\dagger^0 \right\|^2 \right)^{1/2}
    = O_p\left(\epsilon_\dagger  \xi_{\dagger}^{1/2} T^{1/2} M N^{1/2}\right)\|r\|.
\end{align}
Similar to (\ref{equ:gamma1-2}), we can obtain $\sum_{t=1}^T \big\| \widehat{p}^{(t)}_{1(m)} - \widetilde{p}^{(t)}_{1(m)} \big\|^2 = O_p(NMT)$, uniformly in $m$ $(m=1,\dots,M)$. Thus, 
\begin{align}
    \label{equ:gamma1w0-2}
    & \sum_{t=1}^T \left\| \Phi_1^{(t)} w_\dagger^0 \right\|^2 = \sum_{t=1}^T \left\| \sum_{m=1}^{M} w_{\dagger m}^0 \left(\widehat{p}^{(t)}_{1(m)} - \widetilde{p}^{(t)}_{1(m)}\right) \right\|^2  \nonumber \\
    \leq & \sum_{t=1}^T \sum_{m=1}^{M} w_{\dagger m}^0 \left\| \widehat{p}^{(t)}_{1(m)} - \widetilde{p}^{(t)}_{1(m)} \right\|^2 = O_p(NMT),
\end{align}
which indicates that 
\begin{align}
    & \sum_{t=1}^T \left\lvert \epsilon_\dagger w_\dagger^{0\top} \Phi_1^{(t)\top} \Omega_1^{(t)} r \right\lvert 
    \leq \epsilon_\dagger \left( \sum_{t=1}^T \left\| \Phi_1^{(t)} w_\dagger^{0} \right\|^2 \right)^{1/2} \left( \sum_{t=1}^T \left\| \Omega_1^{(t)} r \right\|^2 \right)^{1/2} \nonumber \\
    \leq & \epsilon_\dagger \|r\| \left( \sum_{t=1}^T \left\| \Phi_1^{(t)} w_\dagger^{0} \right\|^2 \right)^{1/2} \lambda^{1/2}_{\max} \left( \sum_{t=1}^T \Omega^{(t)} \right) \nonumber \\
    = & O_p\left(\epsilon_\dagger T M^{1/2} N^{1/2} \lvert \Psi_1 \rvert^{1/2}\right)\|r\| .
    \label{equ:item5-2}
\end{align}
Similarly, utilizing (\ref{equ:gamma1-2}) and (\ref{equ:gamma1w0-2}) again, we obtain
\begin{align}
    &\sum_{t=1}^T \left\lvert \epsilon_\dagger r^\top \Phi_1^{(t)\top} \Phi_1^{(t)} w_\dagger^0 \right\rvert 
    \leq \epsilon_\dagger \left\{ \sum_{t=1}^T \left\|\Phi_1^{(t)} r \right\|^2 \right\}^{1/2} \left\{\sum_{t=1}^T \left\|\Phi_1^{(t)}w_\dagger^0  \right\|^2 \right\}^{1/2} \nonumber \\
    \leq & \epsilon_\dagger \|r\| \left\{ \sum_{t=1}^T \left\|\Phi_1^{(t)} \right\|_2^2 \right\}^{1/2} \left\{\sum_{t=1}^T \left\|\Phi_1^{(t)}w_\dagger^0  \right\|^2 \right\}^{1/2}
    = O_p\left(\epsilon_\dagger T M^{3/2} N\right)\|r\|.
    \label{equ:item6-2}
\end{align}

Recalling $\epsilon_\dagger = \xi_{\dagger}^{1/2} T^{-1/2} \lvert \Psi_1 \rvert^{-1/2+\kappa_\dagger}$, we see that 
\begin{gather}
    \label{equ:tconc1-1} 
    \frac{\epsilon_\dagger \xi_{\dagger}^{1/2} T^{1/2} \lvert \Psi_1 \rvert^{1/2}}{\epsilon_\dagger^2 T \lvert \Psi_1 \rvert} = \frac{1}{\lvert \Psi_1 \rvert^{\kappa_\dagger}}=o(1).
\end{gather}
Besides, Assumption 13 ensures that 
\begin{gather}
    \label{equ:tconc1-2}
    \frac{\epsilon_\dagger T N^{1/2} \max\{D_\dagger, \log N\}^{1/2} \lvert \Psi_1 \rvert^{1/2}}{\epsilon_\dagger^2 T \lvert \Psi_1 \rvert} = \frac{N^{1/2} T^{1/2} \max\{D_\dagger, \log N\}^{1/2}}{\xi_{\dagger}^{1/2} \lvert \Psi_1 \rvert^{\kappa_\dagger}} = o(1), \\
    \label{equ:tconc2}
    \frac{\epsilon_\dagger^2 T M N^{1/2} \lvert \Psi_1 \rvert^{1/2}}{\epsilon_\dagger^2 T \lvert \Psi_1 \rvert} = \frac{MN^{1/2}}{\lvert \Psi_1 \rvert^{1/2}} = o(1), 
\end{gather}
and
\begin{equation}
    \label{equ:tconc3}
    \frac{\epsilon_\dagger T M^{1/2} N^{1/2} \lvert \Psi_1 \rvert^{1/2}}{\epsilon_\dagger^2 T \lvert \Psi_1 \rvert} = \frac{(TMN)^{1/2}}{\xi_{\dagger}^{1/2} \lvert \Psi_1 \rvert^{\kappa_\dagger}} = o(1).
\end{equation}
In addition, by Assumption 13, we obtain that
\begin{gather}
    \label{equ:tconc4}
    \frac{\epsilon_\dagger T M N \max\{D_\dagger, \log N\}^{1/2}}{\epsilon_\dagger T N^{1/2} \max\{D_\dagger, \log N\}^{1/2} \lvert \Psi_1 \rvert^{1/2}} = \frac{MN^{1/2}}{\lvert \Psi_1 \rvert^{1/2}} = o(1), \\
    \label{equ:tconc5}
    \frac{\epsilon_\dagger  \xi_{\dagger}^{1/2} T^{1/2} M N^{1/2}}{\epsilon_\dagger \xi_{\dagger}^{1/2} T^{1/2} \lvert \Psi_1 \rvert^{1/2}} = \frac{MN^{1/2}}{\lvert \Psi_1 \rvert^{1/2}} = o(1), 
\end{gather}
and
\begin{equation}
    \label{equ:tconc6}
    \frac{\epsilon_\dagger T M^{3/2} N}{\epsilon_\dagger T M^{1/2} N^{1/2} \lvert \Psi_1 \rvert^{1/2}} = \frac{MN^{1/2}}{\lvert \Psi_1 \rvert^{1/2}} = o(1).
\end{equation}
Therefore, from (\ref{equ:item1-2}),  (\ref{equ:item2-2}), (\ref{equ:item3-2}), (\ref{equ:item4-2}), (\ref{equ:item5-2}) and (\ref{equ:item6-2}), and using (\ref{equ:positive2}) and  (\ref{equ:tconc1-1})-\eqref{equ:tconc6}, we see that the remaining six terms of (\ref{equ:cv-cv2}) are asymptotically dominated by $\sum_{t=1}^T \epsilon_\dagger^2 r^\top \widehat{\Lambda}^{(t)} r$. This completes the proof of Theorem 6.

\section{Verifications of Assumptions 1, 2 and 3}
\label{s7}

\subsection{Verifications of Assumption 1}


Let $\alpha_{(m)}^* \in \mR^N$ and $Z_{(m)}^* \in \mR^{N \times m}$ denote the limiting values of $\widehat{\alpha}_{(m)}$ and $\widehat{Z}_{(m)}$, respectively. We require that $JZ_{(m)}^* = Z_{(m)}^*$ and $J\widehat{Z}_{(m)} = \widehat{Z}_{(m)}$. Denote $\Theta_{(m)}^* = \alpha_{(m)}^* 1_N^\top + 1_N \alpha_{(m)}^{*\top} + Z_{(m)}^* Z_{(m)}^{*\top}$ and $\widehat{\Theta}_{(m)} = \widehat{\alpha}_{(m)} 1_N^\top + 1_N \widehat{\alpha}_{(m)}^\top + \widehat{Z}_{(m)} \widehat{Z}_{(m)}^\top$. We assume that $\|\Theta\|_{\max} \leq \mu$, $\|\Theta_{(m)}^*\|_{\max} \leq \mu$ and $\|\widehat{\Theta}_{(m)}\|_{\max} \leq \mu$ uniformly for $m=1,\dots,M$, where $\|\cdot\|_{\max}$ denotes the maximum absolute value of entries in a matrix. Since $\widehat{\Theta}_{(m)}^0 = \text{logit}(\widehat{P}_{(m)}^0)$ are obtained by minimizing the following objective function:
$$
\mathcal{L}(\Theta) = -\sum_{i,j} \{A(\Psi_1)_{ij} \Theta_{ij} - f(\Theta_{ij})\},
$$
we have 
\begin{equation}
	\mathcal{L}(\widehat{\Theta}_{(m)}^0) - \mathcal{L}(\Theta_{(m)}^{*,0}) \leq 0,
	\label{L-L<0}
\end{equation}
where $\Theta_{(m)}^{*,0}$ is the limiting value of $\widehat{\Theta}_{(m)}^0$.
Expanding $\mathcal{L}(\widehat{\Theta}_{(m)}^0) - \mathcal{L}(\Theta_{(m)}^{*,0})$, we obtain
\begin{align}
	& \mathcal{L}(\Theta_{(m)}^{*,0}) -\mathcal{L}(\widehat{\Theta}_{(m)}^0)  = \sum_{i,j} \left\{ A(\Psi_1)_{ij} \left(\widehat{\Theta}_{(m),ij}^0 - \Theta_{(m),ij}^{*,0} \right) - \left( f(\widehat{\Theta}_{(m),ij}^0) - f(\Theta_{(m),ij}^{*,0})\right) \right\} \nonumber \\
	=& \sum_{i,j} \left\{ \left( A(\Psi_1)_{ij} - f^\prime (\Theta_{(m),ij}^{*,0})\right) \left(\widehat{\Theta}_{(m),ij}^0 - \Theta_{(m),ij}^{*,0} \right) \right\}  \nonumber \\
	& - \sum_{i,j} \left\{ f(\widehat{\Theta}_{(m),ij}^0) - f(\Theta_{(m),ij}^{*,0})-  f^\prime (\Theta_{(m),ij}^{*,0}) \left(\widehat{\Theta}_{(m),ij}^0 - \Theta_{(m),ij}^{*,0} \right) \right\} \nonumber \\
	= & \sum_{i,j} \left\{ \left( A(\Psi_1)_{ij} - f^\prime (\Theta_{(m),ij}^{*,0})\right) \left(\widehat{\Theta}_{(m),ij}^0 - \Theta_{(m),ij}^{*,0} \right) \right\} - \sum_{i,j} \frac{1}{2}f^{\prime \prime} (\bar{\Theta}_{(m),ij})\left(\widehat{\Theta}_{(m),ij}^0 - \Theta_{(m),ij}^{*,0} \right)^2 \nonumber \\
	\leq & \sum_{i,j} \left\{ \left( A(\Psi_1)_{ij} - f^\prime (\Theta_{(m),ij}^{*,0})\right) \left(\widehat{\Theta}_{(m),ij}^0 - \Theta_{(m),ij}^{*,0} \right) \right\} - \frac{1}{2} \min_{\lvert \omega \rvert \leq \mu} f^{\prime \prime} (\omega) \|\widehat{\Theta}_{(m)}^0 - \Theta^{*,0}_{(m)}\|_F^2,
	\label{L-L}
\end{align}
where $\bar{\Theta}_{(m),ij}$ is a value between $\widehat{\Theta}_{(m)}^0$ and $\Theta^{*,0}_{(m)}$. According to \eqref{L-L<0} and \eqref{L-L}, we have
\begin{align}
	&\|\widehat{\Theta}_{(m)}^0 - \Theta^{*,0}_{(m)}\|_F^2 \leq  \frac{2}{\min_{\lvert \omega \rvert \leq \mu} f^{\prime \prime} (\omega)} \sum_{i,j} \left\{ \left( A(\Psi_1)_{ij} - f^\prime (\Theta_{(m),ij}^{*,0})\right) \left(\widehat{\Theta}_{(m),ij}^0 - \Theta_{(m),ij}^{*,0} \right) \right\} \nonumber \\
	= & \frac{2}{\min_{\lvert \omega \rvert \leq \mu} f^{\prime \prime} (\omega)} \left\langle A(\Psi_1) - f^\prime (\Theta_{(m)}^{*,0}), \widehat{\Theta}_{(m)}^0 -  \Theta_{(m)}^{*,0} \right\rangle,
	\label{Theta-Theta}
\end{align}
where $\langle \cdot, \cdot \rangle$ denotes the inner product of two matrices. For any two matrices $A$ and $B$, we have $\lvert  \langle  A, B \rangle\rvert \leq \|A\|_2 \sqrt{\text{rank}(B)} \|B\|_F$. According to the definition of $\widehat{\Theta}_{(m)}$ and $\Theta_{(m)}^*$, we have 
\begin{align*}
	&\text{rank}(\widehat{\Theta}_{(m)}^0 -  \Theta_{(m)}^{*,0}) \\
	=& \text{rank} \left((\widehat{\alpha}_{(m)}^0-\alpha_{(m)}^{*,0}) 1_N^\top + 1_N (\widehat{\alpha}_{(m)}^0-\alpha_{(m)}^{*,0})^\top + \widehat{Z}_{(m)}^0 \widehat{Z}_{(m)}^{0\top} - Z_{(m)}^{*,0} Z_{(m)}^{*,0\top}\right) \leq 2+2m.
\end{align*}
It implies that 
\begin{align}
	&\left \lvert \left\langle A(\Psi_1) - f^\prime (\Theta_{(m)}^{*,0}), \widehat{\Theta}_{(m)}^0 -  \Theta_{(m)}^{*,0} \right\rangle \right\rvert  \nonumber \\
	\leq & \left \lvert \left\langle A(\Psi_1) - pP , \widehat{\Theta}_{(m)}^0 -  \Theta_{(m)}^{*,0} \right\rangle \right\rvert + \left \lvert \left\langle pP- f^\prime (\Theta_{(m)}^{*,0}), \widehat{\Theta}_{(m)}^0 -  \Theta_{(m)}^{*,0} \right\rangle \right\rvert  \nonumber \\
	\leq& \sqrt{2+2m} \|A(\Psi_1) - pP\|_2 \|\widehat{\Theta}_{(m)}^0 -  \Theta_{(m)}^{*,0} \|_F + \|pP- f^\prime (\Theta_{(m)}^{*,0})\|_F \|\widehat{\Theta}_{(m)}^0 -  \Theta_{(m)}^{*,0}\|_F .
	\label{abs-inner}
\end{align}
Further, \eqref{Theta-Theta} and \eqref{abs-inner} together imply that
\begin{align*}
	\|\widehat{\Theta}_{(m)}^0 -  \Theta_{(m)}^{*,0} \|_F \leq & \frac{2 \sqrt{2+2m}\|A(\Psi_1) - pP\|_2 + 2\|pP- f^\prime  (\Theta_{(m)}^{*,0})\|_F}{\min_{\lvert \omega \rvert \leq \mu} f^{\prime \prime} (\omega)} \nonumber \\
	\leq & \frac{2 \sqrt{2+2m}}{\min_{\lvert \omega \rvert \leq \mu} f^{\prime \prime} (\omega)} \|A(\Psi_1) - pP\|_2 + \frac{\|\widehat{\Theta}_{(m)}^0 -  \Theta_{(m)}^{*,0}\|_F + \|\widehat{\Theta}_{(m)}^0 - \Theta_p\|_F }{8\min_{\lvert \omega \rvert \leq \mu} f^{\prime \prime} (\omega)},
\end{align*}
where $\Theta_p = (\text{logit}(pP_{ij})) \in \mathbb{R}^{N \times N}$. Assume that $\mu$ satisfies $\min_{\omega \leq \mu} f^{\prime \prime} (\omega) > 1/8$, then we have
\begin{align}
	\|\widehat{\Theta}_{(m)}^0 -  \Theta_{(m)}^{*,0} \|_F \leq \frac{16\min_{\lvert \omega \rvert \leq \mu} f^{\prime \prime} (\omega) \sqrt{2+2m}}{(8\min_{\lvert \omega \rvert \leq \mu} f^{\prime \prime} (\omega)-1) \min_{\lvert \omega \rvert \leq \mu} f^{\prime \prime} (\omega)} \|A(\Psi_1) - pP\|_2 + \frac{\|\widehat{\Theta}_{(m)}^0 - \Theta_p\|_F }{8\min_{\lvert \omega \rvert \leq \mu} f^{\prime \prime} (\omega)-1}.
	\label{Theta-Theta-new}
\end{align}
Since $E(A(\Psi_1)_{ij}) = pP_{ij}$, according to Lemma \ref{lemma2-new}, we have 
\begin{align}
	\|A(\Psi_1) - pP\|_2 = O_p(\max\{D, \log N\}^{1/2}).
	\label{A-pP}
\end{align}

Further assume that $\max_{1 \leq i \leq N} \|z_{i}\|$, and $\|\alpha\|_{\max} \leq \mu_1/2$. Denote the condition number of $Z$  (i.e., the ratio of the largest to the smallest singular values) as $\kappa_Z$. When $m \geq d_0$, Theorem 9 of \cite{ma2020universal} implies that, under Assumption 8 of \cite{ma2020universal} and some constraints on $\|Z\|_2^2$, we have 
\begin{align}
	\|\widehat{\Theta}_{(m)}^0 - \Theta_p\|_F^2 \leq C_4 \kappa_Z^2 e^{2 \mu_1} N m \times \max\left\{ e^{\mu}, \frac{\log N}{N}\right\},
	\label{true-model-theta}
\end{align}
with probability at least $1-N^{-C_5}$, where $C_4$ and $C_5$ are some positive constants.

When the dimension of the latent vector is mis-specified, i.e., $m \leq d_0$, Theorem 14 of \cite{ma2020universal} gives the upper bound of the estimation error of $\Theta$ obtained by the projected gradient descent algorithm. Let $U_m D_m U_m^\top$ be the best rank-$m$ approximation to $ZZ^\top$ and $\bar{F}_m = ZZ^\top - U_m D_m U_m^\top$. Denote the condition number of $U_m D_m^{1/2}$ as $\kappa_\star$. According to this theorem, under Assumption 8 of \cite{ma2020universal} and some constraints on $\|ZZ^\top\|_2$, we have 
\begin{align}
	\|\widehat{\Theta}_{(m)}^0 - \Theta_p\|_F^2 \leq C_4 \kappa_\star^2 \left[ e^{2 \mu_1} N m \times \max\left\{ e^{\mu}, \frac{\log N}{N}\right\} + e^{\mu_1} \|\bar{F}_m\|_F^2 \right],
	\label{theta0-thetap}
\end{align}
with probability at least $1-N^{-C_5}$. Furthermore, by Schmidt-Mirsky-Eckart Young Theorem, we have $\|\bar{F}_m\|_F^2 = \lambda_{m+1}^2 + \dots + \lambda_{d_0}^2$, where $\lambda_r$ is the $r$th largest singular value of $ZZ^\top$ for $r=1,\dots,d_0$. Consider a special case where each element in $\bar{F}_m$ follows a standard Gaussian distribution. According to Theorem 2.5 in  \cite{chafai2009singular}, the order of  $\lambda_{m+1}^2$ is $N$. Assuming that $\lambda_{d_0}^2$ and $\lambda_{m+1}^2$ are of the same order, we find that the order of $	\|\bar{F}_m\|_F^2$ is $N(d_0-m)$.
When $\mu,\ \kappa_\star$ are finite constants, and $e^{\mu_1} = M/d_0=O(1)$, according to \eqref{Theta-Theta-new}-\eqref{theta0-thetap}, we have
\begin{align*}
	\|\widehat{\Theta}_{(m)}^0 -  \Theta_{(m)}^{*,0} \|_F \leq O_p (\sqrt{NM} ),
\end{align*}
when $D=O(N)$.
Then we have
\begin{align}
	\label{ass1-result}
	\max_m \sum_{(i,j) \in \Psi_1} \left( \widehat{P}_{(m),ij} - P^*_{(m),ij} \right)^2 \leq & \max_m \|\widehat{P}_{(m)} - P^*_{(m)}\|_F^2 \nonumber \\
	\leq & \frac{1}{16 p^2} \max_m \|\widehat{\Theta}_{(m)}^0 - \Theta^{*,0}_{(m)}\|_F^2 \nonumber \\
	\leq & O_p (NM),
\end{align}
when $p$ is a fixed constant. Finally, \eqref{ass1-result} implies Assumption 1.

\subsection{Verifications of Assumption 2}
\label{supp: very assump3}
Consider the case where $M < d_0$, which means that the candidate models do not include the correct models. 
Let $\mathcal{H} = \alpha 1_N^\top + 1_N \alpha^\top$, $\mathcal{H}^*(w) = \sum_{m=1}^M w_m (\alpha_{(m)}^* 1_N^\top + 1_N \alpha_{(m)}^{*\top})$, $\mathcal{I} = ZZ^\top$, $\widehat{\mathcal{I}}_{(m)} = \widehat{Z}_{(m)} \widehat{Z}_{(m)}^\top$, $\widehat{\mathcal{I}}(w) = \sum_{m=1}^M w_m \widehat{\mathcal{I}}_{(m)}$, $\mathcal{I}^*_{(m)} = Z_{(m)}^* Z_{(m)}^{*\top}$ and $\mathcal{I}^*(w) = \sum_{m=1}^M w_m \mathcal{I}^*_{(m)}$. Besides,  $\Theta_1$ denotes the matrix obtained from $\Theta$ by setting all entries of $\Theta$ with indices in $\Psi_2$ to 0. Similarly, let $\Theta_1^*(w)$, $\mathcal{H}_1$, $\mathcal{H}_1^*(w)$, $\mathcal{I}_1$ and $\mathcal{I}_1^*(w)$ be the matrices obtained from $\Theta^*(w)$, $\mathcal{H}$, $\mathcal{H}^*(w)$, $\mathcal{I}$ and $\mathcal{I}^*(w)$ by setting all entries of them with indices in $\Psi_2$ to 0, respectively. Then, we have
\begin{align}
	\label{Theta*-Theta}
	& \|\Theta_1^*(w) - \Theta_1\|_F^2 \nonumber \\
	= & \|\mathcal{H}_1^*(w) - \mathcal{H}_1 + \mathcal{I}_1^*(w) - \mathcal{I}_1\|_F^2 \nonumber \\
	\geq& \|\mathcal{H}_1^*(w) - \mathcal{H}_1\|_F^2 + \|\mathcal{I}_1^*(w) - \mathcal{I}_1\|_F^2 - 2 \lvert \left \langle \mathcal{H}_1^*(w) - \mathcal{H}_1, \mathcal{I}_1^*(w) - \mathcal{I}_1 \right \rangle \rvert \nonumber \\
	\geq & \|\mathcal{H}_1^*(w) - \mathcal{H}_1\|_F^2 + \|\mathcal{I}_1^*(w) - \mathcal{I}_1\|_F^2 - 2 \|\mathcal{H}_1^*(w) - \mathcal{H}_1\|_F \|\mathcal{I}_1^*(w) - \mathcal{I}_1\|_F \nonumber \\
	\geq & \|\mathcal{H}_1^*(w) - \mathcal{H}_1\|_F^2 + \|\mathcal{I}_1^*(w) - \mathcal{I}_1\|_F^2 - c_1^{-1} \|\mathcal{H}_1^*(w) - \mathcal{H}_1\|_F^2 - c_1 \|\mathcal{I}_1^*(w) - \mathcal{I}_1\|_F^2  \nonumber \\
	= & (1-c_1^{-1}) \|\mathcal{H}_1^*(w) - \mathcal{H}_1\|_F^2 + (1-c_1) \|\mathcal{I}_1^*(w) - \mathcal{I}_1\|_F^2,
\end{align}
where $0<c_1<1$. Next, we introduce some notations. Let $\mathcal{I}^{\star}_{(m)} = U_m D_m U_m^\top$ be the top-$m$ eigen-decomposition of $\mathcal{I}$, which is the best rank-$m$ approximation to $\mathcal{I}$ according to Schmidt-Mirsky-Eckart Young Theorem, and $\mathcal{I}^{\star}(w) = \sum_{m=1}^M w_m \mathcal{I}^{\star}_{(m)}$. 
Theorem 9 in \cite{ma2020universal} assures that $\widehat{\mathcal{I}}_{(m)}$ and $\widehat{\Theta}_{(m)}$ consistently estimates $\mathcal{I}^{\star}_{(m)}$ and $\mathcal{H}+\mathcal{I}^{\star}_{(m)}$ under certain conditions. Therefore, we have $\mathcal{I}^*_{(m)} = \mathcal{I}^{\star}_{(m)}$ and $\mathcal{H}^*_{(m)} = \mathcal{H}$.
Then \eqref{Theta*-Theta} becomes
\begin{align}
	\|\Theta_1^*(w) - \Theta_1\|_F^2 \geq (1-c_1) \|\mathcal{I}_1^\star(w) - \mathcal{I}_1\|_F^2.
	\label{Theta_1^*(w)}
\end{align}
Schmidt-Mirsky-Eckart Young Theorem implies that 
$\|\mathcal{I}^\star_{(m)} - \mathcal{I}\|_F^2 = \lambda_{m+1}^2 + \dots + \lambda_{d_0}^2$. Then, for any $w \in \mathcal{W}$, 
\begin{align}
	\|\mathcal{I}^\star(w) - \mathcal{I}\|_F^2 \geq \|\mathcal{I}_{(M)}^\star - \mathcal{I}\|_F^2 = \lambda_{M+1}^2 + \dots + \lambda_{d_0}^2,
	\label{lambda+lambda}
\end{align}
where the inequality is because $\mathcal{I}_{(M)}^\star$ is the optimal rank-$M$ approximation of $\mathcal{I}$. 
When each element in $\mathcal{I}_{(M)}^\star - \mathcal{I}$ follows a standard Gaussian distribution, the order of  $\lambda_{M+1}^2$ is $N$. Assuming that $\lambda_{d_0}^2$ and $\lambda_{M+1}^2$ are of the same order, the order of $	\|\mathcal{I}^\star(w) - \mathcal{I}\|_F^2$ is at least $N(d_0-M)$.
For any $i,j \in 1,\dots,N$ and $w \in \mathcal{W}$,
\begin{align*}
	\lvert P_{ij}^*(w) - P_{ij} \rvert
	& = \lvert f^{\prime \prime} (\widetilde{\Theta}) \left\{\Theta_{ij}^*(w) - \Theta_{ij}\right\} \rvert \\
	& = f^{\prime \prime} (\widetilde{\Theta}) \lvert \Theta_{ij}^*(w) - \Theta_{ij} \rvert \\
	& \geq \min_{\lvert \omega \rvert \leq \mu} f^{\prime \prime} (\omega) \lvert \Theta_{ij}^*(w) - \Theta_{ij} \rvert,
\end{align*}
where $\widetilde{\Theta}$ is a value between $\Theta_{ij}^*(w)$ and $\Theta_{ij}$. Then, for any $w \in \mathcal{W}$, we have
\begin{align*}
	L^*(w) = & \sum_{(i,j) \in \Psi_1} \left\{ P_{ij}^*(w) - P_{ij} \right\}^2 \\
	\geq & \left\{\min_{\lvert \omega \rvert \leq \mu} f^{\prime \prime} (\omega)\right\}^2 \sum_{(i,j) \in \Psi_1} \left\{\Theta_{ij}^*(w) - \Theta_{ij}\right\}^2  \\
	=& \left\{\min_{\lvert \omega \rvert \leq \mu} f^{\prime \prime} (\omega)\right\}^2 \|\Theta_1^*(w) - \Theta_1\|_F^2.
\end{align*}
When $\mu$ is a fixed constant, assume that $\|\mathcal{I}_1^\star(w) - \mathcal{I}_1\|_F^2$ has the same order of $\|\mathcal{I}^\star(w) - \mathcal{I}\|_F^2$.
Together with \eqref{Theta_1^*(w)} and \eqref{lambda+lambda}, the first part of Assumption 2 holds true under the condition $M = o(d_0-M)$, and the second part of Assumption 2 holds true when $\max\{D,\log N\}=O(d_0-M)$.

\subsection{Verifications of Assumption 3}

In this subsection, we consider the case where $M \geq d_0 $, that is, the candidate models include the correct models. The set $\mathcal{W}^s$ assigns all weights to the misspecified models. It means that the weights of candidate models with latent space dimensions not less than $d_0$ are all 0. Without loss of generality, we assume that $Z_{(m)}^* = U_mD_m^{1/2}$, and $Z_{(m)}^* Z_{(m)}^{*\top} = \sum_{m_0=1}^m Z_{\cdot, m_0} Z_{\cdot, m_0}^\top$, where $ Z_{\cdot, m_0}$ is the $m_0$-th column vector of $Z$ for $m=1,\dots,d_0$. For any $w \in \mathcal{W}^s$, we have 
\begin{align*}
	L^*(w) \geq & \left\{\min_{\lvert \omega \rvert \leq \mu} f^{\prime \prime} (\omega)\right\}^2 \sum_{(i,j) \in \Psi_1} \left\{ \sum_{m=1}^M w_m(\alpha^*_{(m),i} + \alpha^*_{(m),j} + z_{(m),i}^{*\top} z^*_{(m),j}) - \alpha_i - \alpha_j - z_i^\top z_j \right\}^2 \\
	= & \left\{\min_{\lvert \omega \rvert \leq \mu} f^{\prime \prime} (\omega)\right\}^2 \sum_{(i,j) \in \Psi_1} \left\{ \sum_{m=1}^{d_0-1} w_m(\alpha^*_{(m),i} + \alpha^*_{(m),j} + z_{(m),i}^{*\top} z^*_{(m),j}) - \alpha_i - \alpha_j - z_i^\top z_j \right\}^2  \\
	= & \left\{\min_{\lvert \omega \rvert \leq \mu} f^{\prime \prime} (\omega)\right\}^2 \sum_{(i,j) \in \Psi_1} \left\{ \sum_{m=1}^{d_0-1} w_m z_{(m),i}^{*\top} z^*_{(m),j} - z_i^\top z_j \right\}^2 \\
	= & \left\{\min_{\lvert \omega \rvert \leq \mu} f^{\prime \prime} (\omega)\right\}^2 \|\mathcal{I}_{d_0-1,1}^\star(w) - \mathcal{I}_1\|_F^2,
\end{align*}
where $\mathcal{I}_{d_0-1,1}^\star(w)$ denotes the matrices obtained from $\mathcal{I}_{d_0-1}^\star(w) = \sum_{m=1}^{d_0-1} w_m \mathcal{I}_{(m)}^\star$ by setting all entries of $\mathcal{I}_{d_0-1}^\star(w)$ with indices in $\Psi_2$ to 0. Note that 
\begin{align*}
	\|\mathcal{I}_{d_0-1}^\star(w) - \mathcal{I}\|_F^2 \geq \|\mathcal{I}_{d_0-1}^\star - \mathcal{I}\|_F^2 = \|Z_{\cdot, d_0} Z_{\cdot, d_0}^\top\|_F^2.
\end{align*}
When $\mu$ is a fixed constant, assume that the order of  $\|Z_{\cdot, d_0}\|_2^2$ is $N$, and $\|\mathcal{I}_{d_0-1,1}^\star(w) - \mathcal{I}_1\|_F^2$ has the same order of $\|\mathcal{I}_{d_0-1}^\star(w) - \mathcal{I}\|_F^2$, then the order of $L^*(w)$ is at least $N^2$ for any $w \in \mathcal{W}^s$. Assumption 3 holds when $M = o(N)$ and $\max\{D, \log N\}=O(N)$.

\bibliographystyle{agsm}

\bibliography{ref}